\documentclass[prd,eqsecnum,nofootinbib,floatfix,superscriptaddress,preprint,tightenlines,longbibliography]{revtex4-1}
\usepackage{amsfonts}
\usepackage{amsmath}
\usepackage{amssymb}
\usepackage{bm}
\usepackage[pdftex]{color}
\usepackage{graphicx}
\usepackage[sort&compress]{natbib}
\usepackage[colorlinks=true,linkcolor=Blue,filecolor=Blue,urlcolor=Blue,citecolor=Blue,pdftex,plainpages=false]{hyperref}

\definecolor{Gray}          {cmyk}{0,0,0,0.50}
\definecolor{Red}           {cmyk}{0,1,1,0}
\definecolor{NavyBlue}      {cmyk}{0.94,0.54,0,0}
\definecolor{RoyalBlue}     {cmyk}{1,0.50,0,0}
\definecolor{Blue}          {cmyk}{1,1,0,0}
\definecolor{Green}         {cmyk}{1,0,1,0}

\newcommand{\epj}[1]{#1}              

\newcommand{\bmin}{\begin{minipage}{0.5\textwidth}}
\newcommand{\emin}{\end{minipage}}
\newcommand{\bmini}[1]{\begin{minipage}{#1}}
\newcommand{\bc}{\begin{center}}
\newcommand{\ec}{\end{center}}
\newcommand{\co}{{\cal O}}
\newcommand{\coa}{{\cal O}^{(appx)}}
\newcommand{\coag}{{\cal O}^{(appx),g}}

\begin{document}

\title{Status and Future Perspectives for Lattice Gauge Theory Calculations to the Exascale and Beyond}
\author{B\'alint Jo\'o }
\email{Editor, \tt bjoo@jlab.org}
\affiliation{Theory Center, Thomas Jefferson National Accelerator Facility, Newport~News, VA 23606}
\author{Chulwoo Jung}
\email{Editor, \tt chulwoo@bnl.gov}
\affiliation{Physics Department, Brookhaven National Laboratory, Upton, NY 11973}
\author{Norman H. Christ}
\affiliation{Department of Physics, Columbia University, New York, NY 10027}
\author{William Detmold}
\affiliation{Department of Physics, Massachusetts Institute of Technology, Cambridge, MA 02139}
\author{Robert G. Edwards}
\affiliation{Theory Center, Thomas Jefferson National Accelerator Facility, Newport~News, VA 23606}
\author{Martin Savage}
\affiliation{Institute for Nuclear Theory, University of Washington, Seattle, WA 98195-1550}
\author{Phiala Shanahan}
\affiliation{Department of Physics, Massachusetts Institute of Technology, Cambridge, MA 02139}

\collaboration{USQCD Collaboration}
\noaffiliation

\date{\today}

\begin{abstract}
In this and a set of companion whitepapers, the USQCD Collaboration lays out a program of science and computing for lattice gauge 
theory.
These whitepapers describe how calculation using lattice QCD (and other gauge theories) can aid the interpretation of ongoing and 
upcoming experiments in particle and nuclear physics, as well as inspire new ones.
\vfill
\noindent\includegraphics[width=\textwidth]{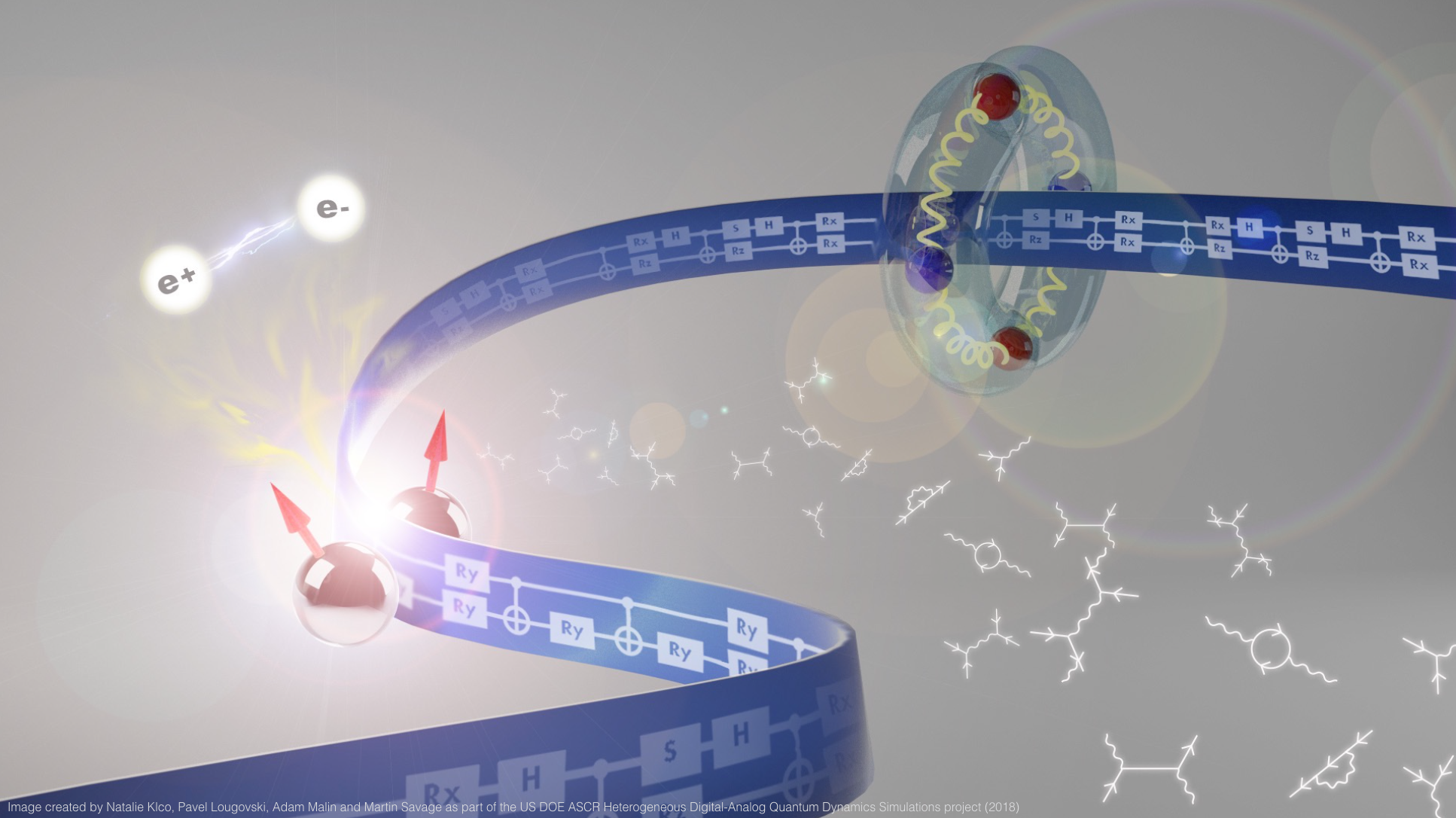}
\end{abstract}

\maketitle
\newpage
\section*{Executive Summary}

In 2018, the USQCD collaboration’s Executive Committee organized several subcommittees to recognize future opportunities and formulate possible goals for lattice field theory calculations in several physics areas.  The conclusions of these studies, along with community input, are presented in seven whitepapers~\cite{Bazavov:2018qcd,Brower:2018qcd,Davoudi:2018qcd,Detmold:2018qcd,Kronfeld:2018qcd,Lehner:2018qcd,Joo:2018qcd}.

Numerical studies of lattice gauge theories in general---and of lattice quantum chromodynamics~(lattice QCD) in
particular---have been a driver of high performance computing (HPC) for nearly 40 years.
Lattice-QCD practitioners have innovated in the algorithmic, hardware, and performance space with an impact that has reached
substantially beyond the field.  Examples include influence on supercomputer architectures such as the IBM Blue Gene line of systems,
development of algorithms that are used in other domains, such as hybrid Monte Carlo, and early adoption of new technologies such as graphics processing units.

The power of computers continues to increase.  At the same time, the adoption of novel algorithmic approaches such as better preconditioners, improved linear and eigensolvers,
more efficient molecular dynamics time integrators, and more powerful
boost the available statistics.
Thus, the scientific output of lattice-QCD calculations has far exceeded the growth one would expect purely from
hardware speed-up alone.

These advances have been supported in two main ways, under the umbrella of the \href{https://www.usqcd.org}{USQCD
Collaboration}.
One consists of software development through successive generations of the United States Department of Energy (DOE) Scientific Discovery through Advanced
Computing (SciDAC) program, and now also by the DOE Exascale Computing Project (ECP).  The other is a series of infrastructure projects, known as LQCD, supported by the DOE Office of High Energy Physics and the Office
of Nuclear Physics.
These efforts have led to large-scale cooperation with colleagues at large-scale and leadership-class computing facilities~(LCFs),
as well as industrial partners.  U.S. lattice-QCD researchers has invested in and fostered a community with strong HPC
expertise, which allows them to be energetic participants in the drive towards exascale computing,
with the 
USQCD Collaboration forming a very active component within the ECP.

Lattice-QCD has been an early adopter of many disruptive technologies and its community expertise in the use of
heterogeneous, multi-core and accelerated architectures at leadership computing facilities may be of benefit to experimental
colleagues.

The community is also working actively to develop approaches which can successfully bring artificial intelligence and machine
learning approaches to our computational toolkit where appropriate, and efforts are beginning to develop quantum computing methods
for lattice-QCD. 

After an introduction to lattice-QCD in section~\ref{sec:intro}, a description of lattice-QCD workflow is given in section~\ref{sec:sim_campaigns}. 
Section \ref{sec:gaugegen}-\ref{sec:contract} give more detailed descriptions of algorithms, techniques and challenges in each major part of lattice-QCD worflow: gauge generation, propagator generation and correlation function construction. 
We discuss the hardware landscape as we see it currently going forward to the Exascale in section~\ref{sec:hw}, and discuss the software technologies and
strategy that underlie our successful exploitation of leadership as well as smaller scale departmental sized computer systems in section~\ref{sec:sfw}.
We conclude the whitepaper with a section on the potential use of machine learning techniques (section \ref{sec:ml}) and an outline of the state of the field in its approach to research in quantum computing (section \ref{sec:quantum}).

\newpage

\section{Lattice QCD and Leadership Computing}\label{sec:intro}

\subsection{Introduction}

Lattice quantum chromodynamics is the only known, model-independent, non--perturbative method available for the evaluation of path
integrals in quantum chromodynamics (QCD), and as such plays a vital role in modern theoretical high energy and nuclear physics.
lattice-QCD calculations underpin vital research, for example testing the limits of the Standard Model of particle
physics, aiding in the understanding of the production and properties of hybrid meson resonances and many other areas.
The focus of this whitepaper is a description of the state of the art in computational technology underpinning such calculations, as
well as further research opportunities in the computational arena to benefit future calculations on forthcoming pre-exascale and
exascale systems.
We also consider rapidly developing new areas of computation such as machine learning (ML) and Big Data approaches and quantum
computing.
We refer the reader to the companion whitepapers for details of research opportunities in various areas of nuclear and high energy
physics \cite{Bazavov:2018qcd,Detmold:2018qcd,Davoudi:2018qcd,Kronfeld:2018qcd,Lehner:2018qcd,Brower:2018qcd}.

In this initial section, we give a very high level overview of the interactions of lattice-QCD and computing and the main
topics of this whitepaper, which are then expanded upon in (occasionally technical) detail in the subsequent sections.
We will detail the mechanics of lattice-QCD workflows in Sec.~\ref{sec:sim_campaigns} outlining the main stages of the
computation and commenting on the primary computational features of the stages.
In Sec.~\ref{sec:gaugegen} we will discuss current lattice-QCD generation algorithms and advances including topics of current
research, which we will follow in Sec.~\ref{sec:solvers} with a discussion of linear solvers and eigensolvers.
Thereafter we will discuss {\em correlation function construction} in Sec.~\ref{sec:contract}, present our view of the current and
upcoming hardware systems in Sec.~\ref{sec:hw} and our software efforts to exploit them in Sec.~\ref{sec:sfw}.
We will round out our overview with details of nascent research into machine learning and quantum computing in Sec.~\ref{sec:ml} and
Sec.~\ref{sec:quantum} respectively.

\subsection{Evaluating lattice QCD path integrals}
Through the process of discretizing four-dimensional spacetime onto a regular lattice, path integrals are turned into \epj{a
high-dimensional} but countable set of integrals.
Due to analytical continuation to Euclidean time, the path integral weight of a
configuration of fields can take on an interpretation as a Boltzmann-like probabilistic weight.
Mathematically, the system to be solved becomes similar in nature to a crystalline system or spin-glass such as one may encounter in
condensed-matter physics, and the path integrals themselves can be evaluated numerically through Monte Carlo methods.
The probabilistic weight due to dynamical quarks corresponds to the determinant of the respective fermion kernels which are
prohibitively expensive to evaluate directly.
As such, these determinants are universally simulated by expressing the determinant as auxiliary integral over so called
pseudofermion fields, which can be evaluated in the same process as the main path integral.
 
\subsection{Exploiting leadership computers and new architectures}
The cost of such calculations is formidable and in the pursuit of ever more realistic calculations, lattice-QCD practitioners
have been working at the forefront of high performance computing (HPC) for over 40 years contributing in aspects as diverse as {\em
building custom supercomputers}, {\em carrying out architecture specific code optimizations}, {\em inventing and contributing to new
simulation algorithms}, adopting {\em software engineering best practices} and most recently applying other practices such as
opportunistic running through workflow systems using similar tools as colleagues in experiment.
Indeed, lattice-QCD has turned into a mature science with complex simulation campaigns carried out in an industrial fashion.
Current sophisticated calculations have been carried out over a long period of time at large-scale facilities such as Oak Ridge
Leadership Computing Facility (OLCF), Argonne Leadership Computing Facility (ALCF) and the National Energy Research Scientific
Computing Center (NERSC), as well as using dedicated cluster facilities at Thomas Jefferson National Accelerator Facility (Jefferson
Lab), Fermi National Accelerator Laboratory (Fermilab) and Brookhaven National Laboratory (BNL) in a coordinated fashion.
An important aspect when using so many resources is the automation of the analysis portion of calculations where lattice-QCD
researchers have formed a partnership with the ATLAS PanDA~\cite{PANDA} team at BNL to bring workflow technologies such as the ones
used in experimental big-data analysis into regular use for lattice-QCD calculations.
 
 \subsection{Lattice-QCD spin-offs into other science and technology domains}
 
The innovations in lattice-QCD research have had several spin-offs in other areas of science. The hybrid Monte-Carlo
algorithm~\cite{Duane:1987de} used to generate lattice gauge configurations with dynamical fermions has found many other
applications, including machine learning~\cite{e2544badcbd543a481cc0bdf45041dc4}, the study of protein structures, e.g.,
\cite{SchuetteFischerHuisingaetal.1999, doi:10.1002/bip.360330815}, and in the analysis of financial time series e.g.,
\cite{2017arXiv171202326D,2008arXiv0807.4394T}. Two of the same authors also developed a precursor of Riemann manifold
HMC~\cite{Duane:1988vr} more than 20 years before it was fully developed in~\cite{RSSB:RSSB765}.

\subsection{Designing and utilizing novel hardware in partnership with industry}
\marginpar{\vspace*{-3.5em}}

Lattice-QCD practitioners have built or were involved in building several custom computers designed to carry out lattice QCD
simulations efficiently including early systems built at Columbia University, the GF11 supercomputer built at
IBM~\cite{Beetem:1985:GS:327070.327139}, the APE series of computers in Europe e.g.~\cite{AGLIETTI1998216} and relatively recently
the QCDSP systems built at Columbia University (CU) and Brookhaven National Laboratory (BNL) and the
QCDOC~\cite{Boyle:2005:OQQ:1665957.1665969} systems built as collaborative projects between CU, BNL, the RIKEN-BNL Research Center
(RBRC) and Edinburgh University.
The QCDOC system in particular was a sister project of the IBM Blue Gene/L~\cite{BGL} supercomputer, with the two systems having
overlapping and collaborating design teams and having shared several pieces of co-designed hardware elements.
Work under contract continued with IBM to develop
the subsequent Blue Gene/P and Blue Gene/Q systems~\cite{Boyle:2012iy}.  Lattice-QCD researchers now work closely with companies such as Intel, Nvidia, HPE, and
SGI in order to maximally exploit up and coming hardware offerings.

Lattice-QCD researchers in Europe started to exploit graphics processing units (GPUs) as early as 2007 writing code
over computer graphics interfaces \cite{Egri:2006zm} and in the U.S. since 2008--2009.  The then newly released CUDA programming API was used to produce the QUDA library \cite{Clark:2009wm,%
Babich:2010mu,Babich:2011np,Clark:2016rdz,QUDADownload}.
QUDA has become the basis for the exploitation of GPUs by USQCD, as discussed below in
Sec.~\ref{sec:sfw}.
Jefferson Lab fielded the first large-scale, GPU-accelerated cluster designed specifically for lattice QCD
in 2009.  Thus, the lattice-QCD community was ready when large-scale GPU resources
 appeared in 2012 on the Titan and BlueWaters systems at the OLCF and at the National Center for Supercomputing
Applications (NCSA).
Since then, the lattice-QCD community has formed strong partnerships with Nvidia (with several lattice-QCD
researchers having found careers there), and with Intel to produce highly optimized codes to run on the emerging Intel Xeon Phi
Knights series of computers working successively with Knights Ferry, Knights Corner and most recently Knights Landing (KNL),
which powers several large-scale systems including the Cori system at NERSC and the Theta system at the ALCF.
Several lattice-QCD researchers received the prestigious Gordon Bell Prize for High Performance Computing in 1998
(price/performance) and in 2006 (Special Achievement Award).
Lattice-QCD software was a Gordon Bell finalist in 2018.

\subsection{Lattice QCD software}

Lattice QCD in the U.S.
has greatly benefited from continuous funding through successive iterations of the {\em Scientific Discovery through Advanced
Computing} (SciDAC) program of the US Department of Energy (DOE), Office of Science.
Through this funding an extremely capable set of codes and libraries have been developed that allow lattice-QCD practitioners
to exploit the architectural diversity of currently available computing hardware.
Most lattice-QCD codes are written in a mixture of C and C++, making use of on-node threading (typically with OpenMP on
multi-core CPUs and CUDA on Nvidia GPU accelerators) and internode communications with message passing between nodes (and sometimes
within them).
Lattice-QCD codes attempt to leverage development best practices including the use of distributed version control systems, and
regular testing.
Indeed, some pieces of software support full continuous integration testing.

\subsection{Towards the exascale}
As we stand at the dawn of the exascale era, the USQCD Collaboration is participating in the U.S.\ Exascale Computing Project (ECP),
to ensure readiness of our codes for the forthcoming generation of computers.
The USQCD contribution encompasses many areas including the improvement of gauge field generation, research in the areas
of linear solver and related algorithms (e.g., eigensolvers, and trace estimation), a re-tooling of the software infrastructure
paying special attention to upcoming architectures, modularization and layering of software components and their interoperability,
novel computing languages, and new programming models to provide productivity and performance portability.

\subsection{Machine learning and quantum computing}

Recent trends in high performance computing have raised to prominence the rapidly developing areas of machine learning, big data
analysis, and quantum computing.
Deep learning has major commercial applications that are influencing both hardware (such as the tensor cores on the Nvidia Volta
V100 GPU, or the increase in support for low precision arithmetic on the Intel Xeon Phi Knights Mill architecture) as well as how
some areas of the scientific enterprise are carried out.
While machine learning has not yet been applied at a large scale to lattice-QCD simulation, it offers several areas where it
can assist in optimizing simulation campaigns.

Finally, although the use of quantum computing for production lattice-QCD calculations is likely still several years away, the
community is already investing resources to be ready for its exploitation, working with early quantum computing systems.

\subsection{Summary} 
In summary, lattice-QCD calculations are, and have historically been, at the very forefront of high performance computing.
Research topics range from areas as diverse as hardware architecture design, applied mathematics, software and workflow
technologies, to machine learning and quantum computing.
lattice-QCD has had several algorithmic spin-offs into other disciplines and maintains close links with industrial partners
for co-design and early adoption and exploitation of new hardware technologies.
Computational lattice-QCD is well poised to utilize the very first exascale and pre-exascale systems and is already
undertaking research into exploiting the newly emerging area of quantum computing.

\section{Lattice-QCD Calculations Need Speed and Throughput}\label{sec:sim_campaigns}
Lattice-QCD calculations are generally comprised of three main parts.
The first step is {\em the generation of ensembles of gauge field configurations}, which is a Monte-Carlo sampling of the strong
force fields in the vacuum, including all the effects of gluonic self interactions and the interactions of gluons and sea quarks.
The ensembles are parameterized by the {\em strong coupling} and the {\em masses of the sea quarks}.
These parameters implicitly set the lattice spacing.
The second step of the process is the computation of quantities of interest (known as {\em observables}) which for quantities
involving quarks generally includes the computation of {\em quark propagators}.
The computation of an observable is also referred to as ``measurement''.
We discuss all these stages in detail in subsequent sections, and will restrict our focus here to their computational
properties.

\subsection{Gauge Generation}
Generally several ensembles need to be generated, with a range of sea quark masses, lattice spacings and volumes, in order to allow
controlled extrapolations between ensembles to the physical quark masses, the continuum limit and infinite volume.
While calculations directly at the physical quark masses have recently become feasible, heavier mass ensembles may be required, for
example, for parameter tuning purposes or to map the quark mass dependence of quantities of interest, depending on the calculation

Within each ensemble successive configurations are generated from preceding ones via a Markov process, so that the probability of a
configuration occurring is proportional to its probabilistic weight in the path integral ({\em importance sampling}).
In this case, measurements can simply be averaged over the configurations in the ensemble to form the
estimators for the path integrals.
The statistical uncertainty of these {\em ensemble averages} depends on the number of independent configurations, $N_{\rm cfg}$, in
the ensemble, and decreases as $1/\sqrt{N_{\rm cfg}}$.
As such the statistical precision can be controlled by increasing the number of configurations in the ensemble, so that given
sufficient computer time high-precision calculations are possible, for example, as are needed to test the
Standard Model.

The primary computational cost of gauge generation is the cost of a single Markov update step during which time the lattice Dirac
equation may need to be solved hundreds of times---as detailed in Sec.~\ref{sec:gaugegen}---and at various quark masses.
Due to the sequential nature of the Markov process, the only parallelism available is the data parallelism arising between the
lattice sites, making gauge generation a {\em strong scaling} challenge.
In current advanced simulations, a single Markov step may take between 10 minutes and 1 hour on on hundreds or thousands of GPU
devices on a leadership system such as OLCF Titan.
One generally saves the state of the fields every couple of steps, resulting in a data access pattern of writing one configuration
of size a few tens of GB every hour or so.

\subsection{Propagator Computation}
The second step in the process is the computation of quark propagators.
This is akin to releasing a test charge in an electric field and following its progress along the field lines, but a color charge is
used instead of an electric one in the background of the strong force (gauge) field, and rather than ``following the movement'' of a
quark, an object known as a quark propagator is generated.
The quark propagator is labeled at each end by indices corresponding to the position, color, and other quantum numbers.
These indices later need to be contracted into ``colorless'' {\em correlation functions} in a step known as {\em correlation function
construction}.
The latter two steps are deeply related, depending on the nature of the observables at hand, since that dictates the number and
nature of propagators which need to be computed.
From a workflow point of view however, it is worth separating them, as the propagator calculation and contraction steps have
somewhat different computational characteristics.

The propagator calculation step involves solving the lattice Dirac equation for a variety of color sources.
Depending on the observables, O(100k--1M) propagator solves may need to be carried out on each configuration in a given ensemble.
Here one can use the data parallelism available from the lattice, but also since each gauge configuration in an ensemble, and each
quark source are treated as being independent, one has access to a large degree of comfortable parallelism.
This step is very much in the vein of {\em industrial computing}.
Rather than being a strong-scaling challenge {\em this step is gated by throughput}.
Typically O(10--100) GPUs are used per solve instead of O(1000), and O(10--100) separate propagator computations can be run in a
single job, on a leadership class system potentially using up to the whole system in ensemble mode.
Since smaller partitions are needed for each solve, {\em mid-range} cluster resources built out of commodity white box components
and having potentially less capable, {\em but substantially cheaper} communication fabrics than the leadership systems have
proved extraordinarily cost effective for this step of calculations.

Due to their size, leadership systems can provide higher instantaneous throughput at any given time, potentially shortening time to
solution.
Ultimately, however, the overall progress on either type of system are gated also by the sizes of computer time allocations.
Typical propagator runs in the area of quark propagators been able to regularly use O(2000-8000) GPU nodes of a
system such as Titan at OLCF, or indeed several 1k--4k partitions of systems like the Intrepid IBM Blue Gene/P system (and its
predecessors) at ALCF.

This phase of the calculation is entirely dominated by a variety of linear and eigensolvers, which will be discussed further in
Sec.~\ref{sec:solvers}.
Further, this stage has a very different I/O pattern from gauge generation.
In the case of Wilson fermions, a propagator component is about one third the size of a gauge field.
Hence in a several hour job, with perhaps O(10k) solves, there is the potential to generate of O(10-100TB) of data.
This is generally not feasible to save outright and typically some amount of {\em in-situ data reduction} is carried out.
Hence this step is characterized by an ensemble style, with a high degree of output I/O that is generally written in a temporally
structured pattern, and database technologies may well be used to store the output data.
Due to the high amount of I/O this stage can also benefit from burst buffer technologies and potentially I/O staging and in-memory
coupling with subsequent analysis steps.

\subsection{Correlation function construction}
The final contraction step generally uses the outputs of the propagator stage, to generate correlation functions, which can then be
subjected to statistical analysis (fitting, error estimation, etc.).
Typically, this stage is also carried out in an ensemble style.
However, it is distinct from the propagator calculation, since the contractions typically need to be performed only over
3-dimensional space for a given value of Euclidean time.
Hence, one can in principle often exploit parallelism among the Euclidean time-slices and codes also need to access 3-dimensional
subsets of the original data.
Since one needs to work generally on one time-slice at a time, the contraction codes are typically single node using at most thread
level parallelism on multi-core processors.
These codes have generally not yet been transferred to GPUs to such an extent as the preceding stages and exploiting more recent
architectures for this work is a critical target of current development projects.
These codes have yet again different I/O patterns, requiring essentially random read access to the products of the propagator
calculation step, and can also benefit from database technologies, burst buffers and potential in-memory coupling to the preceding
propagator calculation steps.

While gauge generation campaigns are typically carried out by simulating only a handful of ensembles concurrently over long periods
of time, propagator and correlation function calculations employ a much higher level of ensemble parallelism and campaigns need to
manage thousands of jobs.
Workflow technologiues such as PANDA~\cite{PANDA} or home-grown ones such as~\cite{Ayyar:2018wwf} present opportunities to have better control over the analysis campaigns, for example to
automatically schedule and distribute jobs, restart failed ones and to exploit backfill cycles where available.

\subsection{Simulations of QCD at finite temperature and density}
An important exception to the pattern discussed above are calculations carried out at finite temperature.
In these calculations, finite temperature is achieved by having shorter Euclidean time extents, so lattices are typically 
smaller than in the zero-temperature calculations, discussed above.
In these situations, one can in many cases carry out both the gauge generation and the analysis part of the calculation using single
nodes, and the key to progress comes from carrying out ensemble jobs to perform parameter sweeps, for example, to find phase
transitions, critical points or to determine quantities as a function of temperature or density.
In turn, finite-temperature codes have made perhaps best use of the hardware available in hybrid, accelerated nodes, by being able
to separately use the GPUs for one part of their work (e.g., propagator calculation) while using the CPUs
concurrently to make progress on other aspects (e.g., gauge generation).
Due to their ensemble nature, and the requirements of using only a single node (or at most a few nodes) per task, these calculations
can also derive a large benefit from workflow engines, and especially if these can exploit the backfill cycles of large systems such
as those at leadership facilities.
As finite-temperature calculations advance, the spatial lattices are expected to grow in size, and it will be necessary to move
beyond large ensembles of single node jobs, to ensembles of multi-node jobs.
Preparing code for this future is a focus of the DOE Office of Nuclear Physics SciDAC-4 project.

\subsection{Impacts of hardware architecture trends}
The current trend in leadership computer systems is for individual compute nodes to attain denser and denser
floating-point capability, resulting in systems with fewer, more floating point capable nodes.
At the same time, interconnect speeds have grown more slowly, which can present challenges for strong-scaling problems such as gauge
generation.
Simply put, a single stream of gauge generation may not be able to strong scale as effectively, to as many nodes as before.
As a result, one can anticipate that gauge generation will also undergo a transformation to a more ensemble style calculation where
(ideally) several 
independent Markov chains run in parallel.  Each would use a larger partition than is used for propagator
calculations, but with a single partition not yet taking up a very large fraction of a leadership system.
Parallelism between Markov chains will be required to use the leadership hardware most efficiently.
In this situation, gauge generation will become a much more complex task to manage in terms of human time, and it therefore will benefit from suitable workflow systems.

Finally, we note that lattice-QCD calculations are voracious, since the statistical precision is limited principally by the
number of configurations in an ensemble and by the number of propagators one can compute on these configurations.
As such, it is highly unlikely that the full needs of very high precision lattice calculations can be met by a single computer
system even in the exascale era.
It is highly likely that just as now, work will proceed through coordinated use of several sites, including ASCR facilities as well
as local, institutional, or collaboration wide resources.
To facilitate this, high speed networks and efficient data transfer tools (such as Globus~\cite{Foster:2011:GOA:1978245.1978305})
are essential and are comparable in need to those of high-energy or nuclear physics experiments.

\section{The Computational Casino: Gauge Generation}\label{sec:gaugegen}
\subsection{Overview}

In this section, we consider details of the gauge generation.
As mentioned previously, the overall gauge generation process is a Markov chain, which generates each successive configuration from
the previous one.
As a large ensemble of independent configurations is required at each set of physical parameters, it is typical to run a few chains
simultaneously.
Successive configurations are typically correlated, and the number of Markov steps that must be taken before two configurations can
be considered independent is characterized by some form of {\em autocorrelation time}, of which we typically consider two kinds.
First is the {\em integrated autocorrelation time} for an observable, which is used to inflate the statistical error for the
observable to take into account the autocorrelations between the configurations.
The second is the {\em exponential autocorrelation time}, which is the maximum of all integrated autocorrelation
times in the Markov process.
They provide a guide how many Markov steps one needs to take, starting from a given position, before one can
consider that one is sampling the desired equilibrium probability distribution properly.

The process of equilibration therefore is a modest up-front cost to generating an ensemble (typically a few hundred to a thousand
updates) which needs to be paid at least once.
While the ideal approach is to generate one single very long stream to minimize the cost of equilibration and the effects from
rapidly increasing integrated autocorrelation times in simulations with ever finer lattice spacings (see Sec.~\ref{sec:CSD}), the
serial nature of ensemble generation can make the effective utilization of very large partitions on current large scale computing
resources challenging due strong scaling constraints and as a result it is not uncommon to generate up to 4--5 streams per set of
physical parameters in parallel.
After an initial chain is deemed to have been equilibrated, other streams are generally split off from it.
Each of these streams must also decorrelate from the main stream before its configurations can be considered independent.

As an example, in the USQCD Wilson-clover program, typically one aims for 6000--10000 Markov updates comprised of configurations in
4--5 ensembles, with an initial equilibration of approximately 1000 updates in the first ensemble and around 200--300 in each
branching ensemble.
This typically yields around 400--500 usable configurations which are still potentially correlated, so that binning and other error
estimation techniques are used to estimate the true statistical uncertainties.

Since the parallelism of Markov chains is limited, one must rely on data parallelism provided by the lattice sites and the fact that
the lattice gauge theories are local.
Within a given chain, it is not possible to change the lattice volume (which would change the integral the Monte Carlo is
evaluating).
Further, while the typical cost of a simulation scales only mildly with the number of lattice sites (the volume), it scales in a
much worse fashion with other physical parameters such as reducing lattice spacing or the light-sea-quark masses.

In order to focus the power of the computing elements onto these more challenging factors, rather than on increasing the volume, the
primary scaling metric for gauge generation is {\em strong scaling}.
In other words, as one scales to an increasing number of compute nodes, the global problem size remains unchanged and
correspondingly, the problem size on each individual compute node decreases.
Since the predominant communication patterns are nearest-neighbor (stencil-like) boundary exchanges and global sums, decreasing
local problem size will engender a worse surface-to-volume ratio for each compute node. This results in decreased opportunity to
{\em overlap local computation with communication} and a greater exposure to bottlenecks arising from communication latencies and
bandwidths.
Further, the evolution of hardware has favored increased single-node
floating-point capability and memory bandwidth, while internode fabric capabilities have not advanced at a comparable rate, making
the problem of strong scaling even worse on recent architectures.

In order to address the strong-scaling concerns, a great deal of current research is focused on communication reduction and
avoidance in one form or another.
Typical approaches include using domain-decomposition oriented preconditioners for linear solvers, communicating
nearest-neighbor boundary data in reduced precision (requiring less bandwidth), and using
communication-reduction oriented (``$s$-step'') solvers which reduce the number of reduction points and, hence latency effects.

One of the recent advances in lattice QCD has been the development of adaptive multi-grid methods for linear solvers for some fermion
actions.
Typically such solvers provide a near order of magnitude improvement over the best available implementation of regular Krylov
solvers such as conjugate gradients~\cite{citeulike:9077321} or BiCGStab~\cite{BiCGStab} for light quark masses.
However, for these solvers to be most effective they require a setup phase that can be computationally expensive, and they are more
constrained by strong scaling on their coarser grids.
In order to exploit multi-grid solvers for gauge generation, both these issues will need to be confronted as we shall describe in
Sec.~\ref{subsec:mgGaugeGen}.

Finally, autocorrelation times are known to increase as the lattice spacing is reduced, leading to a phenomenon of {\em critical
slowing down}.
This is important as one performs calculations on ever finer lattices in order to give a better lever-arm for continuum
extrapolations.
Thus, in addition to combating the effects of strong scaling, a large component of research is focused on new
Monte Carlo techniques that reduce this critical slowing down, as discussed in Sec.~\ref{subsec:criticalSlowingDown}.

In order to put these statements into context, in the following subsections we will describe today's workhorse algorithm of hybrid
Monte Carlo.
We will discuss our state of the art implementation for so called Wilson clover fermions (although the general principles are
the same for any fermion action) including algorithmic improvements and the use of the multi-grid
algorithm and will show the benefit of these improvements on the Summit and Titan systems at OLCF.

\subsection{Hybrid Molecular Dynamics Monte Carlo}
Hybrid Monte Carlo (HMC)~\cite{Duane:1987de}, often known as Hamiltonian Monte Carlo, belongs to a class of algorithms referred to as
hybrid molecular dynamics Monte Carlo (MDMC).
Treating the SU(3) link matrices of a gauge field as canonical coordinates, the methods proceed by ascribing {\em canonically
conjugate momenta} and generating a Hamiltonian system, with 
\begin{equation}
    H(\pi,U) = T(\pi) + S(U),
\end{equation}
where $T(\pi)$ is a kinetic energy term depending on the momenta, and the potential term $S(U)$ is the action to simulate.
By drawing $\pi$ on each lattice link from a Gaussian heat-bath, the kinetic energy term maintains its familiar form 
$T(\pi)=\frac{1}{2}||\pi||^2$.
One can generate new configurations from old ones by performing {\em Hamiltonian molecular dynamics} (MD) time integration
of Hamilton's equations in a fictitious simulation time.
Starting from some state with momenta $\pi'$ and gauge field (coordinate) $U'$, a new state $(\pi,U)$ is generated by MD and is then
either accepted or rejected with a Metropolis~\cite{METR1953} acceptance probability.
If the trial state is rejected the original $(\pi', U')$ becomes the next state in the chain.

In order for the process to work, it must be both {\em ergodic} and have the required equilibrium probability as its {\em fixed
point}.
Since the MD is energy conserving, the energy change along a trajectory and the resulting acceptance probability are determined
solely by the truncation error in the numerical MD integration algorithm, which can be controlled by changing the integration
step-size.
However in this case the system is integrated on a single hypersurface of (nearly constant) energy.
In order to ensure ergodicity, the momenta are refreshed regularly from a heat--bath which has the effect of moving the system to a
different energy hypersurface.
In the regular HMC algorithm, momentum refreshment is done before every trajectory but more elaborate approaches to momentum
refreshment have been proposed (such as generalized hybrid Monte Carlo~\cite{Horowitz:1991rr}).

In order for the desired equilibrium probability to be the {\em fixed point} of the Markov chain, it is sufficient (but not
necessary) to ensure that the {\em detailed balance} is maintained by the algorithm.
This can be ensured by using {\em reversible time-integration} methods which also preserve the integration measure.
The requirements of reversibility and area preservation limit us in our choice of integrators and in practice they are satisfied by
utilizing a reversible combination of {\em symplectic integrator steps}.
Commonly used integrators are the second order leapfrog, second order minimum norm (2MN)~\cite{Takaishi:2005tz}, fourth order
minimum norm (4MN)~\cite{Omelyan, Takaishi:2005tz}, and more recently force-gradient~\cite{Kennedy:2009fe,Clark:2011ir,Yin:2011np} integration schemes, which over a trajectory of unit length, have truncation
errors of $O(\delta \tau^2)$ and $O(\delta \tau^4)$ respectively where $\delta \tau$ is the integrator step-size.
As the lattice volume grows, generally one needs to take smaller steps to maintain a constant acceptance rate.
It can be shown that for an integrator that scales as $O(\delta \tau^n)$ for some integer power $n$, the numerical cost of the
algorithm scales as $O(V^{1+1/2n})$.

In order to include quarks, one employs the method of pseudofermions.
In this case, the Grassman integrals over quark fields are carried out explicitly and each fermion flavor adds a determinant weight
into the equilibrium probability of a given configuration.
Evaluating the full matrix determinant is computationally prohibitive, and instead the determinant terms are expressed as an
integral over bosonic fields, making use of the identity for two degenerate flavors of quark, with fermion kernel $M$:
\begin{equation}
    \det( M^\dagger M ) \propto \int d\phi^\dagger d\phi \ e^{- \phi^\dagger \left( M^\dagger M \right)^{-1} \phi} 
\end{equation}
where the $\phi$ fields are known as pseudofermion fields.
These integrals are folded into the main HMC process, by refreshing the pseudofermion fields from a heatbath at the start of every
update and carrying out the integrals stochastically over the whole simulation.
Typically the quarks are considered in degenerate pairs since the combination of $M^\dagger M$ in the integrals is manifestly
Hermitian positive definite (HPD) and the resulting determinant is manifestly real.
One exception is domain-wall fermion formalism where the determinant for a single flavor can be rewritten to be manisfestly
HPD~\cite{Chen:2014hyy,Jung:2017xef}, without the need for the squared operator.

To simulate flavor combinations that cannot be expressed as $M^\dagger M$ or in other HPD ways, one can take roots of the squared
term which are computed using an approximation such as the {\em optimal rational approximation} expressed in partial fraction form,
{\em e.g.,} for the square root, as:
\begin{equation}
\phi^\dagger \left( M^\dagger M \right)^{-1/2} \phi = A \sum_{i=1}^{N} p_i \phi^\dagger \left( M^\dagger M + q_i \right)^{-1} \phi
\end{equation} 
where $N$ is the order of the approximation and $A$, $p_i$ and $q_i$ are approximation coefficients.
This approach is referred to as rational hybrid Monte Carlo (RHMC)~\cite{RHMC}.
Other approximation schemes are possible for example by using Chebyshev polynomial approximations in the polynomial hybrid Monte
Carlo algorithm (PHMC)~\cite{Frezzotti:1997ym}.
%
%

The molecular dynamics algorithms are composed of updates to the momenta and gauge field combined in a reversible manner.
The momentum update needs to evaluate the force term resulting from the action, which for fermionic terms ultimately results in the
need to solve the Dirac equation in several forms (discussed in Sec.~\ref{sec:solvers}).
These computations require the use of linear solvers, and the vast majority of the time spent in the MD updating is spent in force
computations and in linear solvers.

\subsection{State of the art algorithms}
Current state of the art gauge generation codes employ an impressive battery of algorithmic tricks.
As quark masses approach their physical values the linear system to be solved for the light quarks becomes increasingly more ill
conditioned, and the resulting forces grow in size requiring finer and finer time-steps~\cite{UKAWA2002195}.
To overcome this, simulations employ {\em mass preconditioning}~\epj{\cite{Hasenbusch:2001ne,Hasenbusch:2002ai,Urbach:2005ji}} by breaking up the light quark determinant,
into a chain of auxiliary determinants as
\begin{multline}
    \det ( M^\dagger M ) = \det\left[ \frac{M^\dagger M}{M_0^\dagger M_0} \right] \,
        \det\left[ \frac{M_0^\dagger M_0}{M_1^\dagger M_1} \right] \ldots\\
        \det\left[ \frac{M_{n-1}^\dagger M_{n-1}}{M_n^\dagger M_n} \right] \,
        \det\left[ M_n M_n \right]
\end{multline}
where we use the shorthand notation
\begin{equation}
    \left[ \frac{ M_i^\dagger M_i }{ M_{i+1}^\dagger M_{i+1} }\right] =
        M^{-1}_{i+1} \left[ M^\dagger_i M_i \right] ( M^\dagger_{i+1})^{-1}
\end{equation}
and $M_i$ and $M_{i+1}$ are slightly different in some way, for example by having a slightly different quark mass.
Since matrices in each term are identical except up to a small perturbation, the ratio terms will be close to the identity matrix
with a small perturbation.
As such, the force terms which result from the small perturbation will also be small.
The final term, which cancels off the effects of the chain is a regular two flavor term, but is no longer at a light mass and can be
integrated with reasonable cost.
In order to take advantage of mass preconditioning one needs to use {\em a multiple time-scale integrator}~\cite{Sexton:1992nu}
which allows each term in the action to be integrated on a separate time-scale.
Terms with small forces can be integrated with fewer long steps (and fewer force evaluations), while terms with large forces (such
as the cancellation two flavor piece or the gauge forces) need many fine time-steps, but their evaluation is considerably less
expensive numerically.
Heuristically time step sizes can be selected using the infinity norms of the associated MD forces~\cite{Urbach:2005ji}, and can be
more formally understood and tuned using the technology of {\em Poisson brackets}~\cite{Kennedy:2009fe,Clark:2011ir}.

A recent advance in this area has been the introduction of force-gradient integrators~\cite{Kennedy:2009fe,Clark:2011ir,Yin:2011np} which can give a fourth
order accurate integrator with fewer linear system solutions per time step needed than the previously discussed fourth order minimum
norm integrator.
Force-gradient integrators can be understood using the framework of shadow Hamiltonian methods and Poisson brackets, however a
particularly elegant implementation trick has been discovered~\cite{Yin:2011np}, which makes their implementation surprisingly
straightforward.
Extending them to multiple time scales is also possible.

Finally to reduce the cost of linear systems even further one can utilize so called {\em chronological predictors}.
These components attempt to provide a good initial guess to the current system to be solved based on the solutions from previous
steps along the MD trajectory.
While technically their use violates reversibility, this is done in a soft way which can be controlled by performing an accurate
enough solution.

\subsection{The use of multi-grid solvers in HMC}\label{subsec:mgGaugeGen}
The use of multi-grid solvers in HMC presents numerous opportunities but also has some basic challenges.
The biggest challenge is due to the need to set-up and refresh the near-null space basis on which the multi-grid method relies.
Second, strong scaling is a more difficult challenge when using multi-grid solvers than for typical single grid solvers, since
multi-grid strong scaling is dictated by the coarsest grid with the smallest number of lattice points.
We discuss multi-grid solvers in detail in Sec.~\ref{sec:solvers} and here consider our working code which uses the implementation
in the QUDA~\cite{Clark:2009wm,Babich:2010mu,Clark:2016rdz,QUDADownload} library.
Our work builds on previous efforts, using the QOP-MG library~\cite{QOPQDP} on CPUs, which has been reported in~\cite{MeifengMG} in
HMC and previously in~\cite{Luscher:2007es} in the context of a DD-HMC simulation~\cite{Luscher:2003vf}.

The QUDA multi-grid implementation constructs the null space by solving for the {\em null vectors} $v_i$ by running an iterative
solver on the system $M v_i = 0$ with a random initial guesses for the vectors $v_i$, until some absolute precision, or maximum
number of iterations is reached for each one.
Current calculations use 24--32 null-space vectors per level, and so potentially, the setup cost is equivalent to nearly 24--32
solves which would pose a substantial overhead if it were performed before every solve.
One way to reduce the overhead is to use the same subspace for several MD steps and refresh the subspace occasionally as done in
\cite{Luscher:2007es}.

Previous work has shown that for heavy quark masses a preconditioner may remain effective over a whole trajectory~\cite{MeifengMG}
although for lighter quarks the preconditioner deteriorates~\cite{Luscher:2007es} as the MD proceeds, and the gauge fields and the
resulting Dirac operator evolve.
The deterioration is evidenced by an increase in the iteration count in the multi-grid solver.
A simple strategy is to set an iteration threshold, and to refresh the subspace vectors when this threshold is reached.
A further reduction in the refreshment cost can be achieved by not recomputing the subspaces using completely new random initial
guesses, but by taking the existing vectors and iterating the null space solve on them for a fixed number of iterations to
``polish'' them.
This brings two new parameters into play: the iteration threshold for refreshment, and the number of refresh iterations, both of
which need to be tuned for optimal performance.

Several interesting research avenues remain open in the use of multi-grid solvers in gauge generation.
One can attempt to reduce the cost of subspace creation, for example by using an adaptive process such as described
\cite{Luscher:2003qa,Frommer:2013fsa}.
The strong scaling challenges can also be tackled for example using domain-decomposed preconditioners~\cite{Frommer:2013fsa} and
communication avoiding solvers.
Higher raw performance may potentially be obtained by working with multi-grid algorithms in a block-solver mode solving several
systems at once.
Fitting the latter into an HMC algorithm still requires additional research.

\subsection{A case study: Chroma on Summit}\label{subsec:caseStudy}
As a case study we describe briefly the implementation of gauge generation in the {\em Chroma} code \cite{Edwards:2004sx} for use with GPUs in
general and on the OLCF Summit supercomputer in particular.
Chroma is built on a data parallel framework known as QDP++, and on GPU systems there is an implementation of this known as QDP-JIT
\cite{Winter:2014:FLQ:2650283.2650646} which generates the necessary GPU kernels out of the QDP++ expression templates
``just-in-time'' using the NVPTX back end of the LLVM compiler framework.
Thus using the QDP-JIT framework all the lattice wide operations utilized by Chroma are automatically GPU accelerated.
Chroma in turn implements a variety of HMC integration schemes, including leapfrog, second and fourth order minimum norm
integrators, and most recently a force-gradient integrator~\cite{Kennedy:2009fe,Clark:2011ir,Yin:2011np}.
 
For the linear solvers, Chroma calls out to the implementations in the QUDA library~\cite{Clark:2009wm,Babich:2010mu,Clark:2016rdz,
QUDADownload}, which provides a variety of solvers for the lattice Dirac equation as discussed earlier and in
Sec.~\ref{sec:solvers}.
In particular for two flavor and determinant ratio solves we have used its implementation of the aggregation multi-grid solver,
where multi-grid is used as a preconditioner to a generalized conjugate residual (GCR) iteration.

The multi-grid preconditioner features smoothers using the minimal residual (MR)~\cite{Saad:2003:IMS:829576} algorithm whereas the
bottom solver is a recursively multi-grid preconditioned GCR~\cite{Saad:2003:IMS:829576} except for the lowest level where it is
unpreconditioned.
Recent improvements have also yielded an implementation of communication-avoiding GCR (CA-GCR) which can be used instead of MR in
the smoothers, and instead of GCR at the coarsest level of the multigrid hierarchy, although comparative numbers from Summit are not
yet available for this development.
Subspace creation and operator coarsening can be done entirely on the GPU.
The chronological predictor from QUDA has also been interfaced with Chroma which will give QUDA access to Chroma's chronological
vectors for potential future algorithmic optimizations.
Subspace refreshment is implemented as described earlier by setting an iteration threshold which can trigger refreshment if
exceeded, and the number of refreshment iterations can also be chosen by the user.

\begin{figure}
    \centering
    \includegraphics[width=12cm]{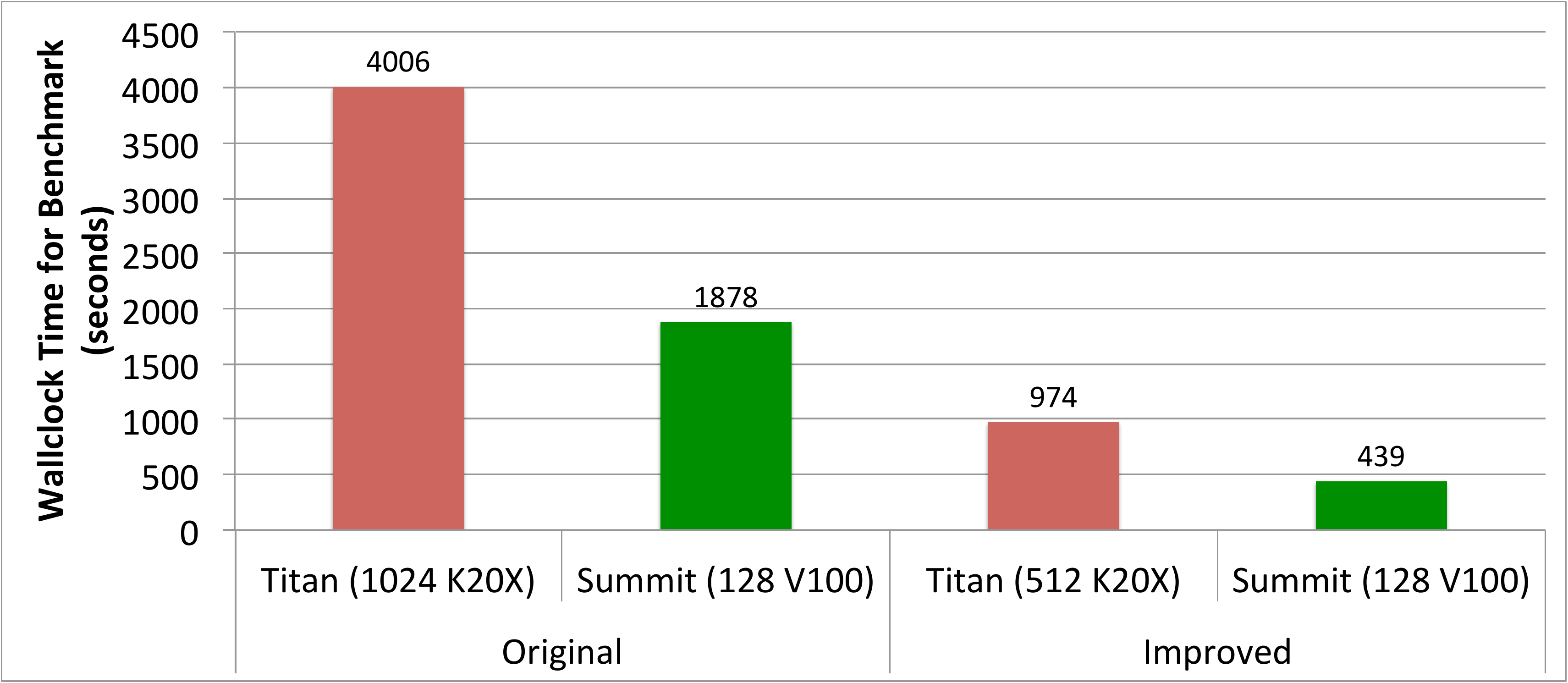}
    \caption{Wallclock times for running a Wilson-clover gauge generation benchmark, on Titan (red bars) and Summit
    (green bars) prior to (left) and after (right) algorithmic and code optimizations}.
    \label{fig:speedups}
\end{figure}
In Fig.~\ref{fig:speedups}, we show the wallclock times of running a gauge generation benchmark, which is a single trajectory on a
lattice of size $64^3\times 128$ sites with light quarks corresponding to a pion mass of $172~{\rm MeV}$.
We show the performance of the original setup on Titan, which used three determinant ratio terms and three single-flavor (rational
approximation) terms, two of which acted as the cancellation terms for the chain of determinant ratios.
The evolution used the fourth order minimum norm time-stepper with five force evaluations (MN5FV).
The two-flavor term used a GCR solver with a domain-decomposed preconditioner (DD+GCR)~\cite{Babich:2011np}.

The optimized algorithm uses instead four determinant ratio terms, with the final term in the chain being canceled by a very heavy 
two-flavor term, and only one single-flavor rational term for the strange quark.
The determinant ratio terms had the choice of masses re-optimized.
All two flavor and determinant ratio solves on Summit used the multi-grid solver, with the subspace generated for the lightest quark
flavor.
Re-optimizing the determinant ratio mass choices was possible because the multi-grid solver really tamed the cost of the ratio terms
featuring light masses.
The optimized setup also used the newly developed force-gradient integrator~\cite{Kennedy:2009fe,Clark:2011ir}, the chronological predictor from QUDA and a host of
QUDA improvements including pipelining of the GCR solver.
 
One can see that simply moving from 1024 nodes of Titan to 32 nodes of Summit (using 4 out of the 6 GPUs per node) without any
algorithmic improvement resulted in a reduction in wallclock time from 4006 to 1878 seconds, or a 2.13$\times$ wall--clock
speedup.
However taking into account the reduction in the number of GPU devices (8$\times$) the integrated speedup is 17$\times$.
Some of this is due to relief from strong scaling effects faced by the 1024 GPU Titan run, and the rest is from hardware
improvements going from Titan to Summit.
When the algorithmic improvements are also folded in the Summit time drops to 439 seconds, an overall wall-clock speedup of
9.13$\times$ compared to the original Titan run and taking into account the reduction in devices an overall improvement of
73$\times$.
The algorithmic improvements fed back to 512 Titan nodes (the maximum number of devices onto which the coarsest grid of the
multi-grid could scale for this problem size) resulted in a run-time of 974 seconds, a wall-clock speedup of 4.11$\times$ and an
integrated improvement of about 8.2$\times$ taking into account the reduction in the number of devices.

\subsection{Critical Slowing Down and Future Prospects} \label{subsec:criticalSlowingDown}
\label{sec:CSD}

As the lattice spacing is reduced in current and future simulations, the range of length scales in the problem (stretching from the
pion Compton wavelength down to the lattice spacing) grows and the current evolution algorithms suffer critical slowing down as
noted earlier: the stiff modes which evolve quickly must be integrated with a small step size while the soft, long-distance modes
will change very little in such a step and require many steps to evolve significantly.

While this problem of critical slowing down has been recognized from the beginning and interesting solutions proposed in the 1980s,
{\it e.g.,} Ref.~\cite{Batrouni:1985jn}, current calculations may now involve a sufficiently large range of scales that substantial
benefit may result from such methods.
The study and development of methods to reduce critical slowing down is a current focus of the lattice-QCD application project
within the DOE Exascale Computing Project.
The most promising approaches are based on Fourier acceleration.
Here the canonical momenta in the HMC evolution are made to depend on the gauge fields in such a way that the stiff, short-distance
modes are given a larger, fictitious mass in the kinetic energy term, so that energy equipartition requires those modes to move with
a small velocity while the soft, long-distance modes are given a small mass and hence a larger velocity.

Introducing such a gauge-field dependent mass faces two difficulties.
First, the resulting kinetic energy 
depends on both the canonical momenta and coordinates, making the problem
non-separable and requiring an implicit integration scheme~\cite{RSSB:RSSB765}.
Second, the usual relation between wavelength and frequency is spoiled by gauge symmetry, requiring either a gauge-invariant operator
such as the lattice Laplacian be used in the mass term or that a gauge-fixed evolution be performed.
At present a number of methods are being developed including the Riemann manifold hybrid Monte Carlo (RMHMC)
algorithm~\cite{RSSB:RSSB765} and the look-ahead HMC (LAHMC) algorithm~\cite{pmlr-v32-sohl-dickstein14}. While there are encouraging signs, a significant amount of statistics is still needed to estabilsh the effect of these algorithms on known observables with longest autocorrelation times.

\subsection{Multi-grid inspired Monte-Carlo methods}
Finally, we note that there has been research into using multi-grid like approaches (distinct from multi-grid linear solvers) to
speed up the decorrelation of HMC simulations.
Recent work~\cite{Endres:2015yca} has shown that one can reduce overall equilibration time for a given fine system by projecting a
fine lattice onto a coarse lattice (restriction), equilibrating that at a faster rate using a suitably matched coarse action, then
prolongating back to the fine level and re-equilibrating on the fine level.
This approach can have several practical applications, from reducing the equilibration time of a single Markov chain, to producing
independent seed configurations for several Markov chains which can then be simulated on their respective fine grids in parallel.
To date this approach has been applied to pure Yang-Mills theories but there are no in principle issues with applying it to
simulations with dynamical fermions.

\section{Linear Systems and Eigensystems in Lattice QCD}\label{sec:solvers}

\subsection{Solvers in lattice QCD}

As we have discussed above, the solution of various forms of Dirac equations constitute a major part of
lattice-QCD calculations.
While colloquially referred to as {\em inversions}, what is required is the application of a matrix inverse onto a source, in other
words, the solution of a linear system of equations.
The solution of linear systems is a very rich field of numerical linear algebra and lattice-QCD calculations can, on the one hand drive
research in this area of applied mathematics and on the other be a successful application of existing techniques.

The linear systems under considerations typically have dimensions proportional to the lattice volume, which for current calculations
can be O($10^8$--$10^9$) sites, depending on the fermion formulation used (Wilson-like and staggered
fermion formulations result in four-dimensional systems, while domain-wall like fermions possess an additional fifth dimension).
In all cases, the linear operators in these systems are {\em complex valued} and {\em sparse} following either a nearest neighbor
(Wilson-like or domain-wall--like) or next-to-nearest neighbor stencil pattern (improved staggered fermions).
An exception is the so called {\em overlap formulation} where the linear operator itself is not nearest neighbor but evaluating it
relies on applying a nearest-neighbor Wilson kernel.
The linear operators themselves are not Hermitian, but typically have a $J$~Hermitian form (some form of $\gamma_5$ Hermiticity,
depending on the action) which is generally maximally indefinite.
The methods of choice solving these systems are typically iterative solvers for sparse linear systems.
As the quark masses in the linear operators approach the physical light quark masses from above, the resulting linear systems become
ill conditioned resulting in poor convergence behavior for the more conventional Krylov subspace solvers, which is another form of
{\em critical slowing down}.

It is standard practice to use a so called {\em even-odd (or red-black) checkerboard preconditioning} where iterative solvers need
to be applied only to half the lattice sites (one checkerboard) using the Schur complement operator.
The system on the other half of the sites is then usually trivial to solve.
In HMC simulations, pseudofermions need only be kept for the checkerboard on which the Schur-complement operator acts, with the
determinant on the other checkerboard being handled explicitly if it is not trivial.
This gives a speedup of 2--3~$\times$ in solves and also reduces the forces in the gauge generation compared to the unpreconditioned
case.

Due to the different spectral properties of the different fermion formulations, the methods of choice for solving the various
resulting Dirac equations differ.
There are several systems of equations to solve: a) $M x = b$ needed for propagators; b) $M^\dagger M x = b$ needed for two flavor
and ratio terms in gauge generation; c) the shifted system $( M^\dagger M + \sigma_i I ) x_i = b$ with solutions $x_i$ and shifts
$\sigma_i$ for systems arising in, for example, rational approximations; and finally d) $ ( M + \delta m_i I ) x_i = b $ is useful,
where $ \delta m_i$ is a shift in the quark mass (staggered and overlap fermions).

\subsection{Commonly used Krylov subspace methods}
The method of choice for forms c) and d) to date have been the use of multi-shift conjugate gradients (M-CG) \cite{Glassner:1996gz,
Jegerlehner:1996pm}.
This method generates all the solutions $x_i$ for the cost of converging the system with the smallest shift.
One downside of the method is that it relies on all initial guesses for the $x_i$ to be parallel (typically the zero vector is used
for all $x_i$), which limits several tricks to aid performance such as restarting methods and residual replacement strategies and
hence also the use of reduced precision.
Communication reduction approaches that rely on using reduced communication preconditioners also cannot be used.
To combat these challenges, typically the solves are done to single precision initially, and then, if double precision is required,
the individual solutions are polished with subsequent non-multishift, potentially mixed precision iterations.
Additionally chronological approaches may be used so that information built up during polishing one solution can be applied to the
next.

We note that in case b) one has a manifestly Hermitian, positive definite (HPD) combination of $M^\dagger M$ meaning that classical
conjugate gradients \cite{citeulike:9077321} should always be able to solve the system, although in ill conditioned cases
convergence may be very slow.
In the case of form a) one can turn to conjugate gradients on the normal equations (CGNE) or on the normal residuals (CGNR).
However, form b) can also be solved for some fermion actions using a two-step process by first solving $M^\dagger y = b$ for an
auxiliary vector $y$ and then solving $M x = y$.
The convergence of each step is thus now gated by the condition number of $M$ or $M^\dagger$, which is the square root of that of
$M^\dagger M$, which can result in an overall gain.
However, since $M$ and $M^\dagger$ are no longer HPD, a solver is needed which can deal with non-HPD matrices.

The Wilson and Wilson-clover formulations have perhaps lent themselves to the richest exploration of algorithms in the area of
solvers for lattice QCD.
Early on it has been discovered that for heavy to medium light quark masses BiCGStab \cite{BiCGStab} is an effective solver for
forms a) and b).
In order to save effort at light quark masses, especially in the analysis part of calculations a variety of methods were developed
including deflated solvers such as Eig-CG \cite{Stathopoulos:2007zi} and Eig-BiCGStab~\cite{AbdelRehim:2009by} which find the basis
for the low modes of the operator during the initial first few solves and then deflate them in subsequent ones.
Another approach to deflation was proposed in Ref.~\cite{Frommer:2012zm}, using FGMRES-DR \cite{GMRESDR} where a subspace developed
during an FGMRES \cite{Saad:1993:FIP:160077.160089} Arnoldi cycle is reused in subsequent {\em augmented} Arnoldi cycles.
Yet another approach to deflation was found in Ref.~\cite{Luscher:2007se}, where a GCR solver was deflated based on a subspace
generated by near zero modes of the operator $M$ supported over blocks of the lattice, an idea that is in essence a two grid variant
of an additive multi-grid algorithm.
Explicit deflation by a pre-computed basis of low-lying eigenvectors is also possible, and may be desirable if the eigenvectors
computed can be used elsewhere, such as in the low mode averaging (LMA) and all mode averaging (AMA) approaches to computing
correlation functions (see Sec.~\ref{sec:contract}).
Recently {\em adaptive multi-grid} algorithms have also been successfully applied to Wilson-clover systems \cite{Babich:2010qb,
Frommer:2013fsa} which we discuss in Sec.~\ref{subsec:AAMG}.

Other formulations have had varied success in employing newer solver technologies to date.
BiCGStab, for example, simply fails with domain wall fermions.
Multi-grid approaches such as HDCG \cite{Boyle:2014rwa} and HDCR \cite{Yamaguchi:2016kop} have been developed for domain wall
fermions but presently use the squared $M^\dagger M$ operator. 
Recently, a similar approach to~\cite{Luscher:2007se}, based on multispliting, has been shown to be effective for domain wall fermions~\cite{Tu:2018wuo}.
In the area of staggered fermions, there has been an exploration of block solvers to make more efficient use of available memory
bandwidth \cite{Clark:2017ekr}, with current development also exploring extended Krylov subspace methods.
In parallel, research is underway as part of ECP to bring the benefits of multi-grid solvers to staggered
fermions~\cite{Brower:2018ymy}.

\subsection{Adaptive aggregation multi-grid methods}\label{subsec:AAMG}

\begin{figure}
\begin{minipage}{0.48\textwidth}
\vspace{-0.12\textwidth}
\includegraphics[width=\textwidth]{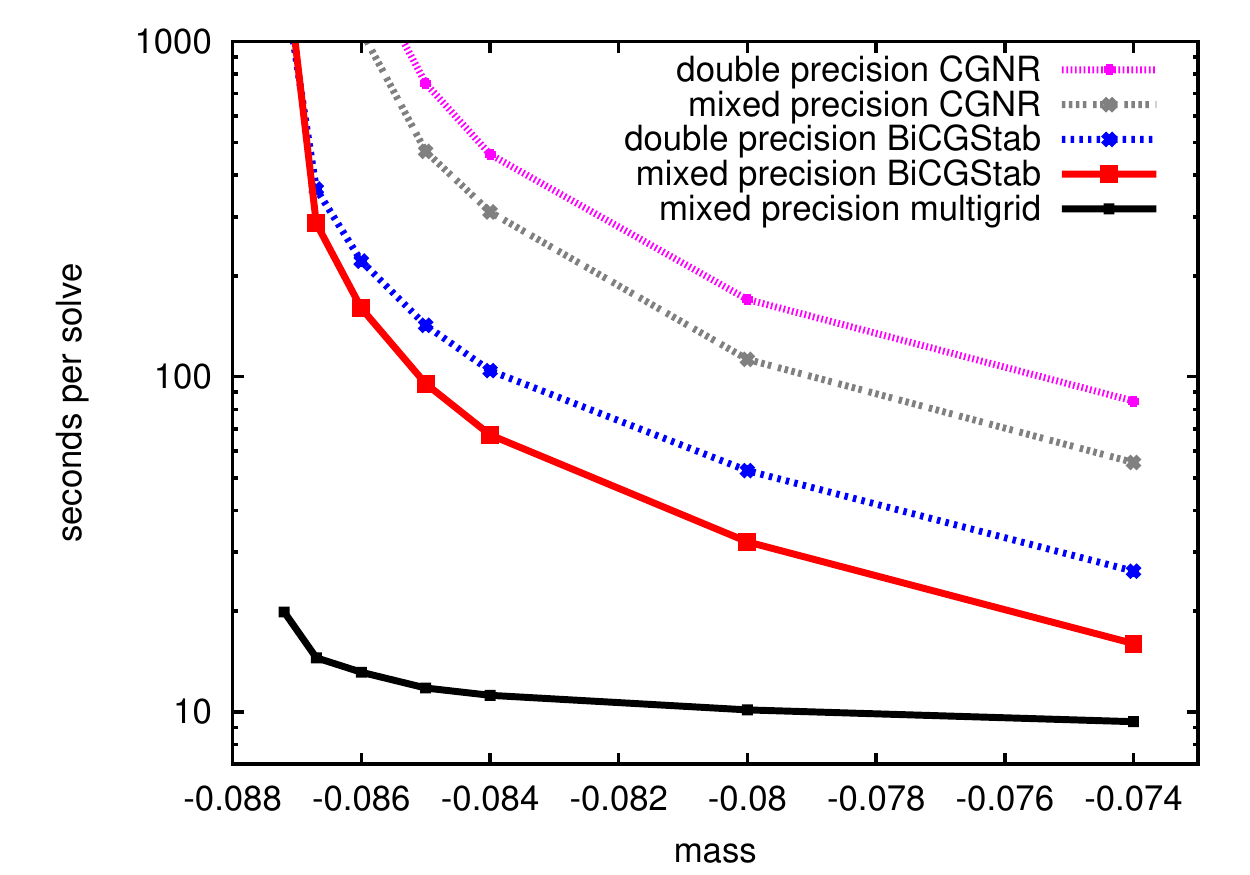}
\caption{Scaling of solvers with bare quark mass parameter.
One can see that CGNR and BiCGStab solvers diverge in terms of solve time as the quark mass is reduced, while the scaling of
multi-gird with decreasing quark mass is comparatively flat.
From Ref.~\cite{Brannick:2007ue}.}
\label{fig:MG_optimality}
\end{minipage}
\hfill
\begin{minipage}{0.48\textwidth}
\vspace{-0.12\textwidth}
\includegraphics[width=\textwidth]{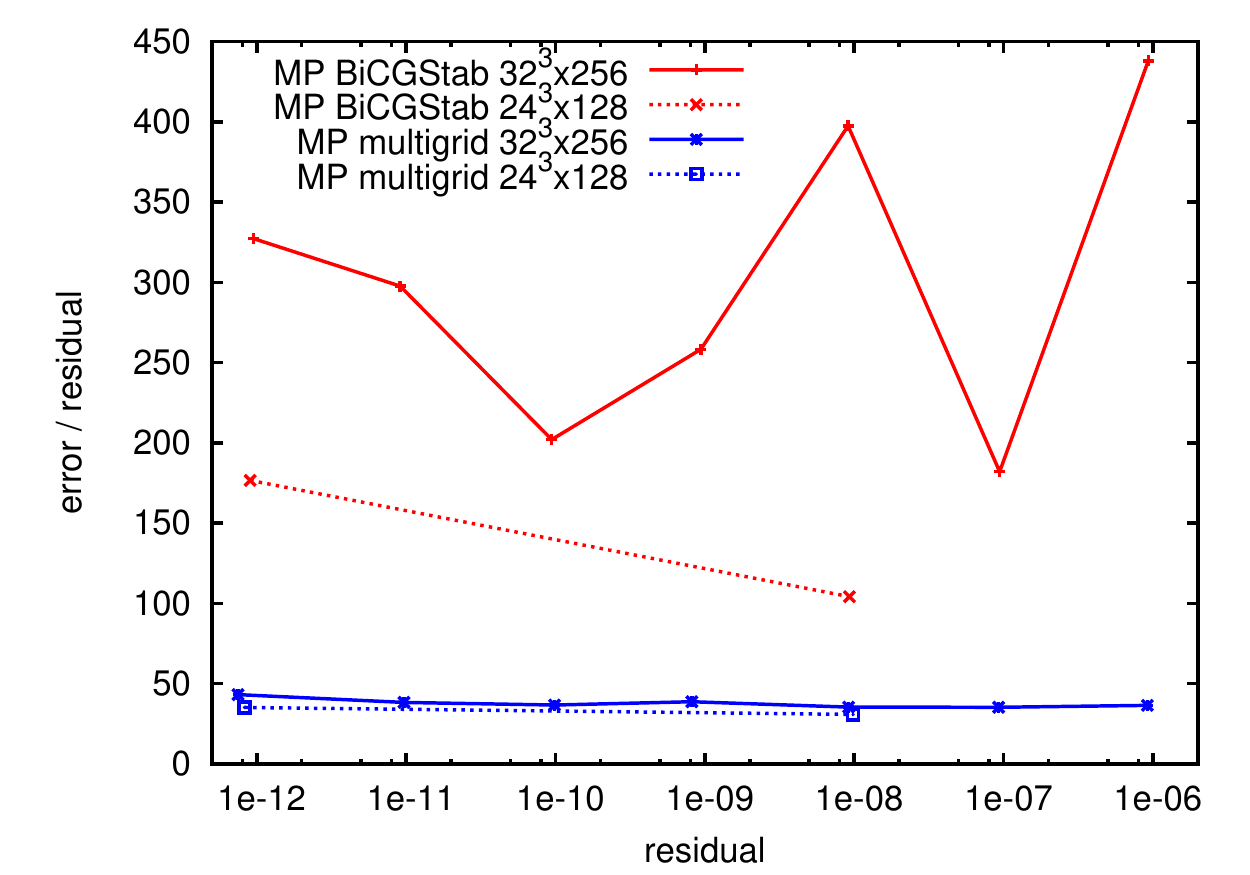}
\caption{A comparison of the ratio of the error to the residuum between BiCGStab and multi-grid solvers.
The multi-grid result is low and does not fluctuate with the solver target residuum, indicating better solution quality.
From Ref.~\cite{Osborn:2010mb}.}
\label{fig:MG_error}
\end{minipage}
\vspace*{-0.7em}
\end{figure}

Multi-grid and related algorithms\epj{, based on local coherence~\cite{Luscher:2007se},} have perhaps been the biggest breakthrough technology in lattice-QCD calculations of
Wilson-clover fermions of recent years \cite{Babich:2010qb,Osborn:2010mb,Frommer:2013fsa,Luscher:2007es,Clark:2016rdz}.
The multi-grid cycles are typically implemented as a preconditioner to a flexible outer solver such as GCR or FGMRES.
The preconditioning step computes an approximate inverse applied to a vector and so is itself a solver.
Multi-grid preconditioners work, colloquially speaking, by separating high and low frequency modes of the linear operator.
The error to the system from the high frequency modes is reduced by a step known as {\em smoothing}, and can be affected by for
example iterating a simple solver like MR \cite{Saad:2003:IMS:829576}, or a Schwarz-alternating process \cite{Frommer:2013fsa}.
To deal with the components of the error from low-modes of the operator, the system is projected onto a {\em coarse lattice}
resulting in a coarse linear system using a coarse linear operator and vectors in a step known as {\em restriction}.
A correction to reduce errors from the low modes is then computed on the coarse grid and the result is then moved back to the fine
grid in a step called {\em prolongation}.
Since this step may bring in some unwanted high frequency modes, the system may undergo {\em smoothing} again to result in a final
multi-grid correction.
The steps of restriction, coarse solve, prolongation and smoothing can be combined into various multi-grid cycles, including
recursively such as the so called $V$~cycle, $K$~cycle, and $F$~cycles.

The multi-grid methods used successfully to date in lattice QCD have been variants of {\em adaptive smoothed
aggregation}~\cite{Brezina04adaptivesmoothed}.
To define the coarse operator and the restriction and prolongation operators, the lattice is split into blocks and fine
degrees of freedom on these blocks are aggregated to form the coarse degrees of freedom.
This relies on the phenomenon of {\em local coherence} \cite{Luscher:2007se}, which is a physical re-statement of the mathematical
{\em weak approximation property}.
Loosely speaking, local coherence means that the long wavelength modes on blocks of the lattice are good representations of long
wavelength modes on the whole lattice, allowing the aggregation over blocks to produce a good coarse operator, which captures
faithfully the low modes of the original operator.

Multi-grid algorithms have several very attractive features: They are {\em optimal} in the sense that their convergence properties
(when properly tuned) should depend only on the volume of the system to be solved and they {\em eliminate the critical slowing down
in terms of quark mass} as can be seen in Fig.~\ref{fig:MG_optimality}.
As a result a well optimized implementation such as the one in the QUDA library for GPUs \cite{Clark:2009wm,Clark:2016rdz} can
provide close to an order of magnitude improvement over the best optimized BiCGStab implementation for Wilson-clover fermions.
This also results in highly increased energy efficiency of the computations.
Since multi-grid works on minimizing the error rather than the residuum of a given system, the solutions produced tend to be of
{\em higher quality} than, say, BiCGStab, in the sense that the error of the solution tends to be smooth and small over all the
lattice sites whereas residuum based methods have error components that can fluctuate a great deal, as shown in
Fig.~\ref{fig:MG_error}.
Due to being used primarily as a preconditioner, multi-grid cycles can make use of the reduced precision capabilities of recent
hardware architectures.
However, as a result of working on a succession of coarser grids, the strong scaling of the method is gated by the volume of the
coarsest level.
This tends to be less of a problem for propagator analysis where one can scale up in ensemble mode running several solves
simultaneously in its own small partition, but can be limiting in strong scaling situations like gauge generation.
There are several ways to reduce strong scaling effects, for example by employing domain decomposition in the smoothers
\cite{Frommer:2013fsa} or turning to {\em communications avoiding algorithms}.
If the coarse system is sufficiently small, it can be replicated and solved (redundantly but faster) by all the nodes of a parallel
calculation.
This is a very active area of research as we head towards the exascale.
 
\subsection{Reduced precision and communication reduction}
Hardware architectural developments have had a strong influence on the development of lattice QCD linear solvers.
The linear operators for the most frequently used fermion formulations have low arithmetic intensities (e.g., less than
1~FLOP/byte in single and less than 0.5~FLOP/byte in double precision) and are memory bandwidth bound on current architectures.
In a parallel system, depending on the size of the halo region the computation may be impacted by both network bandwidth or latency.
Multi-grid approaches on the coarser grid can typically have high surface to volume ratios and may also be latency bound.
As a result a great deal of effort has gone into optimizing memory bandwidth use and reducing communications needs in
implementations.
Further, modern hardware often has a greater capability to deliver single precision FLOPs than double precision.
As such, it is desirable to perform as much computation in reduced precision as possible.

Communication latencies can be reduced through the use of pipelining and {\em communication avoiding} Krylov solvers such as
\emph{$s$-step methods} that reduce the number of potentially latency sensitive {\em global reductions}.
Bandwidth use can be reduced in cases by applying domain specific knowledge such as by employing gauge compression as pioneered in
the QUDA library, which uses the properties of SU(3) matrices to represent them either as two rows (reconstructing the third via
vector-product) or by representing them through the coefficients of the 8 Gell-Mann generator matrices.
These approaches reduce the amount of data transferred through the memory system, trading bandwidth for the freely available FLOPs
required to reconstruct the original data before use.

Mixed precision can be exploited in solvers through using varieties of iterative refinement such as pioneered by the QUDA library in
the form of {\em reliable updates}.
Using a flexible outer solver process, which allows for nonstationary preconditioners, permits multiple
optimizations to be employed in the preconditioner.
First, one can use reduced precision with all the benefits that brings and second one can employ communication reduction/avoidance
techniques such as domain-decomposition, or passing halo boundaries in reduced precision to save on network bandwidth.

Flexible outer solvers can also aid in fault-tolerance, since they can embed the desired regular solver as their preconditioner.
Should the embedded solver complete without issue the surrounding outer flexible solver would not need more than one iteration.
Should soft-faults affect the embedded solve, since it acts as a variable preconditioner, the outer process which should still
converge eventually \cite{Heroux:2013:TRA:2465813.2465814}.

Another way to improve performance of these solvers is to turn to block-solver approaches.
One feature of the block solver approach is that one can improve data reuse by applying the same linear operator to several vectors
at once \cite{Kaczmarek:2014mga,Clark:2017ekr} thus increasing the arithmetic intensity and enabling higher compute efficiencies.
In some cases, this approach can also benefit vectorization (where one can imagine potentially vectorizing over multiple systems).
Since in the analysis step many hundreds of thousands of systems need to be solved with the same gauge field, this approach can find
easy application.

We should note that many of the above techniques can be successfully combined.
For example gauge compression is straightforward to combine with all solvers for all fermion formulations.
Residual replacement techniques such as reliable updates can also be broadly applied, except in special situations like the shifted
conjugate gradients solver.
Variants of these techniques are available to all fermion formulations in the QUDA library, for example.

\subsection{Eigensolvers and deflation} 

\begin{figure}
\begin{minipage}{0.48\textwidth}
\vspace{-0.12\textwidth}
\includegraphics[width=\textwidth,trim={36pt 56pt 22pt 22pt},clip]{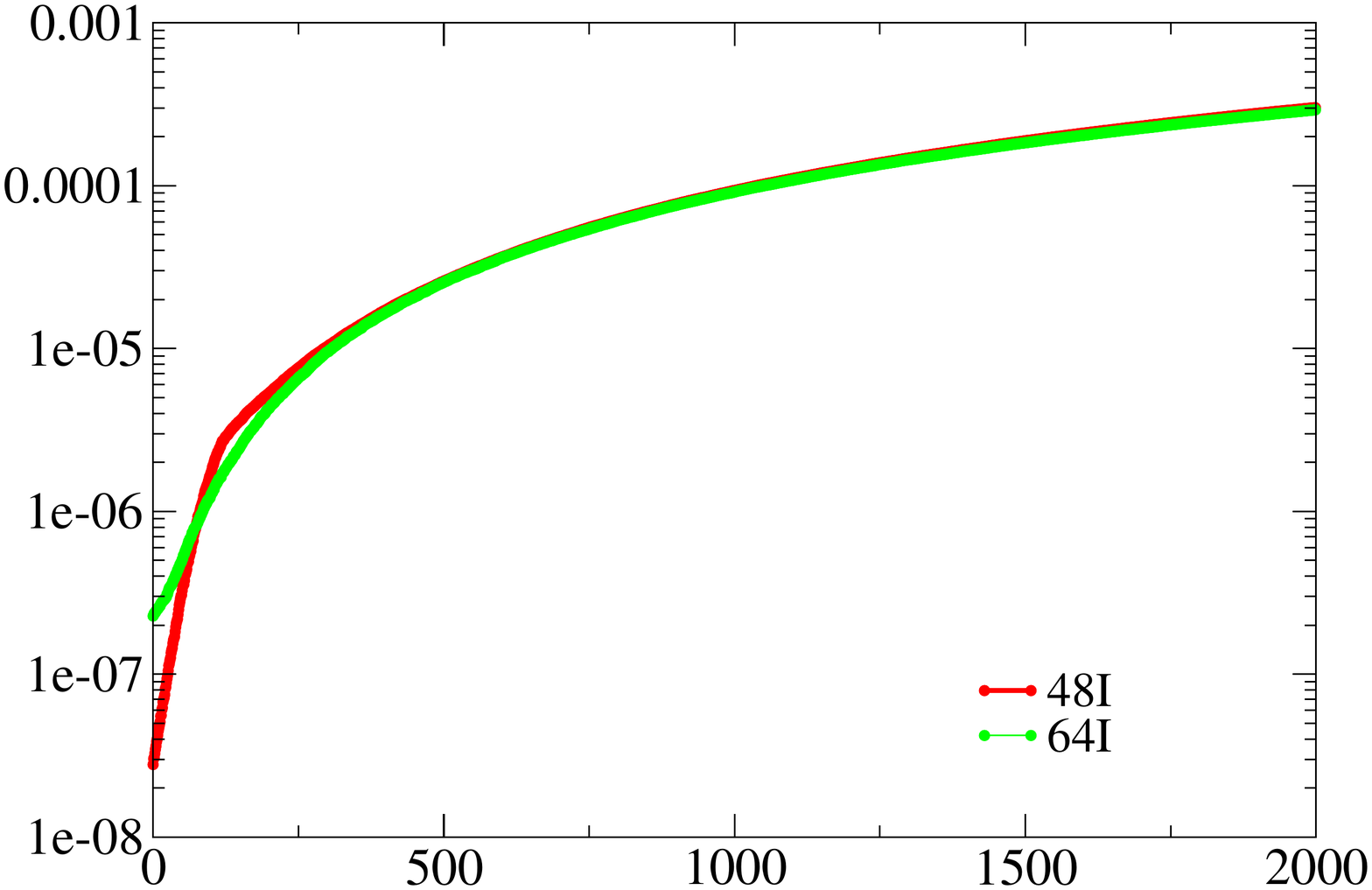}
\caption{ Eigenvalues of the preconditioned M\"obius domain-wall Dirac operator on 2+1 flavor M\"obius ensemble (48I, 64I).
See Ref.~\cite{Blum:2014tka} for details. }
\label{fig:evals}
\end{minipage}
\hfill
\begin{minipage}{0.48\textwidth}
\includegraphics[width=\textwidth]{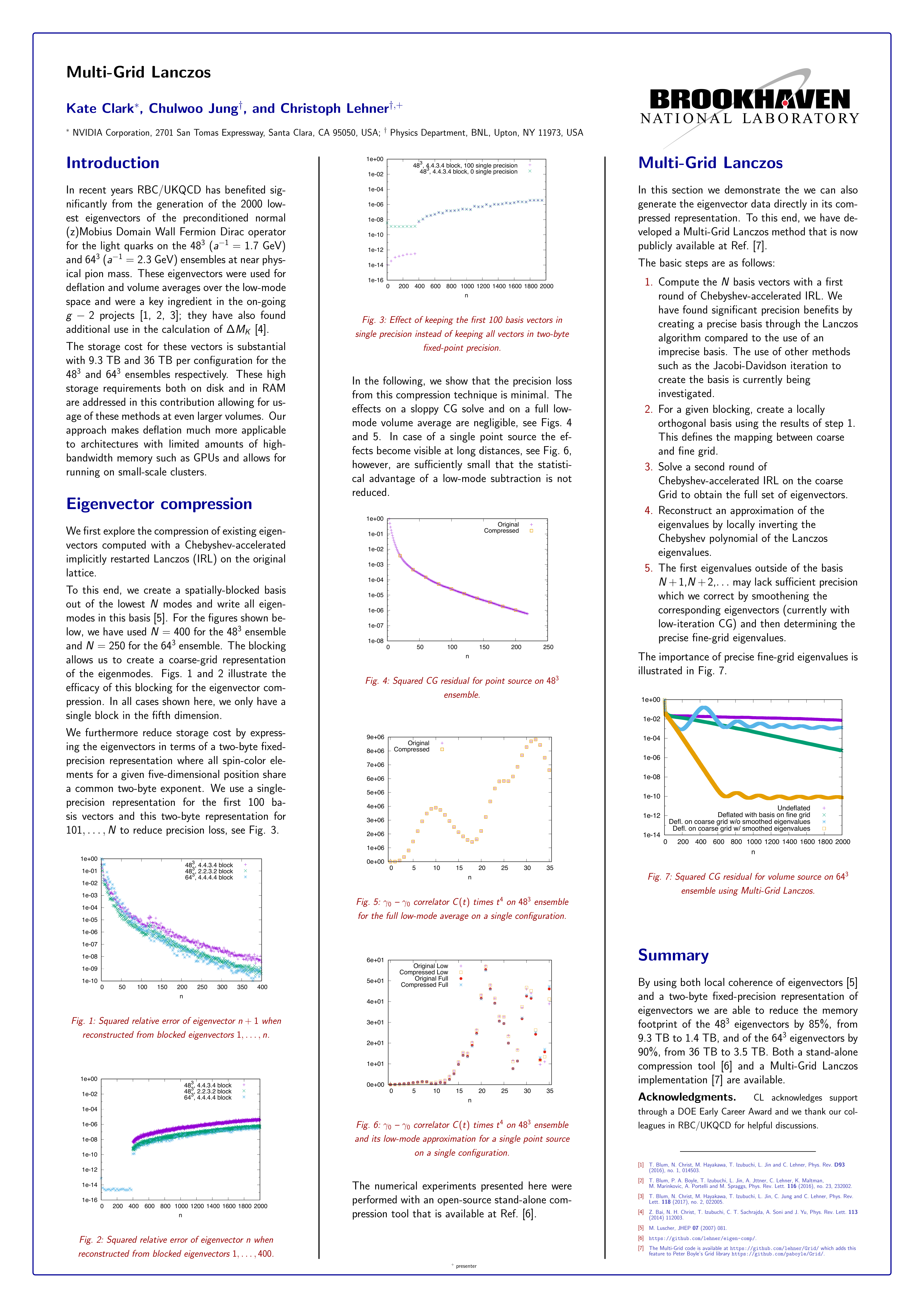}
\caption{
Comparison of residual evolution for undeflated and deflated conjugate gradients, with eigenvectors
generated on fine and coarse grids built with domain decomposed eigenvectors~\cite{Clark:2017wom}.}
\label{fig:mgl}\end{minipage}
\end{figure}

While typical lattice-QCD problems have an enormous number of degrees of freedom, the distribution of eigenvalues, or singular
values for non-Hermitian matrices, offers a path towards a significant reduction in
computational cost.
The spectrum of lattice Dirac operators is typically dense except for a relatively small number of low-lying (small in absolute value)
eigenvalues, where the density is low.
As shown in Fig.~\ref{fig:evals}, there are around $\sim$2000 such low-lying eigenpairs for a physical box of size
$\sim$5~fm, while the total number of degrees of freedom is on the order of $10^9$.
Not only does this allow various exact and approximate eigenvector-based deflated linear solvers to be effective, but
these eigenvectors can also be used to construct effective approximations for various quantities, further reducing the number of
Dirac operator applications necessary to extract a given level of signal from each lattice configuration by as much as {\em two
orders of magnitude}.
Some of these techniques such as all mode averaging (AMA)~\cite{Blum:2012uh,Shintani:2014vja} or all-to-all~\cite{Foley:2005ac}
methods, will be described in Sec.~\ref{sec:contract}.

Naturally, {\em eigensolvers} or {\em singular value decomposition (SVD) solvers} are indispensable tools in such approaches.
Fortunately, the distribution of eigenvalues, except for the smallest ones, tends to be rather stable between different lattice
configurations which makes it possible to use polynomials with predetermined coefficients (typically Chebyshev polynomials) to
create an effective filters for unwanted eigenvalues.
Using such filters, thousands of eigenpairs can be converged by implicitly restarted Lanczos (IRL)~\cite{Calvetti1994AnIR} with the
total number of applications of Dirac operators being typically less than twice the number of desired eigenvectors, minimizing the
number of linear algebra operations.
This allows the generation the necessary eigenvectors, often on the order of thousands, with an amount of Dirac operator
applications comparable to those for 2--30 undeflated inversions, not dissimilar to the amount of work typically used to find the
near-nullspace vectors for multi-grid solvers described in the previous section.

While the techniques developed so far have made the computational cost for eigenvector generation comparable with other (in)exact
techniques, the need for storing these eigenvectors (temporarily or for the long term) poses an additional challenge.
The mixed precision techniques for solvers are also helpful here.
Also, the local coherence property~\cite{Luscher:2007se} or ``smoothness'' of the eigenvectors, which underlies the success of the
multi-grid algorithms in lattice QCD, offers an opportunity for a significant reduction of storage space as well.
Local coherence implies that the number of effective degrees of freedom is significantly smaller than the nominal degrees of freedom
for these eigenvectors.
This feature has been exploited to yield a compression algorithm on domain-decomposed eigenvectors~\cite{Clark:2017wom} resulting in
an order of magnitude or more of data reduction in existing eigenvectors.
Local coherence also leads to a new and more efficient method for generating eigenpairs: by calculating eigenvectors directly on the
subspace defined by domain decomposed lowest eigenvectors, one can generate eigenvectors that are nearly as accurate as the ones
generated in the original vector space for a variety of measurements, as shown in Fig.~\ref{fig:mgl} for deflated CG, while
decreasing the memory footprint at the same time.
lattice-QCD simulations on exascale machines will use finer lattice spacing to control the discretization error.
This means the local coherence-based algorithms will result in even larger relative savings in memory, as the size of the blocks
will increase in units of lattice spacing.

\begin{figure}
\begin{minipage}{0.45\textwidth}
\includegraphics[width=\textwidth]{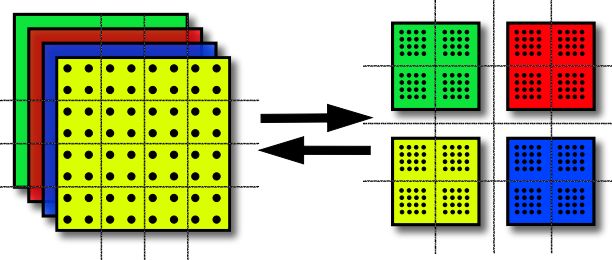}
\caption{Illustration of split grid.}
\label{fig:SplitGrid}
\end{minipage}
\hfill
\vspace{0.05\textwidth}
\begin{minipage}{0.45\textwidth}
\includegraphics[width=\textwidth]{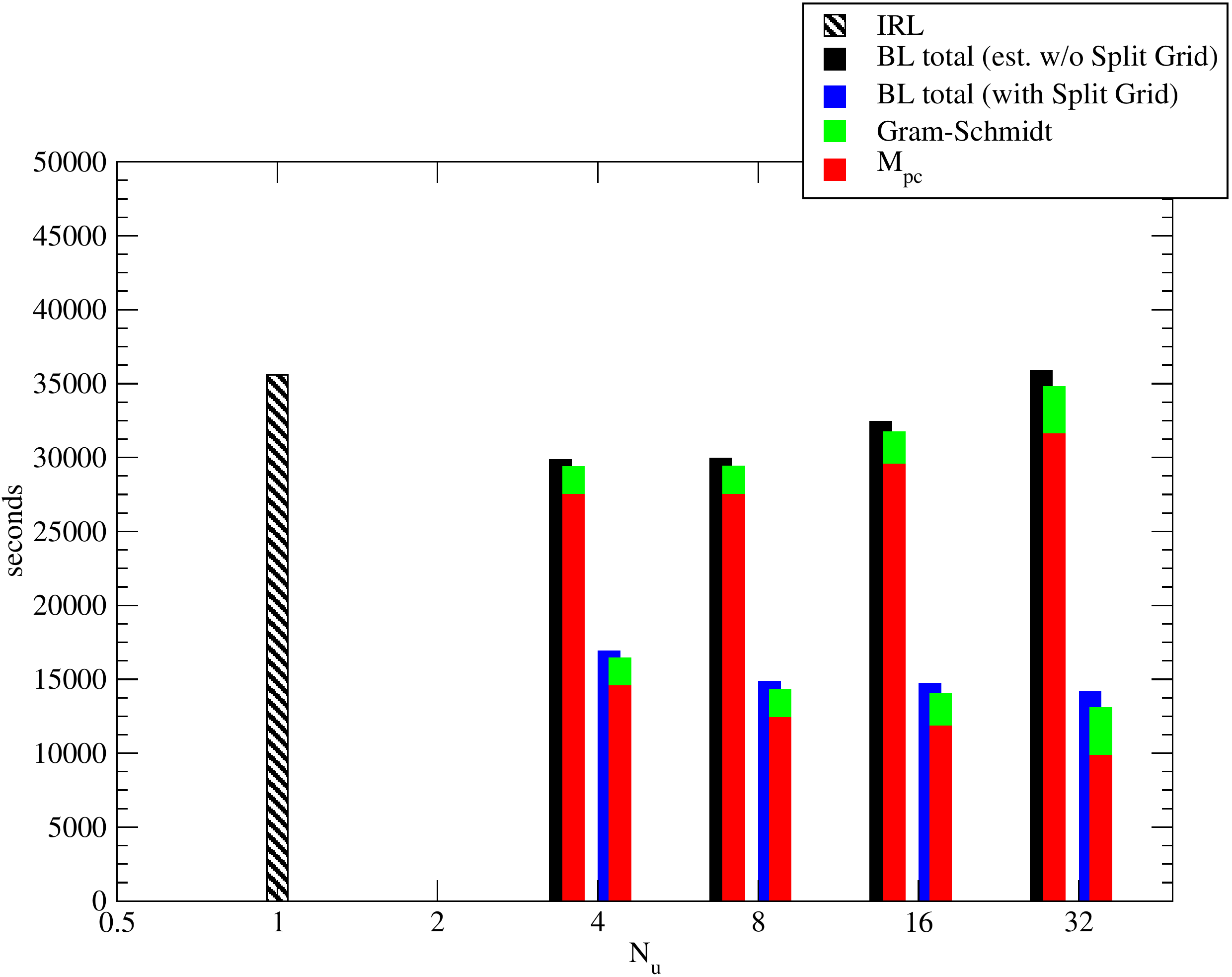}
\caption{Timings of block Lanczos with split grid on 512 nodes of ALCF Theta.
From Ref.~\cite{Jang:2019roq}.}
\label{fig:BL} 
\end{minipage}
\end{figure}

Many lattice-QCD applications must cope with the need to run on sufficiently many nodes to have enough memory for all necessary data, even though running on so many nodes
would result in inefficiencies from strong scaling.
While in some instances, one can split the workflow up into an ensemble approach, running ensemble members in essentially separate
MPI jobs, this is not always desirable or possible.
Inspired by software frameworks developed for big data such as Hadoop~\cite{hadoop}, the split-grid algorithm allows a single
application to switch between one large domain and multiple smaller domains within a single application, so that the performance of
the Dirac operator application within a sub-domain is not in the strong-scaling limit and can benefit from a better surface-to-volume
ratio and network-bandwidth use, while routines without significant internode communications can be run on the larger partition
making use of the aggregate memory to increasing data reuse and eliminate the need for a large amount of disk I/O.
Figure~\ref{fig:SplitGrid} is an illustration of split-grid approach, while Fig.~\ref{fig:BL} is an example of split-grid technique
applied to block Lanczos, which shows a substantial overall speedup despite the fact that the number of Dirac operator applications
is the same as, or larger than, the case for the single-vector~IRL.

\section{Correlation Function Construction}\label{sec:contract}

The penultimate
\footnote{The final stage of the lattice-QCD workflow is statistical analysis, combining information from (ideally) several 
ensembles to produce a physics results with statistical and systematic uncertainty estimates.
As this step poses no special computation problem, we do not cover it in this whitepaper.}
stage of the lattice-QCD workflow is the construction of Euclidean correlation functions.
There are a variety of methods in use for this phase, developed to optimize the calculations of particular observables.
Calculations of two-point correlation functions typically rely on {\em quark smearing} to smooth unphysical ultraviolet fluctuations
that lead to statistical noise.
However, the same methods that can be used to extract the energies of eigenstates are not necessarily applicable to matrix elements,
which also suffer from statistical noise.
Hence one is concerned with two sources of statistical error in correlation functions: a) noise coming from the {\em gauge field
sampling} (gauge noise) which can be controlled by the number of gauge configurations generated, and b) noise coming from
measurements probing a given gauge configuration (measurement noise).
This latter noise can be reduced by carrying out many independent measurements per configuration.
Additionally, many quark operator constructions are needed to study nuclear systems; however, the number of permutations to contract
the quark lines grows factorially.  This challenge is discussed in more detail in Sec.~5.4 of the companion whitepaper ``Hadrons and Nuclei''~\cite{Detmold:2018qcd}.

\subsection{Distillation}\label{subsec:distill}

In a Monte Carlo calculation in lattice gauge theory, the physically relevant signal in a correlation
function falls exponentially as a function of Euclidean time and is rapidly overwhelmed by statistical fluctuations.
Operators that create low-lying energy eigenstates at early values of Euclidean time are therefore invaluable and improve the
quality of data extracted exponentially.
The generalized eigenvector variational procedure~\cite{Michael:1985ne,Luscher:1990ck,Blossier:2009kd} provides a robust method to
project onto the the finite-volume eigenstates in a system, and helps to ameliorate the rapid fall--off of the signal with the
statistical noise.
The method relies on the construction of a large but diverse basis of operators that have varying overlaps onto the ground states as
well as the excited energy levels in a system.
A matrix of correlation functions is constructed with this basis, and the generalized eigenvalue problem is solved.
The time-dependence of the eigenvalues allows the determination of the finite-volume energy levels in the system.

An efficacious approach to constructing a suitable basis of correlation functions is provided by the {\em distillation}
method~\cite{Peardon:2009gh}.
In this method, a low-energy basis of vectors is used to construct a low-rank definition of a quark-smearing operator.
The factorization of the quark smearing leads to the construction of propagators within the low-rank basis, as well as the
construction of multi-quark hadron operators that can be projected onto definite momentum.
A major advantage of the approach is that, {\em a posteriori}, Euclidean correlation functions can be constructed.
In particular, the method is well suited for investigations of multi-hadron correlation functions.
The hadron operator constructions can have relative momentum and projected into an overall definite momentum transforming under
suitable representations of the lattice cubic group of rotations~\cite{Dudek:2012gj}.

\subsection{Low- and all-mode averaging}\label{subsec:ama}

As noted previously, the statistical errors of a given lattice-QCD calculation are limited by the number of measurements, which
involve generating quark propagators from sources placed on different points of the lattice.
The number of propagator inversions on a single configuration can reach into the tens-to-hundreds of thousands or more.
Deflation techniques that exploit the relative sparseness of the lowest eigenvalues of the Dirac operator~$M$, as well as multi-grid
approaches, both of which lower the number of iterations in linear solvers thus bring a substantial reduction in computational cost.
However, there are other properties of $M$ which allow for even further cost reduction.
One such property is that the accuracy of high and intermediate modes in the solutions does not affect the overall error of the
measurement for most of the quantities measured in lattice QCD.
In other words, many ``sloppy'' measurements (with looser solver convergence criteria) with sources placed in different locations,
or measured on different lattice configurations, produce a much smaller error than using fewer, but more accurate solutions (with
tight solver convergence criteria).
All-mode-averaging (AMA)~\cite{Blum:2012uh} exploits this property by defining what is allowed for ``sloppy'' measurement without
introducing bias.

For an observable $\co$ defined in terms of propagators, we can define an approximation $\co^{(appx)}$ which is both numerically
cheaper than calculating $\co$ and remains covariant under lattice symmetry transformations, i.e., for $g \in G$,
$\left<\coa[U^g]\right>=\left<\coag[U]\right>$, where $g$ is a transformation from a larger set $G$, $U^g$ is the transformed gauge
field and $\left< \right>$ denotes an ensemble average.
With these definitions, an improved and unbiased estimate $\co^\text{(imp)}$ can be constructed as
\begin{gather}
    \left<\co^\text{(imp)}\right> = \left<\left( \co - \coa \right)\right> + \frac{1}{N_G} \left<\sum_{g \in G} \coag\right>.
\end{gather}
For low-mode averaging (LMA) the propagators in $\coa$ are constructed purely from the lowest eigenmodes, while in the case of AMA,
one uses the `sloppy' propagators obtained either via a fixed iteration count or with a stopping condition that that is relaxed
compared to the more accurate propagators used to compute $\co$.

This further reduction of necessary $M$ applications (from sloppy solves, and/or efficient eigensolvers to find low modes), on top
of the reduction already obtained from the (in)exact deflation makes it possible for many measurements to achieve statistical errors
comparable to those from fluctuation of gauge configurations (e.g., Ref.~\cite{Blum:2012uh})


\subsection{All-to-all propagators}\label{subsec:a2a}
The all-to-all (A2A) technique is used frequently when the quantities to be measured formally require a number of propagators from
sources on each of the lattice sites, which typically number $10^6$ or more.
A well known example is a class of operators called ``disconnected diagrams,'' in which there are one or more quark loops not
connected to the external legs of the operators.

A2A approximates an arbitrary component of the $M^{-1}$ as a sum of the outer product of pre-calculated vectors:
\begin{multline}
M^{-1} \simeq 
\sum_i^{N_l}   \left| \lambda_i \right> \frac{1}{\lambda_i} \left<\lambda_i\right| \\
+ \sum_{i'}^{N_h} \left( M^{-1} - \sum_i^{N_l}  \left| \lambda_i \right> \frac{1}{\lambda_i} \left<\lambda_i\right| \right)
\left| \eta_{i'} \right> \left< \eta_{i'} \right | \nonumber 
= \sum_{i=1}^{N_l+N_h} \left| v_i \right> \left<w_i\right| ,
\label{eq:A2A} 
\end{multline}
where $\left| \lambda_i \right>$ is the eigenvector with the eigenvalue $\lambda_i$ and $\{ \left| \eta_i \right> \}$ form an
orthonormal basis of stochastic sources.
The idea being that propagators from the stochastic sources include the necessary high frequency modes, while the eigenvector basis
captures the low lying part of the space.

The A2A technique turns the constructions of correlation functions from arbitrary lattice sites into a series of linear algebra
operations without the need for additional Dirac operator inversions or inter-node communication except for the global sums.
A sufficient number of eigenvectors for the low modes with efficiently constructed (stochastic) high modes allows for a reduction of
the measurement error from the A2A procedure to smaller than the gauge noise.
Further, there have been detailed studies in how to choose the pattern for selecting the $\left| \eta_i \right>$ which allows for
much better suppression of statistical error than the naive $1 / \sqrt{N_h}$ factor, by judiciously choosing the grouping of degree
of freedom which eliminates the most significant source of errors from random noise ({\em dilution}), or which allow for progressive
winnowing of random points by utilizing Hadamard vectors ({\em hierarchical probing} \cite{Stathopoulos:2013aci}).

\subsection{Multi--particle contractions}

The evaluation of many-body correlation functions is a challenging aspect of the workflow of lattice-QCD calculations.
One area where these calculations arise is the extraction of a few excited energy levels in many-hadron systems.
The intrinsic finite-volume nature of the calculations is actually an advantage.
The finite-volume energies, and matrix elements, can be related to their Minkowski space infinite volume scattering amplitudes via
the L\"uscher relations~\cite{Luscher:1990ux,Luscher:1991cf,Rummukainen:1995vs}.
The analytic behavior of these $S$-matrix amplitudes can be used to extract resonant properties of states including branching
fraction for decays that can then be directly confronted with experiment~\cite{Briceno:2016mjc}.
The low-energy behavior of scattering amplitudes in nuclear systems provides information on nuclear many-body
forces~\cite{Bedaque:2004kc}.

These calculations must necessarily include many-quark operator constructs that provide overlap onto the finite-volume energy
levels.
With a large basis of quark fields, a particularly challenging aspect is the large combinatoric connections between the operators
leading to large numbers of quark-line graphs that must be evaluated.
A consequence of variational calculations is that there are common intermediate contractions that can be cached.
Improved algorithms can find optimal orderings for constructing the temporaries and the graph evaluations, leading to significant
cost savings.
Research along these directions has been a focus of the Exascale Computing Project for lattice-QCD.
The large combinatorics can be ameliorated by recasting the fermionic integration over the quark fields into the determinant of
systems of equations~\cite{Detmold:2012eu,Vachaspati:2014bda}.
It is clear that more research is needed in these areas.

\section{Exploiting Leadership Computing Systems} \label{sec:hw}

\subsection{Historical perspective and machine building}

Lattice-QCD practitioners have been at the forefront of computing, to the extent of frequently designing special purpose systems
that were tuned to the needs of QCD calculations.
Notable examples of custom lattice-QCD systems in the U.S.\ include the
ACPMAPS~\cite{Mackenzie:1987fc}, QCDSP~\cite{Chen:1998cg}, and QCDOC~\cite{Boyle:2005:OQQ:1665957.1665969}
supercomputers.
ACPMAPS was built from Weitek XL and Intel i860 processors and a custom-built communications backplane such that processing speed,
latency, and data bandwith were all well balanced. QCDSP was built from Texas Instruments DSP processors with a network fabric
made of point-to-point serial links.
QCDOC combined IBM's System-on-a-chip intellectual property with a
PowerPC-440 embedded CPU, a Hummer floating point unit, a communications network built on high-speed serial
sinks (HSSL), and a custom memory controller.
This project was a sister project of IBM's Blue Gene/L system with close collaboration between the design teams and with some of the
co-designed components (such as the memory controller) being shared between the two systems.
The communications fabrics of QCDOC and QCDSP, were designed to provide sufficient
performance to allow the domain-wall  and staggered fermion formulations of quarks to sustain a specific minimum level of
performance even in the strong scaling limit.
Indeed both QCDOC and the Blue Gene line were noted for their excellent strong scaling properties.
USQCD researchers maintained their close links with IBM working as subcontractors in some cases on design elements of successor Blue
Gene/P and Blue Gene/Q architectures.
This work has also developed close ties with ALCF, where high performance lattice-QCD codes were used, for example, to troubleshoot
the Blue Gene racks on delivery.
With such a deep technical background, partnership with vendors, and ready software USQCD made excellent use of the line of Blue
Gene systems at ALCF and to a certain degree at Lawrence Livermore National Laboratory (LLNL).
QCD codes running on QCDSP and on a Blue Gene/L system at LLNL both won Gordon Bell Prizes.

In the space of computational clusters, innovations were made at Fermilab and Jefferson Lab to provide
highly cost-effective systems for analysis, including meshing cluster nodes in a
three-dimensional torus using gigabit ethernet links, Infiniband for adequate latency, and
introducing GPU accelerators.
The first GPU accelerator-based cluster was contructed in Europe~\cite{Egri:2006zm} with gaming GPUs, followed by a
larger-scale installation at Jefferson Lab and systems with scientific-computing GPUs at Fermilab .  Later, GPUs were incorporated into leadership-class systems such as Titan at OLCF or
BlueWaters at NCSA (2012).

\subsubsection{Utilizing GPUs}

The first GPUs used for lattice QCD were programmed using
the OpenGL graphics programming interface \cite{Egri:2006zm}.
The appearance of the more useable CUDA programming system spurred the USQCD development of the QUDA QCD library for GPUs
\cite{Clark:2009wm,Babich:2010mu,QUDADownload}  and its successful interfacing with the Chroma code
\cite{Edwards:2004sx, ChromaWeb}.
The original developers of QUDA found successful careers at Nvidia and maintained their links with the QCD community.  This collaboration led to improvements of QUDA especially in terms of strong scaling, using the Edge analysis
cluster at LLNL to strong scale lattice QCD to over 100 GPUs for the first time~\cite{Babich:2011np}.
In 2012, the era of GPUs in large petascale resources in the U.S.\ began with OLCF deploying Titan and NCSA deploying the BlueWaters
systems, respectively.
Lattice-QCD codes were, thus, ready to exploit these new resources as soon as they became available.
More recently USQCD researchers have been early users on the Summit system at OLCF, in particular providing
test cases and benchmarks to OLCF.
Some of the successful outcomes of this work are described in Sec.~\ref{subsec:caseStudy}.
USQCD researchers also started utilizing tensor cores, first for mixed precision inverters studied in Ref.~\cite{Tu:2018wuo}.

\subsubsection{Utilizing Intel Xeon Phi\texttrademark\ architecture}

Lattice-QCD researchers also developed links with Intel Corporation to develop efficient kernels for the Many Integrated
Core~(MIC) architecture devices, later code-named Xeon~Phi\texttrademark.
The work began on the Intel Knights Ferry system and proceeded to Intel Knights Corner, in partnership with engineers at Intel,
specifically  Intel Parallel Computing Labs at Intel Santa
Clara and Bangalore, India.
High-performance code for the Wilson formulation of QCD available in 2013~\cite{ISC13Phi,QPhiXLocation}.
USQCD also collaborated with European colleagues on Xeon Phi software~\cite{Heybrock:2014:LQD:2683593.2683602} and code for
staggered and domain-wall fermions was being developed contemporaneously and reported shortly after, {\em e.g.} in
Ref.~\cite{Kaczmarek:2014mga}.

With these experiences, USQCD was well positioned to partner with NERSC
during the procurement of the Cori system with sizeable Xeon
Phi Knights Landing (KNL) partition of approximately 9,600 nodes.
Overall three USQCD codes (Chroma, MILC, CPS) ask{were chosen to be} Tier-1 NERSC Exascale Application
Partnership (NESAP) codes, with another (QLua) becoming a Tier-2 code.
As part of this partnership, these codes were further developed and optimized
through hackathons and dungeon sessions with NERSC and Intel~\cite{DeTar:2016ndn,DBLP:conf/supercomputer/JooKKVW16}.
In the context of USQCD, Jefferson Lab deployed a KNL cluster in 2016 and enlarged it in 2018.
Brookhaven deployed one of the first production KNL cluster with dual rail Omni-Path network in 2017, which lead to the discovery
and the development of the solution for the Omni-Path performance issue with KNL-based systems~\cite{Boyle:2017xcy}.
The KNL developments enable USQCD researchers to be productive on Cori (at NERSC) and on Theta (at ALCF).
A~recent summary of KNL code performance worldwide can be found in
Ref.~\cite{Boyle:2017wul}.

\subsection{Current DOE leadership computational landscape}
The current computational landscape in the U.S.\ is in transition.
ALCF, while awaiting the forthcoming exascale Aurora system, still operates the Mira Blue Gene/Q machine, having supplemented it
with Theta, a system based on Intel KNL chips.
NERSC operates Cori, which is a combination of dual socket Haswell servers (data partition) and KNL systems (Cori KNL partition).

Recently, OLCF deployed Summit, which, at the time of writing, is the fastest computer
system in the world according to the Top 500 rankings~\cite{Top500}.
Summit is a GPU-based system featuring the latest Nvidia Tesla V100 (Volta) GPUs.  Each node has six GPU devices, connected with high-speed NVLink connections.  The nodes themselves are connected with Infiniband.
At the same time, OLCF continues to operate Titan, a Cray XT7 system featuring Tesla K20X GPUs and the Cray Gemini interconnect.
USQCD codes can readily exploit Summit by treating it as a regular (albeit large) GPU cluster, however its
Infiniband network can pose challenges to strong scaling.
USQCD has been running on Titan since its arrival.
Communications difficulties in GPU systems have driven software development activity to incorporate communication reduction and
avoidance techniques to alleviate the situation.

\subsection{Exascale and pre-exascale systems}

The ASCR roadmap anticipates three large systems: i) Perlmutter at
NERSC~\cite{Perlmutter}, ii) Aurora at ALCF~\cite{Aurora}, and iii) Frontier at OLCF~\cite{Frontier}.
The Perlmutter system will be delivered by Cray and will feature both a CPU (powered
by Advanced Micro Devices CPUs) and a GPU-enabled partition, utilizing Cray's next-generation ethernet-compatible Slingshot fabric.
From the point of view of existing GPU software, his architecture is an evolutionary step, so USQCD researchers should be well
placed to be able to use the system.

At this point in time, very few public details are available about the Aurora system.
The Early Science Program call for proposals~\cite{AuroraESP} lists a few items of guidance, such as the suggestion that the
architecture will be optimized to support codes with sections of fine grained concurrency, and that OpenMP~5 will likely contain the
constructs necessary to guide the compiler to get optimal performance.
 
Similarly, not much information is yet available regarding the forthcoming Frontier system from OLCF, except that delivery is
expected to begin in 2021 and that the system will support advanced simulation capabilities, high performance data analytics
and artificial intelligence applications.
The architecture might be expected to differ from that of Aurora.
 
As part of the Exascale Computing Project (ECP), some USQCD researchers are involved in co-design activities with ECP industrial
partners.
It is hoped that the combination of co-design partnerships, participation in early science projects (NESAP2, Aurora ESP etc.), and
direct collaboration with ALCF, OLCF, and NERSC will enable the USQCD community to continue to be ready these new pre-exascale and exascale systems 
as soon as they become available.
 
\subsection{General hardware trends}

\subsubsection{On-node parallelism}
The general trend of new and future hardware is towards systems with
increasingly high compute densities.
This manifests itself primarily in ever more powerful nodes, with the main power of the node often coming from on-node parallelism
of some nature, either via many cores on the node or via accelerator devices such as GPUs.
The primary forms of parallelism on node have typically been multi-threading on CPU cores, single-instruction multiple-thread (SIMT)
parallelism on GPUs and single-instruction multiple-data (SIMD) parallelism on CPUs (also known as vector parallelism).

\subsubsection{Mixed precision}
Graphics and machine-learning applications typically tolerate reduced precision compared with traditional
scientific computing applications.  Therefore, recent hardware provide
higher 32-bit and 16-bit processing rates at the expense of double precision.
For example, the tensor cores on the Nvidia Volta architecture enable OLCF's Summit to
already breach the {\em exa-ops} barrier~\cite{2018arXiv181001993K}.
Intel, in turn, has announced the Knights Mill architecture which is an Intel Xeon Phi chip with improved single precision
performance compared to the previous Knights Landing architecture~\cite{KNM} designed specifically to serve the artificial
intelligence market.
This trend can be expected to persist into the future, requiring developers of
high performance libraries and frameworks to invest in codes that allow mixing of precisions and to follow developments in
multi-precision algorithms.

\subsubsection{Memory bandwidth}
Nodes typically host a variety of memory speeds, often including high-bandwidth memory.
On recent GPU systems, this has been on-device HBM/HBM2, whereas in KNL systems it has been on-package MCDRAM.
For applications whose data fit into the fast memory completely, no real management of the fast memory was required.
For others, a variety of solutions became available over time.
Initially, one had to manage the memory explicitly and, in some cases, software cache systems were developed
\cite{Winter:2014:FLQ:2650283.2650646}.
KNL could be run in a {\em cache mode} where the fast MCDRAM acted as a direct mapped level-3 cache that required no user intervention, at the cost of sacrificing some memory bandwidth.
More recently, unified virtual memory and managed memory were developed on the Nvidia GPUs to reduce the burden of explicit
management of transfers between host and device memory spaces.

\subsubsection{Inter-processing-element bandwidth}
While compute nodes have been getting denser and more powerful, the balance of memory bandwidth within the compute devices and
between them has generally deteriorated.
For example, the Blue Gene/Q system featured 42.6~GB/sec bandwidth to DRAM and 10 inter-processor links each running at 2.0~GB/sec
giving a potential maximum off-chip transfer rate of 20~GB/sec which is approximately half the DRAM bandwidth.
On the other hand, Summit contains 6 Nvidia Volta GPUs each having 900~GB/sec bandwidth to stacked high bandwidth HBM2 memory.
Thus, the node provides a total of about 5400~GB/sec of memory bandwidth (without even considering the Power9 CPU sockets).
Each GPU has three NVLink 2.0 bidirectional links running at 50~GB/sec per direction.
Hence, the total off chip bandwidth of a GPU is 300~GB/sec (bidirectional) or 150~GB/sec in one direction (into or out of the GPU).
The 300~GB/sec NVLink bandwidth is one-third of the HBM2 memory bandwidth.
In turn, the maximum bidirectional bandwidth from the network interface card into the dual-rail Infiniband fabric is about 50~GB/sec.
This is now less than 1\% the bandwidth available within the node from the HBM2 memories on the GPUs.

The expectation of future systems is that this trend will continue, and the memory hierarchies will deepen (cache, fast memory
like HBM2, DRAM, NVRAM, remote-node memory via fabric, disk).
The compute elements will have access to limited memories with very high bandwidth, but the deeper one gets into the hierarchy the
lower the bandwidth to the compute node will become.
One can also anticipate islands of relatively high bandwidth connectivity within a node, e.g., GPUs connected with NVLink.
This skewing of the balance of inter-processing-element bandwidth to the fastest available memory bandwidth, can become a serious
bottleneck for strong scaling, where using more compute nodes results in a smaller per-node problem, thus placing a heavier demand
on communications between processing elements, for example for halo exchange.
As such, research into algorithms that can reduce or avoid communications continues to be of the essence.

\section{Software} \label{sec:sfw}
Having a dependable, portable, and efficient software base is key to exploiting both current leadership class systems and future
exascale systems.
The software needs to address a variety of needs.
First and foremost, it needs to be able to exploit the current generation of systems, in order to deliver the science goals of
today.
Second, it should provide a good basis for development to be able to exploit future systems.
It should be modular and extensible to allow the addition or integration of high performance components addressing future
architectures, and it should allow the exploration of new algorithmic directions.

\subsection{Portable and efficient software from SciDAC}
Through funding under multiple iterations of the DOE SciDAC program, USQCD has developed a layered approach to software.
The four primary layers considered are: i) a communications layer called QCD Message Passing (QMP) ii) a data parallel productivity
layer called QCD Data Parallel (QDP) iii) a layer for optimized components, primarily of Dirac operators and linear solvers known as
`'Level~3'', and iv) an application layer which through judicious use of the layers below could encode the physics algorithms and be
used for production.
This last layer features the well-known lattice-QCD application codes MILC, Chroma, the Columbia Physics System (CPS), as well
as newer code frameworks such as QLua and FUEL.

The QMP layer was defined to allow portable communications to systems that may lack a full-blown implementation of MPI, such as
custom QCD systems like the QCDOC with its custom serial communications, and the Jefferson Lab gigabit ethernet clusters, on which
QMP was implemented using M-VIA.
The QDP layer had several implementations, including one over C++ known as QDP++ and one over C called QDP/C.
QDP++ uses C++ expression template mechanisms to provide operator overloading, allowing the majority of code to follow the
mathematics in a straightforward way.

Level~3 introduced a variety of high performance code libraries over the years, including SSE implementations of Wilson-Dirac
operator, custom high performance implementation of the Wilson-Dirac and domain-wall fermion operators via the BAGEL library
\cite{Boyle2009TheBA}, optimized solvers for M\"obius domain-wall fermions in the MDWF
library~\cite{MDWFWeb}, the first publicly available implementation of multi-grid for Wilson-clover fermions
(QOP-MG)~\cite{QOPQDP,Osborn:2010mb}, linear solvers and QCD software components for GPUs in the QUDA
library~\cite{Clark:2009wm} and for Xeon Phi and Xeon architectures in the QPhiX library~\cite{ISC13Phi}, and others.

These high-performance libraries also serve as crucial laboratories for developing new techniques.
A prime example is the QUDA library, which, apart from its advanced solver algorithms, features a host of performance optimizations
for GPUs, including per-kernel performance autotuning.  Reduced precision is used where possible in the Krylov iterative steps and extensively in
preconditioners.
QUDA pioneered the use of gauge-link compression for performance (hitherto it has been used primarily to reduce file sizes on I/O)
and the use of 16-bit precision in lattice QCD.
QUDA supports a variety of inter GPU communications including regular MPI (for between nodes), as well peer-to-peer and GDR for GPUs
within a node or connected by NVLink.
The choice of communications strategy is also tuned automatically.
QUDA was also pioneering in the use of a domain decomposition preconditioner with the aim of improving scalability and continues to
innovate in the area of communications-reduction and avoidance.
Many developments in the investigation of multi-grid algorithms (e.g., multi-grid for staggered fermions, communications avoidance
in smoothers and the bottom solver, and reduced precision on GPUs) take place within the QUDA library.
The library continues to be a testing ground for work in block solvers and extended Krylov-subspace solvers, as discussed in
Sec.~\ref{sec:solvers}.

In turn there have been innovations in other libraries.
BAGEL, for example, pioneered the utilization of reduced precision on halo-swaps, some investigations of features of
programming models (nested parallelism in OpenMP, use of the Kokkos~\cite{CarterEdwards:2014:KOK:2841458.2841785} programming model)
has taken place in the mg\_proto~\cite{MGPROTO} library, while SIMD aware layouts have been pioneered in BAGEL/BFM and are now in
use in the Grid~\cite{Boyle:2015tjk} software.
Recursive-descent, cache-oblivious techniques are used in the MDWF library~\cite{MDWFWeb}.

\subsubsection{Adapting to disruption}
Occasionally, new hardware technologies can disrupt existing programming paradigms.
The arrival of GPUs was a prime example.
One issue with GPUs was that due to the degree of acceleration they could provide to highly optimized code, other parts of an
application which were not accelerated, could become a bottleneck due to Amdahl's law.
Examples of this are various analysis tasks such as the creation of quark sources, which were insignificant before solvers were
accelerated by GPUs and by algorithmic improvements, but have now become bottlenecks.
In the gauge generation phase, code outside linear solvers in the computation of molecular dynamics forces suddenly came to
dominate.
One approach was to re-code the various routines one by one and optimize them for the GPU.

Another way was adopted in QDP++.
In this instance, the expression templates of QDP++, which had hitherto generated code for the lattice-QCD expressions within the
C++ compiler, were rewritten to generate code at run-time in a dynamic manner using a just-in-time (JIT) approach.
This implementation of QDP++ is known as QDP-JIT~\cite{Winter:2014:FLQ:2650283.2650646}.
The current QDP-JIT implementation generates code at runtime using the LLVM compiler framework, which targets GPUs using the NVPTX
back-end and also has targets for most currently available mass market CPUs such as Power9, Intel x86, and compatible chips including Knights Landing and AVX512 support.
By accelerating the entire QDP++ framework, all application code on top of it, including Chroma, is likewise automatically
accelerated, thereby reducing the effects of Amdahl's law.

\subsection{Developments under the Exascale Computing Project}

USQCD software development currently is being undertaken under a SciDAC-4 project supported by the Office of Science, Office of
Nuclear Physics and ASCR and through the Exascale Computing Project funded by ASCR.
Broadly speaking, the purpose of SciDAC funding is to enable partnerships to make the best effective use of the current DOE
leadership facilities, while the Exascale Computing Project (ECP) aims to ensure that high quality scientific software is available
for the forthcoming pre-exascale and exascale systems.
The majority of the software discussion above focuses on developments that were funded primarily by
SciDAC.  Here, other activities in the scope of USQCD's ECP participation are discussed.

As noted above, USQCD has several strands of work under ECP to carry out research in algorithms for improved linear solvers,
to research gauge generation algorithms addressing critical slowing down, and to improve methods of post analysis.
The ECP project also features software development, including the development of production systems, as well as experimentation with
new hardware, and programming models.

Two primary data parallel layers being developed are Grid~\cite{Boyle:2015tjk} and QEX~\cite{Jin:2016ioq}.
Grid is a data-parallel framework similar to and inspired by QDP++.
Unlike QDP-JIT, it utilizes GPUs through static compilation and features a data layout that is highly beneficial for SIMD processing
on CPUs.
QEX uses a new language called Nim which is a higher-level language similar to Python or Julia but featuring multiple dispatch, a
module system similar to Python, incremental multi-stage compilation and an extremely rich approach to metaprogramming.
Nim itself compiles to C directly with an extremely friendly C foreign-function interface to
existing libraries.
Both of these frameworks serve to remedy missing features of the previous QDP++ and QDP software layers (such as the ability to have
multiple grids in scope in QDP++ for use in multi-grid codes) and to provide features the earlier software layers perhaps didn't
anticipate such as the ability to carry out the split-grid approach discussed earlier.

Apart from the re-architecting of frameworks and applications, the software work under ECP also includes experimentation with novel
hardware in partnership with several Path Forward projects with vendors, experimentation with new programming models such as Kokkos,
or by combining and improving existing programming models with C++ features (e.g., examining the interplay of directive based models
like OpenACC and OpenMP with expression template methods).
These exploratory projects typically select a set of kernels and attempt to implement them in the new programming model, to observe
the resulting performance and to note any programming pain points.
In this way, USQCD software work is synergistic with and beneficial to other ECP software technology projects.

\section{Opportunities using Machine Learning}
\label{sec:ml}

Machine learning (ML) is emerging as a powerful and versatile computational tool for scientific
applications~\cite{Mehta:2018dln,Goodfellow-et-al-2016}.
For lattice QCD, ML methods promise to accelerate typical computational workflows and also to enable new directions of exploration
within lattice field theory.
To be effective in the lattice-QCD context, ML methods must be adapted significantly from their use in other fields.
In particular, ML implementations that incorporate the complex exact and approximate symmetries of lattice-QCD datasets must be
developed, and rigorous methods of uncertainty quantification and error propagation must be investigated.
While some promising early explorations of ML for lattice QCD have been undertaken recently, these applications are not
straightforward and to successfully apply them at the scale of state-of-the-art lattice-QCD calculations will need continued
targeted research and development.
\\

\begin{figure}
    \centering
    \includegraphics[width=0.45\linewidth]{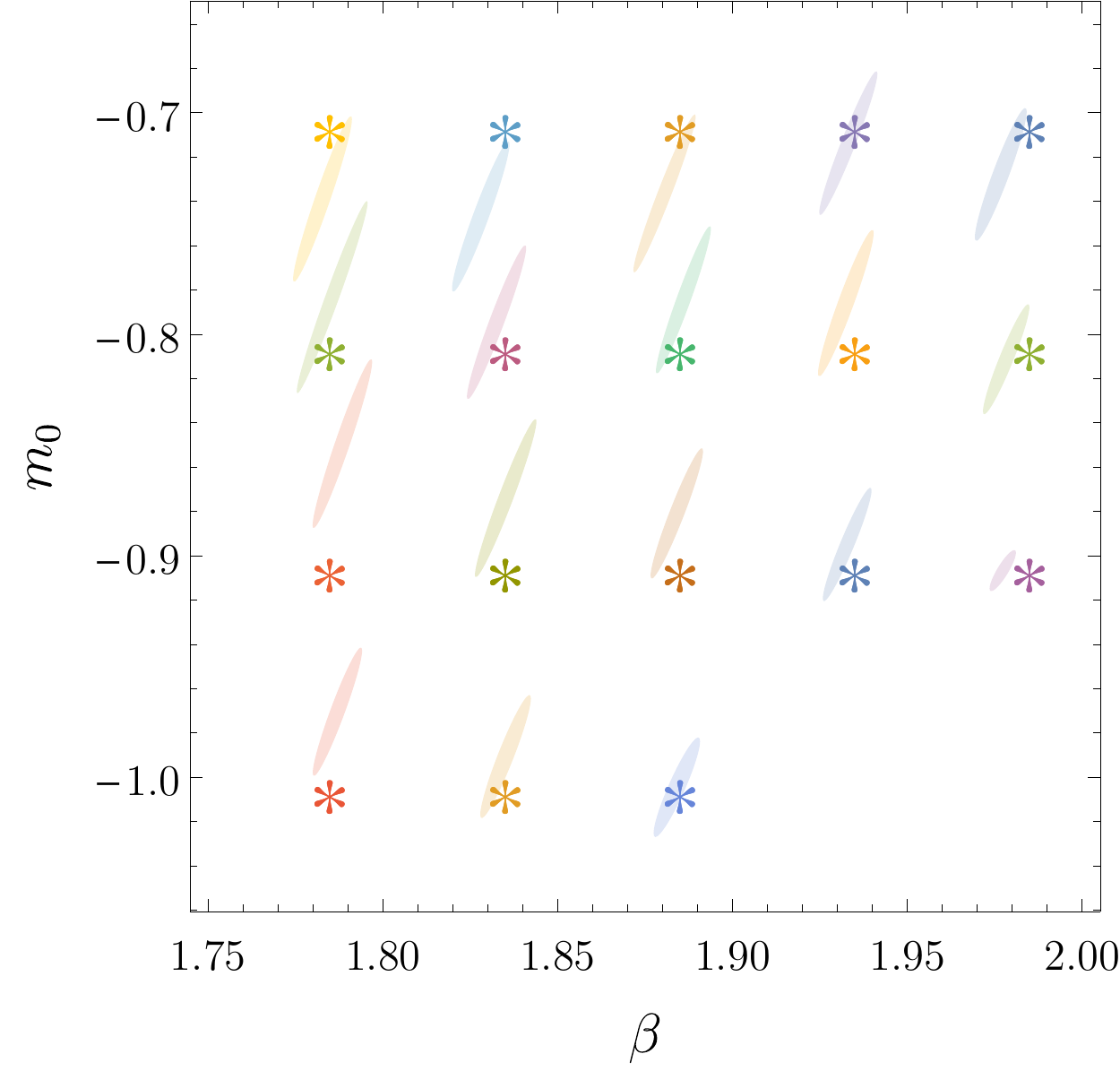}
    \caption{Machine learning applied to parameter regression in Ref.~\cite{Shanahan:2018vcv}.}
    \label{fig:figv116}
\end{figure}

\noindent{\it Gauge field generation}\\
Several ML approaches have been proposed to accelerate the generation of lattice-QCD gauge-field configurations. 
\epj{
Particularly promising are efforts to overcome critical slowing down in HMC as the lattice spacing is 
decreased~\cite{Albergo:2019eim,Shanahan:2018vcv} (Fig.~\ref{fig:figv116}), and ML approaches aimed at reducing autocorrelation within HMC~\cite{Tanaka:2017niz}.
While these studies for lattice QCD are in early stages, for several of these algorithms it has been proven that they maintain the balance and ergodicity properties of HMC and are guaranteed to produce the correct probability distribution in the infinite statistics limit.}
There are also considerable parallel efforts to accelerate Monte Carlo methods in related systems~\cite{Wang:2017mzw,Beyl:2017kwp,Xu:2017vug}; successes in these simpler systems are promising for QCD.
\\
	
\noindent{\it Parameter optimization}\\
Algorithms at all stages of the lattice-QCD workflow involve a significant number of tunable continuous and discrete parameters, whose
optimal values are difficult to determine.
For example, optimization of the many parameters defining HMC gauge generation (such as the concrete composition of the action to
simulate the fermion determinant, and various tunable parameters in the molecular dynamics such as step sizes) is a complex task
with strong correlations between parameters choices.
Similarly, significant tuning is involved in the construction of approximate nullspaces in multi-grid solvers on large lattices, where
the multi-grid scheme can involve a large number of levels and solver parameters at each level.
In cases such as these, the parameter space is too large to feasibly explore using simple grid searches.
ML tools such as Bayesian optimization using Gaussian processes are natural candidates as tools to optimize the parameters.
\\

\noindent{\it Sparse matrix inversions}\\
As discussed in Sec.~\ref{sec:solvers}, sparse matrix inversions are a significant cost in lattice-QCD calculations.
Preliminary explorations in applying ML to this task have recently been reported \cite{2016arXiv160605560Y} with promising results.
Showing that the approach works at the scale of state-of-the-art lattice QCD simulations remains challenging.
\\

\noindent{\it Correlator optimization}\\
ML techniques may provide a method to optimize lattce-QCD interpolating operators to minimize overlap onto unwanted states.
This would reduce the computation required to determined properties of states which can be accessed with current techniques, and
allow states which have previously been inaccessible to be studied.
This is particularly relevant for nuclei, where exponentially more complex operator constructions are required to effectively
suppress unwanted contamination of excited states.
A number of ML approaches well suited to this task have been applied successfully to similar problems in molecular design and drug
discovery.
Recently, an ML approach to determining estimators for lattice QCD has also been explored~\cite{Yoon:2018krb}.
\\
	
\noindent{\it Phase transitions}\\
ML has been successfully applied to detection of phase transitions and discovery of order parameters in a number of condensed matter
systems \cite{2017NatPh..13..431C,2016PhRvB..94p5134T,2017PhRvX...7c1038C,Li:2017xaz}.
A first application of similar methods in the context of quantum field theory has recently been used to study the deconfinement
transition in SU(2) Yang-Mills theory at finite temperature \cite{Wetzel:2017ooo}.
Further refinements may enable new insights into the nature of the QCD deconfinement and chiral symmetry restoration transitions.
\\
	
\noindent{\it Lefschetz thimbles and learnifolds}\\
In recent studies of theories with sign problems, which are difficult to study using standard Monte Carlo methods (such as lattice
QCD at nonzero baryon density), a new approach has emerged based on complexification of the necessary integrations.
Recently, variants of this approach based on optimizing the integration manifold using machine learning have been developed, with
successful applications to the 1+1-dimensional Thirring model~\cite{Alexandru:2017czx} and scalar $\phi^4$ theory in 1+1
dimensions~\cite{Mori:2017pne,Mori:2017nwj}.
While promising, extending these approaches to finite-density QCD is an extremely challenging long-term research problem.


\section{Quantum Computing and Quantum Information Science}\label{sec:quantum}


USQCD has carefully studied its HPC and workforce
requirements during previous self-studies and through DOE initiated review
processes, such as the Exascale Requirements Review that was completed
in 2016. 
We have determined that, with access to exascale computing
resources, we can determine many of the strong-interaction physics
quantities that are important to the DOE missions in nuclear physics and in
high-energy physics with the required precision.  Indeed, that is the focus of this and the companion
whitepapers~\cite{Bazavov:2018qcd,Detmold:2018qcd,Davoudi:2018qcd,Kronfeld:2018qcd,Lehner:2018qcd,Brower:2018qcd}.
That said, we have also
determined that there are other important physics quantities that will not
be able to be determined with sufficient precision, even with access to
exascale computing resources or beyond. 
Precision computations directly
from lattice QCD of modest- and high-density systems, the real-time
evolution of systems, transport in systems, out of equilibrium systems, and
high-energy inelastic processes, such as fragmentation, are simply out of
reach.

\begin{figure}
    \includegraphics[width=0.6\textwidth]{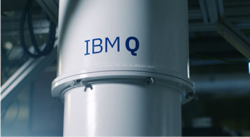} \hfill
    \includegraphics[width=0.32\textwidth]{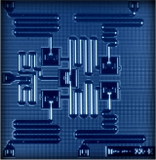}
    \caption{Dilution refrigerator housing the superconducting quantum hardware (left) and the IBM QX2 quantum chip comprised of
    five qubits and radio-frequency couplings (right).}
    \label{fig:qc-ibm}
\end{figure}

As USQCD is an organization that has been at the forefront of HPC architectures and their development.  Examples include ACPMAPS, QCDSP, and QCDOC, which were designed by lattice-gauge theory teams, as well as early adoption of
IBM Blue Gene, GPUs, Intel Xeon Phis.  Through its long history of close collaboration with technology industry and vendors, it is natural for us to seek forward-looking
solutions to our current challenges.
The rapid developments in quantum information science (QIS) and quantum computing (QC)
make it time to explore how quantum
devices can complement classical computing resources.
At present, cloud-accessible quantum devices at IBM (Fig.~\ref{fig:qc-ibm}) and Rigetti are available, as is private access to trapped ion systems.
Further, there us the expectation that a range of such devices will become readily available, along with simulators and technical
support.
Finally, technology companies such as Google, IBM, Intel, and Microsoft expect to have universal quantum computers with
about 50-qubits (without error correction) available now. 
IBM announced an
operational 50-qubit QC near the end of 2017, and currently has a 20-qubit
QC, a 16-qubit QC and two 5-qubit QCs available to users through the IBM
Q Experience web-interface.
The architectures of these QCs encompass
superconducting qubits (IBM, Google and Intel) and topological qubits
(Microsoft).
D-Wave’s quantum devices use quantum annealing to address
minimization problems.
A growing number of other technology companies
are also building programmable QCs, such as Rigetti and IonQ. In addition,
some university- and national-laboratory-based groups are making significant progress in quantum
simulation with recent results announced in quantum many-body systems
obtained with programmable quantum simulators using more than 50 cold
trapped ions as qubits.

Current hardware specifications, such as number of qubits, coherence times,
measurement errors, are presently very restrictive, a circumstance that has been dubbed the 
noisy intermediate scale quantum (NISQ) era by John \epj{Preskill~\cite{Preskill2018quantumcomputingin}}.  It is therefore unlikely that
realistic quantum field theories will be simulated on such
devices in the foreseeable future.
In many ways, the present situation is
reminiscent of the situation lattice QCD found itself in during the 1970s
with classical computing. Then, it was clear that certain classes
of problems would be solvable with sufficient classical computing resources
and with complementary developments in algorithms for linear algebra and
ones specific to quantum field theory. With a quantum advantage for some
problems in atomic and molecular systems expected to be gained with greater
than about 50 qubits supporting a modest size gate depth, a quantum
advantage for QCD is expected to require many orders of magnitude more
and with error correction. However, with such capabilities, we expect to be
able to perform reliable calculations of non-equilibrium systems, of the
fragmentation of hadrons, and high-density strongly interacting systems.

It is conceivable that the first quantum devices to be deployed in a
meaningful way to tackle QCD will be embedded in exascale, or beyond,
classical computing hardware, and will consist of a few-qubit systems
distributed within classical compute nodes, in analogy with the deployment
of GPUs in present and future classical hardware. This scenario suggests
that lattice QCD could benefit from developing hybrid classical-quantum
algorithms to perform some of the tasks currently performed solely on
classical hardware or for new tasks for systems currently out of reach.

Given that this is only one possible scenario, it seems wise for a fraction of the USQCD collaboration to embark on an effort to
understand the potential of QC and QIS for QCD calculations important to high-energy and nuclear physics of the future.  It makes sense to learn how to use available hardware to develop an understanding of this new computing technology, which can then be deployed to address the USQCD mission and begin to develop relevant algorithms.
We anticipate the next several years to involve some retooling for some of the collaboration, a period to develop ties with
technology companies, and to learn from other communities.
USQCD has established a committee to perform ongoing evaluations of QIS and QC as it relates to quantum field theory and quantum
chromodynamics.
This committee is expected to provide regular updates to the USQCD Executive Committee.
Finally, we expect that a funding framework similar in spirit to the SciDAC program would serve us well, as it continues to do so
for classical computing developments.

As outlined in a whitepaper prepared for the US~DOE,
``Quantum Computing for Theoretical Nuclear Physics'' \cite{Carlson:2018doe},
one can imagine that a plausible approach to begin to understand how to compute properties
of quantum field theories with quantum computers is to begin by analyzing
low-dimensional theories, starting with scalar field theory and the
Schwinger model. Then increasing the dimensionality of the systems,
moving to non-Abelian theories, and then finally to 3+1 dimensional QCD.
This would include optimizing the layout on qubits, optimizing quantum
circuits and then running codes on available quantum devices. It is
conceivable that the need for efficient Trotterization of the QCD evolution
operator could drive the design of quantum hardware and communication
fabrics in the same way it did in classical computing. These are obvious
and conceptually simple directions to consider.

\bibliography{bibliography,usqcd-wp}

\begin{thebibliography}{142}%
\makeatletter
\providecommand \@ifxundefined [1]{%
 \@ifx{#1\undefined}
}%
\providecommand \@ifnum [1]{%
 \ifnum #1\expandafter \@firstoftwo
 \else \expandafter \@secondoftwo
 \fi
}%
\providecommand \@ifx [1]{%
 \ifx #1\expandafter \@firstoftwo
 \else \expandafter \@secondoftwo
 \fi
}%
\providecommand \natexlab [1]{#1}%
\providecommand \enquote  [1]{``#1''}%
\providecommand \bibnamefont  [1]{#1}%
\providecommand \bibfnamefont [1]{#1}%
\providecommand \citenamefont [1]{#1}%
\providecommand \href@noop [0]{\@secondoftwo}%
\providecommand \href [0]{\begingroup \@sanitize@url \@href}%
\providecommand \@href[1]{\@@startlink{#1}\@@href}%
\providecommand \@@href[1]{\endgroup#1\@@endlink}%
\providecommand \@sanitize@url [0]{\catcode `\\12\catcode `\$12\catcode
  `\&12\catcode `\#12\catcode `\^12\catcode `\_12\catcode `\%12\relax}%
\providecommand \@@startlink[1]{}%
\providecommand \@@endlink[0]{}%
\providecommand \url  [0]{\begingroup\@sanitize@url \@url }%
\providecommand \@url [1]{\endgroup\@href {#1}{\urlprefix }}%
\providecommand \urlprefix  [0]{URL }%
\providecommand \Eprint [0]{\href }%
\providecommand \doibase [0]{http://dx.doi.org/}%
\providecommand \selectlanguage [0]{\@gobble}%
\providecommand \bibinfo  [0]{\@secondoftwo}%
\providecommand \bibfield  [0]{\@secondoftwo}%
\providecommand \translation [1]{[#1]}%
\providecommand \BibitemOpen [0]{}%
\providecommand \bibitemStop [0]{}%
\providecommand \bibitemNoStop [0]{.\EOS\space}%
\providecommand \EOS [0]{\spacefactor3000\relax}%
\providecommand \BibitemShut  [1]{\csname bibitem#1\endcsname}%
\let\auto@bib@innerbib\@empty
\bibitem [{\citenamefont {Bazavov}\ \emph {et~al.}(2019)\citenamefont
  {Bazavov}, \citenamefont {Karsch}, \citenamefont {Mukherjee},\ and\
  \citenamefont {Petreczky}}]{Bazavov:2018qcd}%
  \BibitemOpen
  \bibfield  {author} {\bibinfo {author} {\bibfnamefont {Alexei}\ \bibnamefont
  {Bazavov}}, \bibinfo {author} {\bibfnamefont {Frithjof}\ \bibnamefont
  {Karsch}}, \bibinfo {author} {\bibfnamefont {Swagato}\ \bibnamefont
  {Mukherjee}}, \ and\ \bibinfo {author} {\bibfnamefont {Peter}\ \bibnamefont
  {Petreczky}} (\bibinfo {collaboration} {USQCD}),\ }\bibfield  {title}
  {\enquote {\bibinfo {title} {Hot-dense lattice {QCD}},}\ }\href@noop {} {\
  (\bibinfo {year} {2019})}\BibitemShut {NoStop}%
\bibitem [{\citenamefont {Brower}\ \emph {et~al.}(2019)\citenamefont {Brower},
  \citenamefont {Hasenfratz}, \citenamefont {Neil} \emph
  {et~al.}}]{Brower:2018qcd}%
  \BibitemOpen
  \bibfield  {author} {\bibinfo {author} {\bibfnamefont {Richard}\ \bibnamefont
  {Brower}}, \bibinfo {author} {\bibfnamefont {Anna}\ \bibnamefont
  {Hasenfratz}}, \bibinfo {author} {\bibfnamefont {Ethan~T.}\ \bibnamefont
  {Neil}},  \emph {et~al.} (\bibinfo {collaboration} {USQCD}),\ }\bibfield
  {title} {\enquote {\bibinfo {title} {Lattice gauge theory for physics beyond
  the {Standard Model}},}\ }\href@noop {} {\  (\bibinfo {year}
  {2019})}\BibitemShut {NoStop}%
\bibitem [{\citenamefont {Cirigliano}\ \emph {et~al.}(2019)\citenamefont
  {Cirigliano}, \citenamefont {Davoudi} \emph {et~al.}}]{Davoudi:2018qcd}%
  \BibitemOpen
  \bibfield  {author} {\bibinfo {author} {\bibfnamefont {Vincenzo}\
  \bibnamefont {Cirigliano}}, \bibinfo {author} {\bibfnamefont {Zohreh}\
  \bibnamefont {Davoudi}},  \emph {et~al.} (\bibinfo {collaboration} {USQCD}),\
  }\bibfield  {title} {\enquote {\bibinfo {title} {The role of lattice {QCD} in
  searches for violations of fundamental symmetries and signals for new
  physics},}\ }\href@noop {} {\  (\bibinfo {year} {2019})}\BibitemShut
  {NoStop}%
\bibitem [{\citenamefont {Detmold}\ \emph {et~al.}(2019)\citenamefont
  {Detmold}, \citenamefont {Edwards} \emph {et~al.}}]{Detmold:2018qcd}%
  \BibitemOpen
  \bibfield  {author} {\bibinfo {author} {\bibfnamefont {William}\ \bibnamefont
  {Detmold}}, \bibinfo {author} {\bibfnamefont {Robert~G.}\ \bibnamefont
  {Edwards}},  \emph {et~al.} (\bibinfo {collaboration} {USQCD}),\ }\bibfield
  {title} {\enquote {\bibinfo {title} {Hadrons and nuclei},}\ }\href@noop {} {\
   (\bibinfo {year} {2019})}\BibitemShut {NoStop}%
\bibitem [{\citenamefont {Kronfeld}\ \emph {et~al.}(2019)\citenamefont
  {Kronfeld}, \citenamefont {Richards} \emph {et~al.}}]{Kronfeld:2018qcd}%
  \BibitemOpen
  \bibfield  {author} {\bibinfo {author} {\bibfnamefont {Andreas~S.}\
  \bibnamefont {Kronfeld}}, \bibinfo {author} {\bibfnamefont {David~G.}\
  \bibnamefont {Richards}},  \emph {et~al.} (\bibinfo {collaboration}
  {USQCD}),\ }\bibfield  {title} {\enquote {\bibinfo {title} {Lattice {QCD} and
  neutrino-nucleus scattering},}\ }\href@noop {} {\  (\bibinfo {year}
  {2019})}\BibitemShut {NoStop}%
\bibitem [{\citenamefont {Lehner}\ \emph {et~al.}(2019)\citenamefont {Lehner},
  \citenamefont {Meinel} \emph {et~al.}}]{Lehner:2018qcd}%
  \BibitemOpen
  \bibfield  {author} {\bibinfo {author} {\bibfnamefont {Christoph}\
  \bibnamefont {Lehner}}, \bibinfo {author} {\bibfnamefont {Stefan}\
  \bibnamefont {Meinel}},  \emph {et~al.} (\bibinfo {collaboration} {USQCD}),\
  }\bibfield  {title} {\enquote {\bibinfo {title} {Opportunities for lattice
  {QCD} in quark and lepton flavor physics},}\ }\href@noop {} {\  (\bibinfo
  {year} {2019})}\BibitemShut {NoStop}%
\bibitem [{\citenamefont {{Jo\'o}}\ \emph {et~al.}(2019)\citenamefont
  {{Jo\'o}}, \citenamefont {Jung} \emph {et~al.}}]{Joo:2018qcd}%
  \BibitemOpen
  \bibfield  {author} {\bibinfo {author} {\bibfnamefont {{B\'alint}}\
  \bibnamefont {{Jo\'o}}}, \bibinfo {author} {\bibfnamefont {Chulwoo}\
  \bibnamefont {Jung}},  \emph {et~al.} (\bibinfo {collaboration} {USQCD}),\
  }\bibfield  {title} {\enquote {\bibinfo {title} {Status and future
  perspectives for lattice gauge theory calculations to the exascale and
  beyond},}\ }\href@noop {} {\  (\bibinfo {year} {2019})}\BibitemShut {NoStop}%
\bibitem [{PAN()}]{PANDA}%
  \BibitemOpen
  \href@noop {} {\enquote {\bibinfo {title} {{PANDA}},}\ }\bibinfo
  {howpublished}
  {\href{http://news.pandawms.org/panda.html}{http://news.pandawms.org/panda.html}}\BibitemShut
  {NoStop}%
\bibitem [{\citenamefont {Duane}\ \emph {et~al.}(1987)\citenamefont {Duane},
  \citenamefont {Kennedy}, \citenamefont {Pendleton},\ and\ \citenamefont
  {Roweth}}]{Duane:1987de}%
  \BibitemOpen
  \bibfield  {author} {\bibinfo {author} {\bibfnamefont {S.}~\bibnamefont
  {Duane}}, \bibinfo {author} {\bibfnamefont {A.~D.}\ \bibnamefont {Kennedy}},
  \bibinfo {author} {\bibfnamefont {B.~J.}\ \bibnamefont {Pendleton}}, \ and\
  \bibinfo {author} {\bibfnamefont {D.}~\bibnamefont {Roweth}},\ }\bibfield
  {title} {\enquote {\bibinfo {title} {{Hybrid Monte Carlo}},}\ }\href
  {\doibase 10.1016/0370-2693(87)91197-X} {\bibfield  {journal} {\bibinfo
  {journal} {Phys. Lett.}\ }\textbf {\bibinfo {volume} {B195}},\ \bibinfo
  {pages} {216--222} (\bibinfo {year} {1987})}\BibitemShut {NoStop}%
\bibitem [{\citenamefont {Andrieu}\ \emph {et~al.}(2003)\citenamefont
  {Andrieu}, \citenamefont {{De Freitas}}, \citenamefont {Doucet},\ and\
  \citenamefont {Jordan}}]{e2544badcbd543a481cc0bdf45041dc4}%
  \BibitemOpen
  \bibfield  {author} {\bibinfo {author} {\bibfnamefont {Christophe}\
  \bibnamefont {Andrieu}}, \bibinfo {author} {\bibfnamefont {Nando}\
  \bibnamefont {{De Freitas}}}, \bibinfo {author} {\bibfnamefont {Arnaud}\
  \bibnamefont {Doucet}}, \ and\ \bibinfo {author} {\bibfnamefont {Michael~I.}\
  \bibnamefont {Jordan}},\ }\bibfield  {title} {\enquote {\bibinfo {title} {An
  introduction to {MCMC} for machine learning},}\ }\href {\doibase
  10.1023/A:1020281327116} {\bibfield  {journal} {\bibinfo  {journal} {Machine
  Learning}\ }\textbf {\bibinfo {volume} {50}},\ \bibinfo {pages} {5--43}
  (\bibinfo {year} {2003})}\BibitemShut {NoStop}%
\bibitem [{\citenamefont {Sch{\"u}tte}\ \emph {et~al.}(1999)\citenamefont
  {Sch{\"u}tte}, \citenamefont {Fischer}, \citenamefont {Huisinga},\ and\
  \citenamefont {Deuflhard}}]{SchuetteFischerHuisingaetal.1999}%
  \BibitemOpen
  \bibfield  {author} {\bibinfo {author} {\bibfnamefont {C.}~\bibnamefont
  {Sch{\"u}tte}}, \bibinfo {author} {\bibfnamefont {A.}~\bibnamefont
  {Fischer}}, \bibinfo {author} {\bibfnamefont {W.}~\bibnamefont {Huisinga}}, \
  and\ \bibinfo {author} {\bibfnamefont {P.}~\bibnamefont {Deuflhard}},\
  }\bibfield  {title} {\enquote {\bibinfo {title} {A direct approach to
  conformational dynamics based on hybrid {Monte Carlo}},}\ }\href
  {http://www.zib.de/PaperWeb/abstracts/SC-98-45} {\bibfield  {journal}
  {\bibinfo  {journal} {J. Comput. Phys.}\ }\textbf {\bibinfo {volume} {151}},\
  \bibinfo {pages} {146 -- 168} (\bibinfo {year} {1999})}\BibitemShut {NoStop}%
\bibitem [{\citenamefont {Brass}\ \emph {et~al.}(1993)\citenamefont {Brass},
  \citenamefont {Pendleton}, \citenamefont {Chen},\ and\ \citenamefont
  {Robson}}]{doi:10.1002/bip.360330815}%
  \BibitemOpen
  \bibfield  {author} {\bibinfo {author} {\bibfnamefont {A.}~\bibnamefont
  {Brass}}, \bibinfo {author} {\bibfnamefont {B.~J.}\ \bibnamefont
  {Pendleton}}, \bibinfo {author} {\bibfnamefont {Y.}~\bibnamefont {Chen}}, \
  and\ \bibinfo {author} {\bibfnamefont {B.}~\bibnamefont {Robson}},\
  }\bibfield  {title} {\enquote {\bibinfo {title} {Hybrid monte carlo
  simulations theory and initial comparison with molecular dynamics},}\ }\href
  {\doibase 10.1002/bip.360330815} {\bibfield  {journal} {\bibinfo  {journal}
  {Biopolymers}\ }\textbf {\bibinfo {volume} {33}},\ \bibinfo {pages}
  {1307--1315} (\bibinfo {year} {1993})},\ \Eprint
  {http://arxiv.org/abs/https://onlinelibrary.wiley.com/doi/pdf/10.1002/bip.360330815}
  {https://onlinelibrary.wiley.com/doi/pdf/10.1002/bip.360330815} \BibitemShut
  {NoStop}%
\bibitem [{\citenamefont {{Dias}}\ and\ \citenamefont
  {{Ehlers}}(2017)}]{2017arXiv171202326D}%
  \BibitemOpen
  \bibfield  {author} {\bibinfo {author} {\bibfnamefont {D.~S.}\ \bibnamefont
  {{Dias}}}\ and\ \bibinfo {author} {\bibfnamefont {R.~S.}\ \bibnamefont
  {{Ehlers}}},\ }\bibfield  {title} {\enquote {\bibinfo {title} {Stochastic
  volatily models using {Hamiltonian Monte Carlo} methods and {Stan}},}\
  }\href@noop {} {\  (\bibinfo {year} {2017})},\ \Eprint
  {http://arxiv.org/abs/1712.02326} {arXiv:1712.02326 [stat.AP]} \BibitemShut
  {NoStop}%
\bibitem [{\citenamefont {{Takaishi}}(2008)}]{2008arXiv0807.4394T}%
  \BibitemOpen
  \bibfield  {author} {\bibinfo {author} {\bibfnamefont {T.}~\bibnamefont
  {{Takaishi}}},\ }\bibfield  {title} {\enquote {\bibinfo {title} {Financial
  time series analysis of {SV} model by hybrid {Monte Carlo}},}\ }\href@noop {}
  {\  (\bibinfo {year} {2008})},\ \Eprint {http://arxiv.org/abs/0807.4394}
  {arXiv:0807.4394 [q-fin.ST]} \BibitemShut {NoStop}%
\bibitem [{\citenamefont {Duane}\ and\ \citenamefont
  {Pendleton}(1988)}]{Duane:1988vr}%
  \BibitemOpen
  \bibfield  {author} {\bibinfo {author} {\bibfnamefont {Simon}\ \bibnamefont
  {Duane}}\ and\ \bibinfo {author} {\bibfnamefont {Brian~J.}\ \bibnamefont
  {Pendleton}},\ }\bibfield  {title} {\enquote {\bibinfo {title} {Gauge
  invariant {Fourier} acceleration},}\ }\href {\doibase
  10.1016/0370-2693(88)91270-1} {\bibfield  {journal} {\bibinfo  {journal}
  {Phys. Lett.}\ }\textbf {\bibinfo {volume} {B206}},\ \bibinfo {pages}
  {101--106} (\bibinfo {year} {1988})}\BibitemShut {NoStop}%
\bibitem [{\citenamefont {Girolami}\ and\ \citenamefont
  {Calderhead}(2011)}]{RSSB:RSSB765}%
  \BibitemOpen
  \bibfield  {author} {\bibinfo {author} {\bibfnamefont {Mark}\ \bibnamefont
  {Girolami}}\ and\ \bibinfo {author} {\bibfnamefont {Ben}\ \bibnamefont
  {Calderhead}},\ }\bibfield  {title} {\enquote {\bibinfo {title} {{Riemann
  manifold Langevin and Hamiltonian Monte Carlo methods}},}\ }\href {\doibase
  10.1111/j.1467-9868.2010.00765.x} {\bibfield  {journal} {\bibinfo  {journal}
  {J. Royal Stat. Soc.}\ }\textbf {\bibinfo {volume} {B73}},\ \bibinfo {pages}
  {123--214} (\bibinfo {year} {2011})}\BibitemShut {NoStop}%
\bibitem [{\citenamefont {Beetem}\ \emph {et~al.}(1985)\citenamefont {Beetem},
  \citenamefont {Denneau},\ and\ \citenamefont
  {Weingarten}}]{Beetem:1985:GS:327070.327139}%
  \BibitemOpen
  \bibfield  {author} {\bibinfo {author} {\bibfnamefont {John}\ \bibnamefont
  {Beetem}}, \bibinfo {author} {\bibfnamefont {Monty}\ \bibnamefont {Denneau}},
  \ and\ \bibinfo {author} {\bibfnamefont {Don}\ \bibnamefont {Weingarten}},\
  }\bibfield  {title} {\enquote {\bibinfo {title} {The {GF11} supercomputer},}\
  }\href {\doibase 10.1145/327070.327139} {\bibfield  {journal} {\bibinfo
  {journal} {SIGARCH Comput. Archit. News}\ }\textbf {\bibinfo {volume} {13}},\
  \bibinfo {pages} {108--115} (\bibinfo {year} {1985})}\BibitemShut {NoStop}%
\bibitem [{\citenamefont {Vicini}\ \emph {et~al.}(1998)\citenamefont {Vicini}
  \emph {et~al.}}]{AGLIETTI1998216}%
  \BibitemOpen
  \bibfield  {author} {\bibinfo {author} {\bibfnamefont {P.}~\bibnamefont
  {Vicini}} \emph {et~al.},\ }\bibfield  {title} {\enquote {\bibinfo {title}
  {{The teraflop supercomputer APEmille: Architecture review and project status
  report}},}\ }\href {\doibase 10.1016/S0010-4655(97)00180-X} {\bibfield
  {journal} {\bibinfo  {journal} {Comput. Phys. Commun.}\ }\textbf {\bibinfo
  {volume} {110}},\ \bibinfo {pages} {216--219} (\bibinfo {year}
  {1998})}\BibitemShut {NoStop}%
\bibitem [{\citenamefont {Boyle}\ \emph {et~al.}(2005)\citenamefont {Boyle},
  \citenamefont {Chen}, \citenamefont {Christ}, \citenamefont {Clark},
  \citenamefont {Cohen}, \citenamefont {Cristian}, \citenamefont {Dong},
  \citenamefont {Gara}, \citenamefont {Jo\'o}, \citenamefont {Jung},
  \citenamefont {Kim}, \citenamefont {Levkova}, \citenamefont {Liao},
  \citenamefont {Liu}, \citenamefont {Mawhinney}, \citenamefont {Ohta},
  \citenamefont {Petrov}, \citenamefont {Wettig},\ and\ \citenamefont
  {Yamaguchi}}]{Boyle:2005:OQQ:1665957.1665969}%
  \BibitemOpen
  \bibfield  {author} {\bibinfo {author} {\bibfnamefont {P.~A.}\ \bibnamefont
  {Boyle}}, \bibinfo {author} {\bibfnamefont {D.}~\bibnamefont {Chen}},
  \bibinfo {author} {\bibfnamefont {N.~H.}\ \bibnamefont {Christ}}, \bibinfo
  {author} {\bibfnamefont {M.~A.}\ \bibnamefont {Clark}}, \bibinfo {author}
  {\bibfnamefont {S.~D.}\ \bibnamefont {Cohen}}, \bibinfo {author}
  {\bibfnamefont {C.}~\bibnamefont {Cristian}}, \bibinfo {author}
  {\bibfnamefont {Z.}~\bibnamefont {Dong}}, \bibinfo {author} {\bibfnamefont
  {A.}~\bibnamefont {Gara}}, \bibinfo {author} {\bibfnamefont {B.}~\bibnamefont
  {Jo\'o}}, \bibinfo {author} {\bibfnamefont {C.}~\bibnamefont {Jung}},
  \bibinfo {author} {\bibfnamefont {C.}~\bibnamefont {Kim}}, \bibinfo {author}
  {\bibfnamefont {L.~A.}\ \bibnamefont {Levkova}}, \bibinfo {author}
  {\bibfnamefont {X.}~\bibnamefont {Liao}}, \bibinfo {author} {\bibfnamefont
  {G.}~\bibnamefont {Liu}}, \bibinfo {author} {\bibfnamefont {R.~D.}\
  \bibnamefont {Mawhinney}}, \bibinfo {author} {\bibfnamefont {S.}~\bibnamefont
  {Ohta}}, \bibinfo {author} {\bibfnamefont {K.}~\bibnamefont {Petrov}},
  \bibinfo {author} {\bibfnamefont {T.}~\bibnamefont {Wettig}}, \ and\ \bibinfo
  {author} {\bibfnamefont {A.}~\bibnamefont {Yamaguchi}},\ }\bibfield  {title}
  {\enquote {\bibinfo {title} {Overview of the {QCDSP} and {QCDOC}
  computers},}\ }\href {\doibase 10.1147/rd.492.0351} {\bibfield  {journal}
  {\bibinfo  {journal} {IBM J. Res. Dev.}\ }\textbf {\bibinfo {volume} {49}},\
  \bibinfo {pages} {351--365} (\bibinfo {year} {2005})}\BibitemShut {NoStop}%
\bibitem [{\citenamefont {Adiga}\ \emph {et~al.}(2002)\citenamefont {Adiga},
  \citenamefont {Almasi}, \citenamefont {Almasi}, \citenamefont {Aridor},
  \citenamefont {Barik}, \citenamefont {Beece}, \citenamefont {Bellofatto},
  \citenamefont {Bhanot}, \citenamefont {Bickford}, \citenamefont {Blumrich},
  \citenamefont {Bright}, \citenamefont {Brunheroto}, \citenamefont {Cascaval},
  \citenamefont {Castanos}, \citenamefont {Chan}, \citenamefont {Ceze},
  \citenamefont {Coteus}, \citenamefont {Chatterjee}, \citenamefont {Chen},\
  and\ \citenamefont {Yates}}]{BGL}%
  \BibitemOpen
  \bibfield  {author} {\bibinfo {author} {\bibfnamefont {N.~R.}\ \bibnamefont
  {Adiga}}, \bibinfo {author} {\bibfnamefont {G.}~\bibnamefont {Almasi}},
  \bibinfo {author} {\bibfnamefont {G.~S.}\ \bibnamefont {Almasi}}, \bibinfo
  {author} {\bibfnamefont {Yariv}\ \bibnamefont {Aridor}}, \bibinfo {author}
  {\bibfnamefont {Rajkishore}\ \bibnamefont {Barik}}, \bibinfo {author}
  {\bibfnamefont {D.}~\bibnamefont {Beece}}, \bibinfo {author} {\bibfnamefont
  {Ralph}\ \bibnamefont {Bellofatto}}, \bibinfo {author} {\bibfnamefont {Gyan}\
  \bibnamefont {Bhanot}}, \bibinfo {author} {\bibfnamefont {Randy}\
  \bibnamefont {Bickford}}, \bibinfo {author} {\bibfnamefont {M.}~\bibnamefont
  {Blumrich}}, \bibinfo {author} {\bibfnamefont {A.~A.}\ \bibnamefont
  {Bright}}, \bibinfo {author} {\bibfnamefont {José}\ \bibnamefont
  {Brunheroto}}, \bibinfo {author} {\bibfnamefont {Calin}\ \bibnamefont
  {Cascaval}}, \bibinfo {author} {\bibfnamefont {J.}~\bibnamefont {Castanos}},
  \bibinfo {author} {\bibfnamefont {W.}~\bibnamefont {Chan}}, \bibinfo {author}
  {\bibfnamefont {Luis}\ \bibnamefont {Ceze}}, \bibinfo {author} {\bibfnamefont
  {Paul}\ \bibnamefont {Coteus}}, \bibinfo {author} {\bibfnamefont
  {S.}~\bibnamefont {Chatterjee}}, \bibinfo {author} {\bibfnamefont
  {D.}~\bibnamefont {Chen}}, \ and\ \bibinfo {author} {\bibfnamefont
  {K.}~\bibnamefont {Yates}},\ }\bibfield  {title} {\enquote {\bibinfo {title}
  {An overview of the {Blue Gene/L} supercomputer},}\ }in\ \href {\doibase
  10.1109/SC.2002.10017} {\emph {\bibinfo {booktitle} {SC '02: Proceedings of
  the 2002 ACM/IEEE Conference on Supercomputing}}}\ (\bibinfo  {publisher}
  {IEEE},\ \bibinfo {year} {2002})\ pp.\ \bibinfo {pages} {1-- 22}\BibitemShut
  {NoStop}%
\bibitem [{\citenamefont {Boyle}(2012)}]{Boyle:2012iy}%
  \BibitemOpen
  \bibfield  {author} {\bibinfo {author} {\bibfnamefont {P.~A.}\ \bibnamefont
  {Boyle}},\ }\bibfield  {title} {\enquote {\bibinfo {title} {{The BlueGene/Q
  supercomputer}},}\ }\href {\doibase 10.22323/1.164.0020} {\bibfield
  {journal} {\bibinfo  {journal} {PoS}\ }\textbf {\bibinfo {volume}
  {LATTICE2012}},\ \bibinfo {pages} {020} (\bibinfo {year} {2012})}\BibitemShut
  {NoStop}%
\bibitem [{\citenamefont {Egri}\ \emph {et~al.}(2007)\citenamefont {Egri},
  \citenamefont {Fodor}, \citenamefont {Hoelbling}, \citenamefont {Katz},
  \citenamefont {Nogradi},\ and\ \citenamefont {Szabo}}]{Egri:2006zm}%
  \BibitemOpen
  \bibfield  {author} {\bibinfo {author} {\bibfnamefont {Gyozo~I.}\
  \bibnamefont {Egri}}, \bibinfo {author} {\bibfnamefont {Zoltan}\ \bibnamefont
  {Fodor}}, \bibinfo {author} {\bibfnamefont {Christian}\ \bibnamefont
  {Hoelbling}}, \bibinfo {author} {\bibfnamefont {Sandor~D.}\ \bibnamefont
  {Katz}}, \bibinfo {author} {\bibfnamefont {Daniel}\ \bibnamefont {Nogradi}},
  \ and\ \bibinfo {author} {\bibfnamefont {Kalman~K.}\ \bibnamefont {Szabo}},\
  }\bibfield  {title} {\enquote {\bibinfo {title} {{Lattice QCD as a video
  game}},}\ }\href {\doibase 10.1016/j.cpc.2007.06.005} {\bibfield  {journal}
  {\bibinfo  {journal} {Comput. Phys. Commun.}\ }\textbf {\bibinfo {volume}
  {177}},\ \bibinfo {pages} {631--639} (\bibinfo {year} {2007})},\ \Eprint
  {http://arxiv.org/abs/hep-lat/0611022} {arXiv:hep-lat/0611022 [hep-lat]}
  \BibitemShut {NoStop}%
\bibitem [{\citenamefont {Clark}\ \emph {et~al.}(2010)\citenamefont {Clark}
  \emph {et~al.}}]{Clark:2009wm}%
  \BibitemOpen
  \bibfield  {author} {\bibinfo {author} {\bibfnamefont {M.~A.}\ \bibnamefont
  {Clark}} \emph {et~al.},\ }\bibfield  {title} {\enquote {\bibinfo {title}
  {{Solving lattice QCD systems of equations using mixed precision solvers on
  GPUs}},}\ }\href {\doibase 10.1016/j.cpc.2010.05.002} {\bibfield  {journal}
  {\bibinfo  {journal} {Comput. Phys. Commun.}\ }\textbf {\bibinfo {volume}
  {181}},\ \bibinfo {pages} {1517--1528} (\bibinfo {year} {2010})},\ \Eprint
  {http://arxiv.org/abs/0911.3191} {arXiv:0911.3191 [hep-lat]} \BibitemShut
  {NoStop}%
\bibitem [{\citenamefont {Babich}\ \emph
  {et~al.}(2010{\natexlab{a}})\citenamefont {Babich}, \citenamefont {Clark},\
  and\ \citenamefont {Jo\'o}}]{Babich:2010mu}%
  \BibitemOpen
  \bibfield  {author} {\bibinfo {author} {\bibfnamefont {Ronald}\ \bibnamefont
  {Babich}}, \bibinfo {author} {\bibfnamefont {Michael~A.}\ \bibnamefont
  {Clark}}, \ and\ \bibinfo {author} {\bibfnamefont {B\'alint}\ \bibnamefont
  {Jo\'o}},\ }\bibfield  {title} {\enquote {\bibinfo {title} {Parallelizing the
  {QUDA} library for multi-{GPU} calculations in lattice quantum
  chromodynamics},}\ }\href {\doibase 10.1109/SC.2010.40} {\bibfield  {journal}
  {\bibinfo  {journal} {{ACM/IEEE Int. Conf. High Performance Computing,
  Networking, Storage and Analysis, New Orleans}}\ } (\bibinfo {year}
  {2010}{\natexlab{a}}),\ 10.1109/SC.2010.40},\ \Eprint
  {http://arxiv.org/abs/1011.0024} {arXiv:1011.0024 [hep-lat]} \BibitemShut
  {NoStop}%
\bibitem [{\citenamefont {Babich}\ \emph {et~al.}(2011)\citenamefont {Babich},
  \citenamefont {Clark}, \citenamefont {Jo\'o}, \citenamefont {Shi},
  \citenamefont {Brower},\ and\ \citenamefont {Gottlieb}}]{Babich:2011np}%
  \BibitemOpen
  \bibfield  {author} {\bibinfo {author} {\bibfnamefont {R.}~\bibnamefont
  {Babich}}, \bibinfo {author} {\bibfnamefont {M.~A.}\ \bibnamefont {Clark}},
  \bibinfo {author} {\bibfnamefont {B.}~\bibnamefont {Jo\'o}}, \bibinfo
  {author} {\bibfnamefont {G.}~\bibnamefont {Shi}}, \bibinfo {author}
  {\bibfnamefont {R.~C.}\ \bibnamefont {Brower}}, \ and\ \bibinfo {author}
  {\bibfnamefont {S.}~\bibnamefont {Gottlieb}},\ }\bibfield  {title} {\enquote
  {\bibinfo {title} {{Scaling Lattice QCD beyond 100 GPUs}},}\ }in\ \href
  {\doibase 10.1145/2063384.2063478} {\emph {\bibinfo {booktitle} {{SC11
  International Conference for High Performance Computing, Networking, Storage
  and Analysis Seattle, Washington, November 12-18, 2011}}}}\ (\bibinfo
  {publisher} {ACM},\ \bibinfo {year} {2011})\ \Eprint
  {http://arxiv.org/abs/1109.2935} {arXiv:1109.2935 [hep-lat]} \BibitemShut
  {NoStop}%
\bibitem [{\citenamefont {Clark}\ \emph {et~al.}(2016)\citenamefont {Clark},
  \citenamefont {Jo\'o}, \citenamefont {Strelchenko}, \citenamefont {Cheng},
  \citenamefont {Gambhir},\ and\ \citenamefont {Brower}}]{Clark:2016rdz}%
  \BibitemOpen
  \bibfield  {author} {\bibinfo {author} {\bibfnamefont {M.~A.}\ \bibnamefont
  {Clark}}, \bibinfo {author} {\bibfnamefont {B\'alint}\ \bibnamefont {Jo\'o}},
  \bibinfo {author} {\bibfnamefont {Alexei}\ \bibnamefont {Strelchenko}},
  \bibinfo {author} {\bibfnamefont {Michael}\ \bibnamefont {Cheng}}, \bibinfo
  {author} {\bibfnamefont {Arjun}\ \bibnamefont {Gambhir}}, \ and\ \bibinfo
  {author} {\bibfnamefont {Richard}\ \bibnamefont {Brower}},\ }\bibfield
  {title} {\enquote {\bibinfo {title} {{Accelerating Lattice QCD Multigrid on
  GPUs Using Fine-Grained Parallelization}},}\ }\href {\doibase
  10.1109/SC.2010.40} {\bibfield  {journal} {\bibinfo  {journal} {{ACM/IEEE
  Int. Conf. High Performance Computing, Networking, Storage and Analysis, Salt
  Lake City, Utah}}\ } (\bibinfo {year} {2016}),\ 10.1109/SC.2010.40},\ \Eprint
  {http://arxiv.org/abs/1612.07873} {arXiv:1612.07873 [hep-lat]} \BibitemShut
  {NoStop}%
\bibitem [{\citenamefont {Clark}\ and\ \citenamefont
  {Babich}()}]{QUDADownload}%
  \BibitemOpen
  \bibfield  {author} {\bibinfo {author} {\bibfnamefont {M.A}\ \bibnamefont
  {Clark}}\ and\ \bibinfo {author} {\bibfnamefont {R.}~\bibnamefont {Babich}},\
  }\href@noop {} {\enquote {\bibinfo {title} {{QUDA: A library for QCD on
  GPUs}},}\ }\bibinfo {howpublished}
  {\url{http://lattice.github.io/quda/}}\BibitemShut {NoStop}%
\bibitem [{\citenamefont {Ayyar}\ \emph {et~al.}(2018)\citenamefont {Ayyar},
  \citenamefont {Hackett}, \citenamefont {Jay},\ and\ \citenamefont
  {Neil}}]{Ayyar:2018wwf}%
  \BibitemOpen
  \bibfield  {author} {\bibinfo {author} {\bibfnamefont {Venkitesh}\
  \bibnamefont {Ayyar}}, \bibinfo {author} {\bibfnamefont {Daniel~C.}\
  \bibnamefont {Hackett}}, \bibinfo {author} {\bibfnamefont {William~I.}\
  \bibnamefont {Jay}}, \ and\ \bibinfo {author} {\bibfnamefont {Ethan~T.}\
  \bibnamefont {Neil}},\ }\bibfield  {title} {\enquote {\bibinfo {title}
  {{Automated lattice data generation}},}\ }\bibfield  {booktitle} {\emph
  {\bibinfo {booktitle} {{Proceedings, 35th International Symposium on Lattice
  Field Theory (Lattice 2017): Granada, Spain, June 18-24, 2017}}},\ }\href
  {\doibase 10.1051/epjconf/201817509009} {\bibfield  {journal} {\bibinfo
  {journal} {EPJ Web Conf.}\ }\textbf {\bibinfo {volume} {175}},\ \bibinfo
  {pages} {09009} (\bibinfo {year} {2018})},\ \Eprint
  {http://arxiv.org/abs/1802.00851} {arXiv:1802.00851 [hep-lat]} \BibitemShut
  {NoStop}%
\bibitem [{\citenamefont {Foster}(2011)}]{Foster:2011:GOA:1978245.1978305}%
  \BibitemOpen
  \bibfield  {author} {\bibinfo {author} {\bibfnamefont {Ian}\ \bibnamefont
  {Foster}},\ }\bibfield  {title} {\enquote {\bibinfo {title} {Globus online:
  Accelerating and democratizing science through cloud-based services},}\
  }\href {\doibase 10.1109/MIC.2011.64} {\bibfield  {journal} {\bibinfo
  {journal} {IEEE Internet Computing}\ }\textbf {\bibinfo {volume} {15}},\
  \bibinfo {pages} {70--73} (\bibinfo {year} {2011})}\BibitemShut {NoStop}%
\bibitem [{\citenamefont {Hestenes}\ and\ \citenamefont
  {Stiefel}(1952)}]{citeulike:9077321}%
  \BibitemOpen
  \bibfield  {author} {\bibinfo {author} {\bibfnamefont {Magnus~R.}\
  \bibnamefont {Hestenes}}\ and\ \bibinfo {author} {\bibfnamefont {Eduard}\
  \bibnamefont {Stiefel}},\ }\bibfield  {title} {\enquote {\bibinfo {title}
  {Methods of conjugate gradients for solving linear systems},}\ }\href
  {\doibase 10.6028/jres.049.044} {\bibfield  {journal} {\bibinfo  {journal}
  {Journal of Research of the National Bureau of Standards}\ }\textbf {\bibinfo
  {volume} {49}},\ \bibinfo {pages} {409--436} (\bibinfo {year}
  {1952})}\BibitemShut {NoStop}%
\bibitem [{\citenamefont {van~der Vorst}(1992)}]{BiCGStab}%
  \BibitemOpen
  \bibfield  {author} {\bibinfo {author} {\bibfnamefont {H.~A.}\ \bibnamefont
  {van~der Vorst}},\ }\bibfield  {title} {\enquote {\bibinfo {title}
  {{BI-CGSTAB: a fast and smoothly converging variant of BI-CG for the solution
  of nonsymmetric linear systems}},}\ }\href {\doibase 10.1137/0913035}
  {\bibfield  {journal} {\bibinfo  {journal} {SIAM J. Sci. Stat. Comput.}\
  }\textbf {\bibinfo {volume} {13}},\ \bibinfo {pages} {631--644} (\bibinfo
  {year} {1992})}\BibitemShut {NoStop}%
\bibitem [{\citenamefont {Metropolis}\ \emph {et~al.}(1953)\citenamefont
  {Metropolis}, \citenamefont {Rosenbluth}, \citenamefont {Rosenbluth},
  \citenamefont {Teller},\ and\ \citenamefont {Teller}}]{METR1953}%
  \BibitemOpen
  \bibfield  {author} {\bibinfo {author} {\bibfnamefont {N.}~\bibnamefont
  {Metropolis}}, \bibinfo {author} {\bibfnamefont {A.W.}\ \bibnamefont
  {Rosenbluth}}, \bibinfo {author} {\bibfnamefont {M.N.}\ \bibnamefont
  {Rosenbluth}}, \bibinfo {author} {\bibfnamefont {A.H.}\ \bibnamefont
  {Teller}}, \ and\ \bibinfo {author} {\bibfnamefont {E.}~\bibnamefont
  {Teller}},\ }\bibfield  {title} {\enquote {\bibinfo {title} {Equations of
  state calculations by fast computing machines},}\ }\href@noop {} {\bibfield
  {journal} {\bibinfo  {journal} {Journal of Chemical Physics}\ }\textbf
  {\bibinfo {volume} {21}},\ \bibinfo {pages} {1087--1091} (\bibinfo {year}
  {1953})}\BibitemShut {NoStop}%
\bibitem [{\citenamefont {Horowitz}(1991)}]{Horowitz:1991rr}%
  \BibitemOpen
  \bibfield  {author} {\bibinfo {author} {\bibfnamefont {Alan~M.}\ \bibnamefont
  {Horowitz}},\ }\bibfield  {title} {\enquote {\bibinfo {title} {{A generalized
  guided Monte Carlo algorithm}},}\ }\href {\doibase
  10.1016/0370-2693(91)90812-5} {\bibfield  {journal} {\bibinfo  {journal}
  {Phys. Lett.}\ }\textbf {\bibinfo {volume} {B268}},\ \bibinfo {pages}
  {247--252} (\bibinfo {year} {1991})}\BibitemShut {NoStop}%
\bibitem [{\citenamefont {Takaishi}\ and\ \citenamefont
  {de~Forcrand}(2006)}]{Takaishi:2005tz}%
  \BibitemOpen
  \bibfield  {author} {\bibinfo {author} {\bibfnamefont {Tetsuya}\ \bibnamefont
  {Takaishi}}\ and\ \bibinfo {author} {\bibfnamefont {Philippe}\ \bibnamefont
  {de~Forcrand}},\ }\bibfield  {title} {\enquote {\bibinfo {title} {{Testing
  and tuning new symplectic integrators for hybrid Monte Carlo algorithm in
  lattice QCD}},}\ }\href {\doibase 10.1103/PhysRevE.73.036706} {\bibfield
  {journal} {\bibinfo  {journal} {Phys. Rev.}\ }\textbf {\bibinfo {volume}
  {E73}},\ \bibinfo {pages} {036706} (\bibinfo {year} {2006})},\ \Eprint
  {http://arxiv.org/abs/hep-lat/0505020} {arXiv:hep-lat/0505020 [hep-lat]}
  \BibitemShut {NoStop}%
\bibitem [{\citenamefont {{Omelyan}}\ \emph {et~al.}(2002)\citenamefont
  {{Omelyan}}, \citenamefont {{Mryglod}},\ and\ \citenamefont
  {{Folk}}}]{Omelyan}%
  \BibitemOpen
  \bibfield  {author} {\bibinfo {author} {\bibfnamefont {I.~P.}\ \bibnamefont
  {{Omelyan}}}, \bibinfo {author} {\bibfnamefont {I.~M.}\ \bibnamefont
  {{Mryglod}}}, \ and\ \bibinfo {author} {\bibfnamefont {R.}~\bibnamefont
  {{Folk}}},\ }\bibfield  {title} {\enquote {\bibinfo {title} {{Optimized
  Forest-Ruth- and Suzuki-like algorithms for integration of motion in
  many-body systems}},}\ }\href {\doibase 10.1016/S0010-4655(02)00451-4}
  {\bibfield  {journal} {\bibinfo  {journal} {Computer Physics Communications}\
  }\textbf {\bibinfo {volume} {146}},\ \bibinfo {pages} {188--202} (\bibinfo
  {year} {2002})},\ \Eprint {http://arxiv.org/abs/cond-mat/0110585}
  {cond-mat/0110585} \BibitemShut {NoStop}%
\bibitem [{\citenamefont {Kennedy}\ \emph {et~al.}(2009)\citenamefont
  {Kennedy}, \citenamefont {Clark},\ and\ \citenamefont
  {Silva}}]{Kennedy:2009fe}%
  \BibitemOpen
  \bibfield  {author} {\bibinfo {author} {\bibfnamefont {A.~D.}\ \bibnamefont
  {Kennedy}}, \bibinfo {author} {\bibfnamefont {M.~A.}\ \bibnamefont {Clark}},
  \ and\ \bibinfo {author} {\bibfnamefont {P.~J.}\ \bibnamefont {Silva}},\
  }\bibfield  {title} {\enquote {\bibinfo {title} {Force gradient
  integrators},}\ }\href@noop {} {\bibfield  {journal} {\bibinfo  {journal}
  {PoS}\ }\textbf {\bibinfo {volume} {LAT2009}},\ \bibinfo {pages} {021}
  (\bibinfo {year} {2009})},\ \Eprint {http://arxiv.org/abs/0910.2950}
  {arXiv:0910.2950 [hep-lat]} \BibitemShut {NoStop}%
\bibitem [{\citenamefont {Clark}\ \emph {et~al.}(2011)\citenamefont {Clark},
  \citenamefont {Jo\'o}, \citenamefont {Kennedy},\ and\ \citenamefont
  {Silva}}]{Clark:2011ir}%
  \BibitemOpen
  \bibfield  {author} {\bibinfo {author} {\bibfnamefont {M.~A.}\ \bibnamefont
  {Clark}}, \bibinfo {author} {\bibfnamefont {B\'alint}\ \bibnamefont {Jo\'o}},
  \bibinfo {author} {\bibfnamefont {A.~D.}\ \bibnamefont {Kennedy}}, \ and\
  \bibinfo {author} {\bibfnamefont {P.~J.}\ \bibnamefont {Silva}},\ }\bibfield
  {title} {\enquote {\bibinfo {title} {{Improving dynamical lattice QCD
  simulations through integrator tuning using Poisson brackets and a
  force-gradient integrator}},}\ }\href {\doibase 10.1103/PhysRevD.84.071502}
  {\bibfield  {journal} {\bibinfo  {journal} {Phys. Rev.}\ }\textbf {\bibinfo
  {volume} {D84}},\ \bibinfo {pages} {071502} (\bibinfo {year} {2011})},\
  \Eprint {http://arxiv.org/abs/1108.1828} {arXiv:1108.1828 [hep-lat]}
  \BibitemShut {NoStop}%
\bibitem [{\citenamefont {Yin}\ and\ \citenamefont
  {Mawhinney}(2011)}]{Yin:2011np}%
  \BibitemOpen
  \bibfield  {author} {\bibinfo {author} {\bibfnamefont {Hantao}\ \bibnamefont
  {Yin}}\ and\ \bibinfo {author} {\bibfnamefont {Robert~D.}\ \bibnamefont
  {Mawhinney}},\ }\bibfield  {title} {\enquote {\bibinfo {title} {Improving
  {DWF} simulations: the force gradient integrator and the {M\'obius}
  accelerated {DWF} solver},}\ }\href@noop {} {\bibfield  {journal} {\bibinfo
  {journal} {PoS}\ }\textbf {\bibinfo {volume} {LATTICE2011}},\ \bibinfo
  {pages} {051} (\bibinfo {year} {2011})},\ \Eprint
  {http://arxiv.org/abs/1111.5059} {arXiv:1111.5059 [hep-lat]} \BibitemShut
  {NoStop}%
\bibitem [{\citenamefont {Chen}\ and\ \citenamefont
  {Chiu}(2014)}]{Chen:2014hyy}%
  \BibitemOpen
  \bibfield  {author} {\bibinfo {author} {\bibfnamefont {Yu-Chih}\ \bibnamefont
  {Chen}}\ and\ \bibinfo {author} {\bibfnamefont {Ting-Wai}\ \bibnamefont
  {Chiu}} (\bibinfo {collaboration} {TWQCD}),\ }\bibfield  {title} {\enquote
  {\bibinfo {title} {Exact pseudofermion action for {Monte Carlo} simulation of
  domain-wall fermion},}\ }\href {\doibase 10.1016/j.physletb.2014.09.016}
  {\bibfield  {journal} {\bibinfo  {journal} {Phys. Lett.}\ }\textbf {\bibinfo
  {volume} {B738}},\ \bibinfo {pages} {55--60} (\bibinfo {year} {2014})},\
  \Eprint {http://arxiv.org/abs/1403.1683} {arXiv:1403.1683 [hep-lat]}
  \BibitemShut {NoStop}%
\bibitem [{\citenamefont {Jung}\ \emph {et~al.}(2018)\citenamefont {Jung},
  \citenamefont {Kelly}, \citenamefont {Mawhinney},\ and\ \citenamefont
  {Murphy}}]{Jung:2017xef}%
  \BibitemOpen
  \bibfield  {author} {\bibinfo {author} {\bibfnamefont {C.}~\bibnamefont
  {Jung}}, \bibinfo {author} {\bibfnamefont {C.}~\bibnamefont {Kelly}},
  \bibinfo {author} {\bibfnamefont {R.~D.}\ \bibnamefont {Mawhinney}}, \ and\
  \bibinfo {author} {\bibfnamefont {D.~J.}\ \bibnamefont {Murphy}},\ }\bibfield
   {title} {\enquote {\bibinfo {title} {Domain wall fermion {QCD} with the
  exact one flavor algorithm},}\ }\href {\doibase 10.1103/PhysRevD.97.054503}
  {\bibfield  {journal} {\bibinfo  {journal} {Phys. Rev.}\ }\textbf {\bibinfo
  {volume} {D97}},\ \bibinfo {pages} {054503} (\bibinfo {year} {2018})},\
  \Eprint {http://arxiv.org/abs/1706.05843} {arXiv:1706.05843 [hep-lat]}
  \BibitemShut {NoStop}%
\bibitem [{\citenamefont {Clark}\ and\ \citenamefont {Kennedy}(2007)}]{RHMC}%
  \BibitemOpen
  \bibfield  {author} {\bibinfo {author} {\bibfnamefont {M.~A.}\ \bibnamefont
  {Clark}}\ and\ \bibinfo {author} {\bibfnamefont {A.~D.}\ \bibnamefont
  {Kennedy}},\ }\bibfield  {title} {\enquote {\bibinfo {title} {Accelerating
  dynamical-fermion computations using the rational hybrid {Monte Carlo}
  algorithm with multiple pseudofermion fields},}\ }\href {\doibase
  10.1103/PhysRevLett.98.051601} {\bibfield  {journal} {\bibinfo  {journal}
  {Phys. Rev. Lett.}\ }\textbf {\bibinfo {volume} {98}},\ \bibinfo {pages}
  {051601} (\bibinfo {year} {2007})}\BibitemShut {NoStop}%
\bibitem [{\citenamefont {Frezzotti}\ and\ \citenamefont
  {Jansen}(1997)}]{Frezzotti:1997ym}%
  \BibitemOpen
  \bibfield  {author} {\bibinfo {author} {\bibfnamefont {Roberto}\ \bibnamefont
  {Frezzotti}}\ and\ \bibinfo {author} {\bibfnamefont {Karl}\ \bibnamefont
  {Jansen}},\ }\bibfield  {title} {\enquote {\bibinfo {title} {A polynomial
  hybrid {Monte Carlo} algorithm},}\ }\href {\doibase
  10.1016/S0370-2693(97)00475-9} {\bibfield  {journal} {\bibinfo  {journal}
  {Phys. Lett.}\ }\textbf {\bibinfo {volume} {B402}},\ \bibinfo {pages}
  {328--334} (\bibinfo {year} {1997})},\ \Eprint
  {http://arxiv.org/abs/hep-lat/9702016} {arXiv:hep-lat/9702016 [hep-lat]}
  \BibitemShut {NoStop}%
\bibitem [{\citenamefont {Ukawa}(2002)}]{UKAWA2002195}%
  \BibitemOpen
  \bibfield  {author} {\bibinfo {author} {\bibfnamefont {A.}~\bibnamefont
  {Ukawa}} (\bibinfo {collaboration} {CP-PACS, JLQCD}),\ }\bibfield  {title}
  {\enquote {\bibinfo {title} {{Computational cost of full QCD simulations
  experienced by CP-PACS and JLQCD Collaborations}},}\ }\href {\doibase
  10.1016/S0920-5632(01)01662-0} {\bibfield  {journal} {\bibinfo  {journal}
  {Nucl. Phys. Proc. Suppl.}\ }\textbf {\bibinfo {volume} {106}},\ \bibinfo
  {pages} {195--196} (\bibinfo {year} {2002})}\BibitemShut {NoStop}%
\bibitem [{\citenamefont {Hasenbusch}(2001)}]{Hasenbusch:2001ne}%
  \BibitemOpen
  \bibfield  {author} {\bibinfo {author} {\bibfnamefont {Martin}\ \bibnamefont
  {Hasenbusch}},\ }\bibfield  {title} {\enquote {\bibinfo {title} {Speeding up
  the hybrid-monte-carlo algorithm for dynamical fermions},}\ }\href@noop {}
  {\bibfield  {journal} {\bibinfo  {journal} {Phys. Lett.}\ }\textbf {\bibinfo
  {volume} {B519}},\ \bibinfo {pages} {177--182} (\bibinfo {year} {2001})},\
  \Eprint {http://arxiv.org/abs/hep-lat/0107019} {hep-lat/0107019} \BibitemShut
  {NoStop}%
\bibitem [{\citenamefont {Hasenbusch}\ and\ \citenamefont
  {Jansen}(2003)}]{Hasenbusch:2002ai}%
  \BibitemOpen
  \bibfield  {author} {\bibinfo {author} {\bibfnamefont {M.}~\bibnamefont
  {Hasenbusch}}\ and\ \bibinfo {author} {\bibfnamefont {K.}~\bibnamefont
  {Jansen}},\ }\bibfield  {title} {\enquote {\bibinfo {title} {{Speeding up
  lattice QCD simulations with clover improved Wilson fermions}},}\ }\href
  {\doibase 10.1016/S0550-3213(03)00227-X} {\bibfield  {journal} {\bibinfo
  {journal} {Nucl. Phys.}\ }\textbf {\bibinfo {volume} {B659}},\ \bibinfo
  {pages} {299--320} (\bibinfo {year} {2003})},\ \Eprint
  {http://arxiv.org/abs/hep-lat/0211042} {arXiv:hep-lat/0211042 [hep-lat]}
  \BibitemShut {NoStop}%
\bibitem [{\citenamefont {Urbach}\ \emph {et~al.}(2006)\citenamefont {Urbach},
  \citenamefont {Jansen}, \citenamefont {Shindler},\ and\ \citenamefont
  {Wenger}}]{Urbach:2005ji}%
  \BibitemOpen
  \bibfield  {author} {\bibinfo {author} {\bibfnamefont {C.}~\bibnamefont
  {Urbach}}, \bibinfo {author} {\bibfnamefont {K.}~\bibnamefont {Jansen}},
  \bibinfo {author} {\bibfnamefont {A.}~\bibnamefont {Shindler}}, \ and\
  \bibinfo {author} {\bibfnamefont {U.}~\bibnamefont {Wenger}},\ }\bibfield
  {title} {\enquote {\bibinfo {title} {{HMC algorithm with multiple time scale
  integration and mass preconditioning}},}\ }\href {\doibase
  10.1016/j.cpc.2005.08.006} {\bibfield  {journal} {\bibinfo  {journal}
  {Comput. Phys. Commun.}\ }\textbf {\bibinfo {volume} {174}},\ \bibinfo
  {pages} {87--98} (\bibinfo {year} {2006})},\ \Eprint
  {http://arxiv.org/abs/hep-lat/0506011} {arXiv:hep-lat/0506011 [hep-lat]}
  \BibitemShut {NoStop}%
\bibitem [{\citenamefont {Sexton}\ and\ \citenamefont
  {Weingarten}(1992)}]{Sexton:1992nu}%
  \BibitemOpen
  \bibfield  {author} {\bibinfo {author} {\bibfnamefont {J.~C.}\ \bibnamefont
  {Sexton}}\ and\ \bibinfo {author} {\bibfnamefont {D.~H.}\ \bibnamefont
  {Weingarten}},\ }\bibfield  {title} {\enquote {\bibinfo {title} {{Hamiltonian
  evolution for the hybrid Monte Carlo algorithm}},}\ }\href {\doibase
  10.1016/0550-3213(92)90263-B} {\bibfield  {journal} {\bibinfo  {journal}
  {Nucl. Phys.}\ }\textbf {\bibinfo {volume} {B380}},\ \bibinfo {pages}
  {665--677} (\bibinfo {year} {1992})}\BibitemShut {NoStop}%
\bibitem [{\citenamefont {Osborn}()}]{QOPQDP}%
  \BibitemOpen
  \bibfield  {author} {\bibinfo {author} {\bibfnamefont {J.~C.}\ \bibnamefont
  {Osborn}},\ }\href@noop {} {\enquote {\bibinfo {title} {{QOPQDP} software
  library},}\ }\bibinfo {howpublished}
  {\url{http://usqcd-software.github.io/qopqdp/}}\BibitemShut {NoStop}%
\bibitem [{\citenamefont {Lin}(2013)}]{MeifengMG}%
  \BibitemOpen
  \bibfield  {author} {\bibinfo {author} {\bibfnamefont {Meifeng}\ \bibnamefont
  {Lin}},\ }\href@noop {} {\enquote {\bibinfo {title} {{Multigrid in HMC}},}\
  }\bibinfo {howpublished}
  {\url{https://indico.fnal.gov/event/7435/session/1/contribution/9/material/slides/0.pdf}}
  (\bibinfo {year} {2013})\BibitemShut {NoStop}%
\bibitem [{\citenamefont {L\"uscher}(2007{\natexlab{a}})}]{Luscher:2007es}%
  \BibitemOpen
  \bibfield  {author} {\bibinfo {author} {\bibfnamefont {Martin}\ \bibnamefont
  {L\"uscher}},\ }\bibfield  {title} {\enquote {\bibinfo {title} {{Deflation
  acceleration of lattice QCD simulations}},}\ }\href {\doibase
  10.1088/1126-6708/2007/12/011} {\bibfield  {journal} {\bibinfo  {journal}
  {JHEP}\ }\textbf {\bibinfo {volume} {12}},\ \bibinfo {pages} {011} (\bibinfo
  {year} {2007}{\natexlab{a}})},\ \Eprint {http://arxiv.org/abs/0710.5417}
  {arXiv:0710.5417 [hep-lat]} \BibitemShut {NoStop}%
\bibitem [{\citenamefont {L\"uscher}(2003)}]{Luscher:2003vf}%
  \BibitemOpen
  \bibfield  {author} {\bibinfo {author} {\bibfnamefont {Martin}\ \bibnamefont
  {L\"uscher}},\ }\bibfield  {title} {\enquote {\bibinfo {title} {Lattice qcd
  and the schwarz alternating procedure},}\ }\href {\doibase
  10.1088/1126-6708/2003/05/052} {\bibfield  {journal} {\bibinfo  {journal}
  {JHEP}\ }\textbf {\bibinfo {volume} {05}},\ \bibinfo {pages} {052} (\bibinfo
  {year} {2003})},\ \Eprint {http://arxiv.org/abs/hep-lat/0304007}
  {arXiv:hep-lat/0304007 [hep-lat]} \BibitemShut {NoStop}%
\bibitem [{\citenamefont {L\"uscher}(2004)}]{Luscher:2003qa}%
  \BibitemOpen
  \bibfield  {author} {\bibinfo {author} {\bibfnamefont {Martin}\ \bibnamefont
  {L\"uscher}},\ }\bibfield  {title} {\enquote {\bibinfo {title} {{Solution of
  the Dirac equation in lattice QCD using a domain decomposition method}},}\
  }\href {\doibase 10.1016/S0010-4655(03)00486-7} {\bibfield  {journal}
  {\bibinfo  {journal} {Comput. Phys. Commun.}\ }\textbf {\bibinfo {volume}
  {156}},\ \bibinfo {pages} {209--220} (\bibinfo {year} {2004})},\ \Eprint
  {http://arxiv.org/abs/hep-lat/0310048} {arXiv:hep-lat/0310048 [hep-lat]}
  \BibitemShut {NoStop}%
\bibitem [{\citenamefont {Frommer}\ \emph {et~al.}(2014)\citenamefont
  {Frommer}, \citenamefont {Kahl}, \citenamefont {Krieg}, \citenamefont
  {Leder},\ and\ \citenamefont {Rottmann}}]{Frommer:2013fsa}%
  \BibitemOpen
  \bibfield  {author} {\bibinfo {author} {\bibfnamefont {Andreas}\ \bibnamefont
  {Frommer}}, \bibinfo {author} {\bibfnamefont {Karsten}\ \bibnamefont {Kahl}},
  \bibinfo {author} {\bibfnamefont {Stefan}\ \bibnamefont {Krieg}}, \bibinfo
  {author} {\bibfnamefont {Bj{\"o}rn}\ \bibnamefont {Leder}}, \ and\ \bibinfo
  {author} {\bibfnamefont {Matthias}\ \bibnamefont {Rottmann}},\ }\bibfield
  {title} {\enquote {\bibinfo {title} {Adaptive aggregation based domain
  decomposition multigrid for the lattice {Wilson-Dirac} operator},}\ }\href
  {\doibase 10.1137/130919507} {\bibfield  {journal} {\bibinfo  {journal} {SIAM
  J. Sci. Comput.}\ }\textbf {\bibinfo {volume} {36}},\ \bibinfo {pages}
  {A1581--A1608} (\bibinfo {year} {2014})},\ \Eprint
  {http://arxiv.org/abs/1303.1377} {arXiv:1303.1377 [hep-lat]} \BibitemShut
  {NoStop}%
\bibitem [{\citenamefont {Edwards}\ and\ \citenamefont
  {Jo\'o}(2005)}]{Edwards:2004sx}%
  \BibitemOpen
  \bibfield  {author} {\bibinfo {author} {\bibfnamefont {Robert~G.}\
  \bibnamefont {Edwards}}\ and\ \bibinfo {author} {\bibfnamefont {B\'alint}\
  \bibnamefont {Jo\'o}},\ }\bibfield  {title} {\enquote {\bibinfo {title} {{The
  Chroma software system for lattice QCD}},}\ }\href@noop {} {\bibfield
  {journal} {\bibinfo  {journal} {Nucl. Phys. B. Proc. Suppl.}\ }\textbf
  {\bibinfo {volume} {140}},\ \bibinfo {pages} {832} (\bibinfo {year}
  {2005})}\BibitemShut {NoStop}%
\bibitem [{\citenamefont {Winter}\ \emph {et~al.}(2014)\citenamefont {Winter},
  \citenamefont {Clark}, \citenamefont {Edwards},\ and\ \citenamefont
  {Jo\'{o}}}]{Winter:2014:FLQ:2650283.2650646}%
  \BibitemOpen
  \bibfield  {author} {\bibinfo {author} {\bibfnamefont {F.~T.}\ \bibnamefont
  {Winter}}, \bibinfo {author} {\bibfnamefont {M.~A.}\ \bibnamefont {Clark}},
  \bibinfo {author} {\bibfnamefont {R.~G.}\ \bibnamefont {Edwards}}, \ and\
  \bibinfo {author} {\bibfnamefont {B.}~\bibnamefont {Jo\'{o}}},\ }\bibfield
  {title} {\enquote {\bibinfo {title} {A framework for lattice {QCD}
  calculations on {GPUs}},}\ }in\ \href {\doibase 10.1109/IPDPS.2014.112}
  {\emph {\bibinfo {booktitle} {Proceedings of the 2014 IEEE 28th International
  Parallel and Distributed Processing Symposium}}},\ \bibinfo {series and
  number} {IPDPS '14}\ (\bibinfo  {publisher} {IEEE Computer Society},\
  \bibinfo {address} {Washington, DC, USA},\ \bibinfo {year} {2014})\ pp.\
  \bibinfo {pages} {1073--1082}\BibitemShut {NoStop}%
\bibitem [{\citenamefont {Saad}(2003)}]{Saad:2003:IMS:829576}%
  \BibitemOpen
  \bibfield  {author} {\bibinfo {author} {\bibfnamefont {Y.}~\bibnamefont
  {Saad}},\ }\href@noop {} {\emph {\bibinfo {title} {Iterative Methods for
  Sparse Linear Systems}}},\ \bibinfo {edition} {2nd}\ ed.\ (\bibinfo
  {publisher} {Society for Industrial and Applied Mathematics},\ \bibinfo
  {address} {Philadelphia, PA, USA},\ \bibinfo {year} {2003})\BibitemShut
  {NoStop}%
\bibitem [{\citenamefont {Batrouni}\ \emph {et~al.}(1985)\citenamefont
  {Batrouni}, \citenamefont {Katz}, \citenamefont {Kronfeld}, \citenamefont
  {Lepage}, \citenamefont {Svetitsky},\ and\ \citenamefont
  {Wilson}}]{Batrouni:1985jn}%
  \BibitemOpen
  \bibfield  {author} {\bibinfo {author} {\bibfnamefont {G.~G.}\ \bibnamefont
  {Batrouni}}, \bibinfo {author} {\bibfnamefont {G.~R.}\ \bibnamefont {Katz}},
  \bibinfo {author} {\bibfnamefont {Andreas~S.}\ \bibnamefont {Kronfeld}},
  \bibinfo {author} {\bibfnamefont {G.~P.}\ \bibnamefont {Lepage}}, \bibinfo
  {author} {\bibfnamefont {B.}~\bibnamefont {Svetitsky}}, \ and\ \bibinfo
  {author} {\bibfnamefont {K.~G.}\ \bibnamefont {Wilson}},\ }\bibfield  {title}
  {\enquote {\bibinfo {title} {{Langevin} simulations of lattice field
  theories},}\ }\href {\doibase 10.1103/PhysRevD.32.2736} {\bibfield  {journal}
  {\bibinfo  {journal} {Phys. Rev.}\ }\textbf {\bibinfo {volume} {D32}},\
  \bibinfo {pages} {2736} (\bibinfo {year} {1985})}\BibitemShut {NoStop}%
\bibitem [{\citenamefont {Sohl-Dickstein}\ \emph {et~al.}(2014)\citenamefont
  {Sohl-Dickstein}, \citenamefont {Mudigonda},\ and\ \citenamefont
  {DeWeese}}]{pmlr-v32-sohl-dickstein14}%
  \BibitemOpen
  \bibfield  {author} {\bibinfo {author} {\bibfnamefont {Jascha}\ \bibnamefont
  {Sohl-Dickstein}}, \bibinfo {author} {\bibfnamefont {Mayur}\ \bibnamefont
  {Mudigonda}}, \ and\ \bibinfo {author} {\bibfnamefont {Michael}\ \bibnamefont
  {DeWeese}},\ }\bibfield  {title} {\enquote {\bibinfo {title} {Hamiltonian
  monte carlo without detailed balance},}\ }in\ \href
  {http://proceedings.mlr.press/v32/sohl-dickstein14.html} {\emph {\bibinfo
  {booktitle} {Proceedings of the 31st International Conference on Machine
  Learning}}},\ \bibinfo {series} {Proceedings of Machine Learning Research},
  Vol.~\bibinfo {volume} {32},\ \bibinfo {editor} {edited by\ \bibinfo {editor}
  {\bibfnamefont {Eric~P.}\ \bibnamefont {Xing}}\ and\ \bibinfo {editor}
  {\bibfnamefont {Tony}\ \bibnamefont {Jebara}}}\ (\bibinfo  {publisher}
  {PMLR},\ \bibinfo {address} {Bejing, China},\ \bibinfo {year} {2014})\ pp.\
  \bibinfo {pages} {719--726}\BibitemShut {NoStop}%
\bibitem [{\citenamefont {Endres}\ \emph {et~al.}(2015)\citenamefont {Endres},
  \citenamefont {Brower}, \citenamefont {Detmold}, \citenamefont {Orginos},\
  and\ \citenamefont {Pochinsky}}]{Endres:2015yca}%
  \BibitemOpen
  \bibfield  {author} {\bibinfo {author} {\bibfnamefont {Michael~G.}\
  \bibnamefont {Endres}}, \bibinfo {author} {\bibfnamefont {Richard~C.}\
  \bibnamefont {Brower}}, \bibinfo {author} {\bibfnamefont {William}\
  \bibnamefont {Detmold}}, \bibinfo {author} {\bibfnamefont {Kostas}\
  \bibnamefont {Orginos}}, \ and\ \bibinfo {author} {\bibfnamefont {Andrew~V.}\
  \bibnamefont {Pochinsky}},\ }\bibfield  {title} {\enquote {\bibinfo {title}
  {{Multiscale Monte Carlo equilibration: Pure Yang-Mills theory}},}\ }\href
  {\doibase 10.1103/PhysRevD.92.114516} {\bibfield  {journal} {\bibinfo
  {journal} {Phys. Rev.}\ }\textbf {\bibinfo {volume} {D92}},\ \bibinfo {pages}
  {114516} (\bibinfo {year} {2015})},\ \Eprint
  {http://arxiv.org/abs/1510.04675} {arXiv:1510.04675 [hep-lat]} \BibitemShut
  {NoStop}%
\bibitem [{\citenamefont {Glassner}\ \emph {et~al.}(1996)\citenamefont
  {Glassner}, \citenamefont {Gusken}, \citenamefont {Lippert}, \citenamefont
  {Ritzenhofer}, \citenamefont {Schilling},\ and\ \citenamefont
  {Frommer}}]{Glassner:1996gz}%
  \BibitemOpen
  \bibfield  {author} {\bibinfo {author} {\bibfnamefont {U.}~\bibnamefont
  {Glassner}}, \bibinfo {author} {\bibfnamefont {S.}~\bibnamefont {Gusken}},
  \bibinfo {author} {\bibfnamefont {T.}~\bibnamefont {Lippert}}, \bibinfo
  {author} {\bibfnamefont {G.}~\bibnamefont {Ritzenhofer}}, \bibinfo {author}
  {\bibfnamefont {K.}~\bibnamefont {Schilling}}, \ and\ \bibinfo {author}
  {\bibfnamefont {A.}~\bibnamefont {Frommer}},\ }\bibfield  {title} {\enquote
  {\bibinfo {title} {{How to compute Green's functions for entire mass
  trajectories within Krylov solvers}},}\ }\href {\doibase
  10.1142/S0129183196000533} {\bibfield  {journal} {\bibinfo  {journal} {Int.
  J. Mod. Phys.}\ }\textbf {\bibinfo {volume} {C7}},\ \bibinfo {pages} {635}
  (\bibinfo {year} {1996})},\ \Eprint {http://arxiv.org/abs/hep-lat/9605008}
  {arXiv:hep-lat/9605008 [hep-lat]} \BibitemShut {NoStop}%
\bibitem [{\citenamefont {Jegerlehner}(1996)}]{Jegerlehner:1996pm}%
  \BibitemOpen
  \bibfield  {author} {\bibinfo {author} {\bibfnamefont {Beat}\ \bibnamefont
  {Jegerlehner}},\ }\bibfield  {title} {\enquote {\bibinfo {title} {{Krylov
  space solvers for shifted linear systems}},}\ }\href@noop {} {\  (\bibinfo
  {year} {1996})},\ \Eprint {http://arxiv.org/abs/hep-lat/9612014}
  {arXiv:hep-lat/9612014 [hep-lat]} \BibitemShut {NoStop}%
\bibitem [{\citenamefont {Stathopoulos}\ and\ \citenamefont
  {Orginos}(2007)}]{Stathopoulos:2007zi}%
  \BibitemOpen
  \bibfield  {author} {\bibinfo {author} {\bibfnamefont {Andreas}\ \bibnamefont
  {Stathopoulos}}\ and\ \bibinfo {author} {\bibfnamefont {Kostas}\ \bibnamefont
  {Orginos}},\ }\bibfield  {title} {\enquote {\bibinfo {title} {{Computing and
  deflating eigenvalues while solving multiple right hand side linear systems
  in Quantum Chromodynamics}},}\ }\href@noop {} {\  (\bibinfo {year} {2007})},\
  \Eprint {http://arxiv.org/abs/0707.0131} {arXiv:0707.0131 [hep-lat]}
  \BibitemShut {NoStop}%
\bibitem [{\citenamefont {Abdel-Rehim}\ \emph {et~al.}(2009)\citenamefont
  {Abdel-Rehim}, \citenamefont {Orginos},\ and\ \citenamefont
  {Stathopoulos}}]{AbdelRehim:2009by}%
  \BibitemOpen
  \bibfield  {author} {\bibinfo {author} {\bibfnamefont {Abdou}\ \bibnamefont
  {Abdel-Rehim}}, \bibinfo {author} {\bibfnamefont {Kostas}\ \bibnamefont
  {Orginos}}, \ and\ \bibinfo {author} {\bibfnamefont {Andreas}\ \bibnamefont
  {Stathopoulos}},\ }\bibfield  {title} {\enquote {\bibinfo {title} {{Extending
  the eigCG algorithm to non-symmetric linear systems with multiple right-hand
  sides}},}\ }\href {\doibase 10.22323/1.091.0036} {\bibfield  {journal}
  {\bibinfo  {journal} {PoS}\ }\textbf {\bibinfo {volume} {LAT2009}},\ \bibinfo
  {pages} {036} (\bibinfo {year} {2009})},\ \Eprint
  {http://arxiv.org/abs/0911.2285} {arXiv:0911.2285 [hep-lat]} \BibitemShut
  {NoStop}%
\bibitem [{\citenamefont {Frommer}\ \emph {et~al.}(2012)\citenamefont
  {Frommer}, \citenamefont {Nobile},\ and\ \citenamefont
  {Zingler}}]{Frommer:2012zm}%
  \BibitemOpen
  \bibfield  {author} {\bibinfo {author} {\bibfnamefont {Andreas}\ \bibnamefont
  {Frommer}}, \bibinfo {author} {\bibfnamefont {Andrea}\ \bibnamefont
  {Nobile}}, \ and\ \bibinfo {author} {\bibfnamefont {Paul}\ \bibnamefont
  {Zingler}},\ }\bibfield  {title} {\enquote {\bibinfo {title} {Deflation and
  flexible {SAP}-preconditioning of {GMRES} in lattice {QCD} simulation},}\
  }\href@noop {} {\  (\bibinfo {year} {2012})},\ \Eprint
  {http://arxiv.org/abs/1204.5463} {arXiv:1204.5463 [hep-lat]} \BibitemShut
  {NoStop}%
\bibitem [{\citenamefont {Morgan}(2002)}]{GMRESDR}%
  \BibitemOpen
  \bibfield  {author} {\bibinfo {author} {\bibfnamefont {Ronald}\ \bibnamefont
  {Morgan}},\ }\bibfield  {title} {\enquote {\bibinfo {title} {Gmres with
  deflation restarting},}\ }\href {\doibase 10.1137/S1064827599364659}
  {\bibfield  {journal} {\bibinfo  {journal} {Siam Journal on Scientific
  Computing}\ }\textbf {\bibinfo {volume} {24}},\ \bibinfo {pages} {20--37}
  (\bibinfo {year} {2002})}\BibitemShut {NoStop}%
\bibitem [{\citenamefont {Saad}(1993)}]{Saad:1993:FIP:160077.160089}%
  \BibitemOpen
  \bibfield  {author} {\bibinfo {author} {\bibfnamefont {Youcef}\ \bibnamefont
  {Saad}},\ }\bibfield  {title} {\enquote {\bibinfo {title} {A flexible
  inner-outer preconditioned gmres algorithm},}\ }\href {\doibase
  10.1007/BF01396750} {\bibfield  {journal} {\bibinfo  {journal} {SIAM J. Sci.
  Comput.}\ }\textbf {\bibinfo {volume} {14}},\ \bibinfo {pages} {461--469}
  (\bibinfo {year} {1993})}\BibitemShut {NoStop}%
\bibitem [{\citenamefont {L\"uscher}(2007{\natexlab{b}})}]{Luscher:2007se}%
  \BibitemOpen
  \bibfield  {author} {\bibinfo {author} {\bibfnamefont {Martin}\ \bibnamefont
  {L\"uscher}},\ }\bibfield  {title} {\enquote {\bibinfo {title} {Local
  coherence and deflation of the low quark modes in lattice qcd},}\ }\href
  {\doibase 10.1088/1126-6708/2007/07/081} {\bibfield  {journal} {\bibinfo
  {journal} {JHEP}\ }\textbf {\bibinfo {volume} {07}},\ \bibinfo {pages} {081}
  (\bibinfo {year} {2007}{\natexlab{b}})},\ \Eprint
  {http://arxiv.org/abs/0706.2298} {arXiv:0706.2298 [hep-lat]} \BibitemShut
  {NoStop}%
\bibitem [{\citenamefont {Babich}\ \emph
  {et~al.}(2010{\natexlab{b}})\citenamefont {Babich}, \citenamefont {Brannick},
  \citenamefont {Brower}, \citenamefont {Clark}, \citenamefont {Manteuffel}
  \emph {et~al.}}]{Babich:2010qb}%
  \BibitemOpen
  \bibfield  {author} {\bibinfo {author} {\bibfnamefont {R.}~\bibnamefont
  {Babich}}, \bibinfo {author} {\bibfnamefont {J.}~\bibnamefont {Brannick}},
  \bibinfo {author} {\bibfnamefont {R.C.}\ \bibnamefont {Brower}}, \bibinfo
  {author} {\bibfnamefont {M.A.}\ \bibnamefont {Clark}}, \bibinfo {author}
  {\bibfnamefont {T.A.}\ \bibnamefont {Manteuffel}},  \emph {et~al.},\
  }\bibfield  {title} {\enquote {\bibinfo {title} {{Adaptive multigrid
  algorithm for the lattice Wilson-Dirac operator}},}\ }\href {\doibase
  10.1103/PhysRevLett.105.201602} {\bibfield  {journal} {\bibinfo  {journal}
  {Phys.Rev.Lett.}\ }\textbf {\bibinfo {volume} {105}},\ \bibinfo {pages}
  {201602} (\bibinfo {year} {2010}{\natexlab{b}})},\ \Eprint
  {http://arxiv.org/abs/1005.3043} {arXiv:1005.3043 [hep-lat]} \BibitemShut
  {NoStop}%
\bibitem [{\citenamefont {Boyle}(2014)}]{Boyle:2014rwa}%
  \BibitemOpen
  \bibfield  {author} {\bibinfo {author} {\bibfnamefont {P~A}\ \bibnamefont
  {Boyle}},\ }\bibfield  {title} {\enquote {\bibinfo {title} {{Hierarchically
  deflated conjugate gradient}},}\ }\href@noop {} {\  (\bibinfo {year}
  {2014})},\ \Eprint {http://arxiv.org/abs/1402.2585} {arXiv:1402.2585
  [hep-lat]} \BibitemShut {NoStop}%
\bibitem [{\citenamefont {Yamaguchi}\ and\ \citenamefont
  {Boyle}(2016)}]{Yamaguchi:2016kop}%
  \BibitemOpen
  \bibfield  {author} {\bibinfo {author} {\bibfnamefont {Azusa}\ \bibnamefont
  {Yamaguchi}}\ and\ \bibinfo {author} {\bibfnamefont {Peter}\ \bibnamefont
  {Boyle}},\ }\bibfield  {title} {\enquote {\bibinfo {title} {{Hierarchically
  deflated conjugate residual}},}\ }\href {\doibase 10.22323/1.256.0374}
  {\bibfield  {journal} {\bibinfo  {journal} {PoS}\ }\textbf {\bibinfo {volume}
  {LATTICE2016}},\ \bibinfo {pages} {374} (\bibinfo {year} {2016})},\ \Eprint
  {http://arxiv.org/abs/1611.06944} {arXiv:1611.06944 [hep-lat]} \BibitemShut
  {NoStop}%
\bibitem [{\citenamefont {Tu}(2018)}]{Tu:2018wuo}%
  \BibitemOpen
  \bibfield  {author} {\bibinfo {author} {\bibfnamefont {Jiqun}\ \bibnamefont
  {Tu}},\ }\bibfield  {title} {\enquote {\bibinfo {title} {{Solving DWF Dirac
  Equation Using Multisplitting Preconditioned Conjugate Gradient}},}\ \
  }(\bibinfo {year} {2018})\ \Eprint {http://arxiv.org/abs/1811.08488}
  {arXiv:1811.08488 [hep-lat]} \BibitemShut {NoStop}%
\bibitem [{\citenamefont {Clark}\ \emph
  {et~al.}(2018{\natexlab{a}})\citenamefont {Clark}, \citenamefont
  {Strelchenko}, \citenamefont {Vaquero}, \citenamefont {Wagner},\ and\
  \citenamefont {Weinberg}}]{Clark:2017ekr}%
  \BibitemOpen
  \bibfield  {author} {\bibinfo {author} {\bibfnamefont {M.~A.}\ \bibnamefont
  {Clark}}, \bibinfo {author} {\bibfnamefont {Alexei}\ \bibnamefont
  {Strelchenko}}, \bibinfo {author} {\bibfnamefont {Alejandro}\ \bibnamefont
  {Vaquero}}, \bibinfo {author} {\bibfnamefont {Mathias}\ \bibnamefont
  {Wagner}}, \ and\ \bibinfo {author} {\bibfnamefont {Evan}\ \bibnamefont
  {Weinberg}},\ }\bibfield  {title} {\enquote {\bibinfo {title} {Pushing memory
  bandwidth limitations through efficient implementations of block-{Krylov}
  space solvers on {GPUs}},}\ }\href {\doibase 10.1016/j.cpc.2018.06.019}
  {\bibfield  {journal} {\bibinfo  {journal} {Comput. Phys. Commun.}\ }\textbf
  {\bibinfo {volume} {233}},\ \bibinfo {pages} {29--40} (\bibinfo {year}
  {2018}{\natexlab{a}})},\ \Eprint {http://arxiv.org/abs/1710.09745}
  {arXiv:1710.09745 [hep-lat]} \BibitemShut {NoStop}%
\bibitem [{\citenamefont {Brower}\ \emph {et~al.}(2018)\citenamefont {Brower},
  \citenamefont {Clark}, \citenamefont {Strelchenko},\ and\ \citenamefont
  {Weinberg}}]{Brower:2018ymy}%
  \BibitemOpen
  \bibfield  {author} {\bibinfo {author} {\bibfnamefont {Richard~C.}\
  \bibnamefont {Brower}}, \bibinfo {author} {\bibfnamefont {M.~A.}\
  \bibnamefont {Clark}}, \bibinfo {author} {\bibfnamefont {Alexei}\
  \bibnamefont {Strelchenko}}, \ and\ \bibinfo {author} {\bibfnamefont {Evan}\
  \bibnamefont {Weinberg}},\ }\bibfield  {title} {\enquote {\bibinfo {title}
  {{Multigrid algorithm for staggered lattice fermions}},}\ }\href {\doibase
  10.1103/PhysRevD.97.114513} {\bibfield  {journal} {\bibinfo  {journal} {Phys.
  Rev.}\ }\textbf {\bibinfo {volume} {D97}},\ \bibinfo {pages} {114513}
  (\bibinfo {year} {2018})},\ \Eprint {http://arxiv.org/abs/1801.07823}
  {arXiv:1801.07823 [hep-lat]} \BibitemShut {NoStop}%
\bibitem [{\citenamefont {Brannick}\ \emph {et~al.}(2008)\citenamefont
  {Brannick}, \citenamefont {Brower}, \citenamefont {Clark}, \citenamefont
  {Osborn},\ and\ \citenamefont {Rebbi}}]{Brannick:2007ue}%
  \BibitemOpen
  \bibfield  {author} {\bibinfo {author} {\bibfnamefont {J.}~\bibnamefont
  {Brannick}}, \bibinfo {author} {\bibfnamefont {R.~C.}\ \bibnamefont
  {Brower}}, \bibinfo {author} {\bibfnamefont {M.~A.}\ \bibnamefont {Clark}},
  \bibinfo {author} {\bibfnamefont {J.~C.}\ \bibnamefont {Osborn}}, \ and\
  \bibinfo {author} {\bibfnamefont {C.}~\bibnamefont {Rebbi}},\ }\bibfield
  {title} {\enquote {\bibinfo {title} {Adaptive multigrid algorithm for lattice
  {QCD}},}\ }\href {\doibase 10.1103/PhysRevLett.100.041601} {\bibfield
  {journal} {\bibinfo  {journal} {Phys. Rev. Lett.}\ }\textbf {\bibinfo
  {volume} {100}},\ \bibinfo {pages} {041601} (\bibinfo {year} {2008})},\
  \Eprint {http://arxiv.org/abs/0707.4018} {arXiv:0707.4018 [hep-lat]}
  \BibitemShut {NoStop}%
\bibitem [{\citenamefont {Osborn}\ \emph {et~al.}(2010)\citenamefont {Osborn},
  \citenamefont {Babich}, \citenamefont {Brannick}, \citenamefont {Brower},
  \citenamefont {Clark} \emph {et~al.}}]{Osborn:2010mb}%
  \BibitemOpen
  \bibfield  {author} {\bibinfo {author} {\bibfnamefont {J.C.}\ \bibnamefont
  {Osborn}}, \bibinfo {author} {\bibfnamefont {R.}~\bibnamefont {Babich}},
  \bibinfo {author} {\bibfnamefont {J.}~\bibnamefont {Brannick}}, \bibinfo
  {author} {\bibfnamefont {R.C.}\ \bibnamefont {Brower}}, \bibinfo {author}
  {\bibfnamefont {M.A.}\ \bibnamefont {Clark}},  \emph {et~al.},\ }\bibfield
  {title} {\enquote {\bibinfo {title} {{Multigrid solver for clover
  fermions}},}\ }\href@noop {} {\bibfield  {journal} {\bibinfo  {journal}
  {PoS}\ }\textbf {\bibinfo {volume} {LATTICE2010}},\ \bibinfo {pages} {037}
  (\bibinfo {year} {2010})},\ \Eprint {http://arxiv.org/abs/1011.2775}
  {arXiv:1011.2775 [hep-lat]} \BibitemShut {NoStop}%
\bibitem [{\citenamefont {Brezina}\ \emph {et~al.}(2004)\citenamefont
  {Brezina}, \citenamefont {Falgout}, \citenamefont {MacLachlan}, \citenamefont
  {Manteuffel}, \citenamefont {McCormick},\ and\ \citenamefont
  {Ruge}}]{Brezina04adaptivesmoothed}%
  \BibitemOpen
  \bibfield  {author} {\bibinfo {author} {\bibfnamefont {M.}~\bibnamefont
  {Brezina}}, \bibinfo {author} {\bibfnamefont {R.}~\bibnamefont {Falgout}},
  \bibinfo {author} {\bibfnamefont {S.}~\bibnamefont {MacLachlan}}, \bibinfo
  {author} {\bibfnamefont {T.}~\bibnamefont {Manteuffel}}, \bibinfo {author}
  {\bibfnamefont {S.}~\bibnamefont {McCormick}}, \ and\ \bibinfo {author}
  {\bibfnamefont {J.}~\bibnamefont {Ruge}},\ }\bibfield  {title} {\enquote
  {\bibinfo {title} {Adaptive smoothed aggregation {($\alpha$SA)}},}\
  }\href@noop {} {\bibfield  {journal} {\bibinfo  {journal} {SIAM J. Sci.
  Comp.}\ }\textbf {\bibinfo {volume} {25}},\ \bibinfo {pages} {2004} (\bibinfo
  {year} {2004})}\BibitemShut {NoStop}%
\bibitem [{\citenamefont {Heroux}(2013)}]{Heroux:2013:TRA:2465813.2465814}%
  \BibitemOpen
  \bibfield  {author} {\bibinfo {author} {\bibfnamefont {Michael~A.}\
  \bibnamefont {Heroux}},\ }\bibfield  {title} {\enquote {\bibinfo {title}
  {Toward resilient algorithms and applications},}\ }in\ \href {\doibase
  10.1145/2465813.2465814} {\emph {\bibinfo {booktitle} {Proceedings of the 3rd
  Workshop on Fault-tolerance for HPC at Extreme Scale}}},\ \bibinfo {series
  and number} {FTXS '13}\ (\bibinfo  {publisher} {ACM},\ \bibinfo {address}
  {New York, NY, USA},\ \bibinfo {year} {2013})\ pp.\ \bibinfo {pages}
  {1--2}\BibitemShut {NoStop}%
\bibitem [{\citenamefont {Kaczmarek}\ \emph {et~al.}(2015)\citenamefont
  {Kaczmarek}, \citenamefont {Schmidt}, \citenamefont {Steinbrecher},\ and\
  \citenamefont {Wagner}}]{Kaczmarek:2014mga}%
  \BibitemOpen
  \bibfield  {author} {\bibinfo {author} {\bibfnamefont {O.}~\bibnamefont
  {Kaczmarek}}, \bibinfo {author} {\bibfnamefont {C.}~\bibnamefont {Schmidt}},
  \bibinfo {author} {\bibfnamefont {P.}~\bibnamefont {Steinbrecher}}, \ and\
  \bibinfo {author} {\bibfnamefont {M.}~\bibnamefont {Wagner}},\ }\bibfield
  {title} {\enquote {\bibinfo {title} {{Conjugate gradient solvers on Intel
  Xeon Phi and NVIDIA GPUs}},}\ }in\ \href {\doibase
  10.3204/DESY-PROC-2014-05/28} {\emph {\bibinfo {booktitle} {{Proceedings, GPU
  Computing in High-Energy Physics (GPUHEP2014): Pisa, Italy, September 10-12,
  2014}}}}\ (\bibinfo  {publisher} {Verlag Deutsches Elektronen-Synchrotron},\
  \bibinfo {year} {2015})\ pp.\ \bibinfo {pages} {157--162},\ \Eprint
  {http://arxiv.org/abs/1411.4439} {arXiv:1411.4439 [physics.comp-ph]}
  \BibitemShut {NoStop}%
\bibitem [{\citenamefont {Blum}\ \emph {et~al.}(2013)\citenamefont {Blum},
  \citenamefont {Izubuchi},\ and\ \citenamefont {Shintani}}]{Blum:2012uh}%
  \BibitemOpen
  \bibfield  {author} {\bibinfo {author} {\bibfnamefont {Thomas}\ \bibnamefont
  {Blum}}, \bibinfo {author} {\bibfnamefont {Taku}\ \bibnamefont {Izubuchi}}, \
  and\ \bibinfo {author} {\bibfnamefont {Eigo}\ \bibnamefont {Shintani}},\
  }\bibfield  {title} {\enquote {\bibinfo {title} {{New class of
  variance-reduction techniques using lattice symmetries}},}\ }\href {\doibase
  10.1103/PhysRevD.88.094503} {\bibfield  {journal} {\bibinfo  {journal} {Phys.
  Rev.}\ }\textbf {\bibinfo {volume} {D88}},\ \bibinfo {pages} {094503}
  (\bibinfo {year} {2013})},\ \Eprint {http://arxiv.org/abs/1208.4349}
  {arXiv:1208.4349 [hep-lat]} \BibitemShut {NoStop}%
\bibitem [{\citenamefont {Shintani}\ \emph {et~al.}(2015)\citenamefont
  {Shintani}, \citenamefont {Arthur}, \citenamefont {Blum}, \citenamefont
  {Izubuchi}, \citenamefont {Jung},\ and\ \citenamefont
  {Lehner}}]{Shintani:2014vja}%
  \BibitemOpen
  \bibfield  {author} {\bibinfo {author} {\bibfnamefont {Eigo}\ \bibnamefont
  {Shintani}}, \bibinfo {author} {\bibfnamefont {Rudy}\ \bibnamefont {Arthur}},
  \bibinfo {author} {\bibfnamefont {Thomas}\ \bibnamefont {Blum}}, \bibinfo
  {author} {\bibfnamefont {Taku}\ \bibnamefont {Izubuchi}}, \bibinfo {author}
  {\bibfnamefont {Chulwoo}\ \bibnamefont {Jung}}, \ and\ \bibinfo {author}
  {\bibfnamefont {Christoph}\ \bibnamefont {Lehner}},\ }\bibfield  {title}
  {\enquote {\bibinfo {title} {{Covariant approximation averaging}},}\ }\href
  {\doibase 10.1103/PhysRevD.91.114511} {\bibfield  {journal} {\bibinfo
  {journal} {Phys. Rev.}\ }\textbf {\bibinfo {volume} {D91}},\ \bibinfo {pages}
  {114511} (\bibinfo {year} {2015})},\ \Eprint {http://arxiv.org/abs/1402.0244}
  {arXiv:1402.0244 [hep-lat]} \BibitemShut {NoStop}%
\bibitem [{\citenamefont {Foley}\ \emph {et~al.}(2005)\citenamefont {Foley}
  \emph {et~al.}}]{Foley:2005ac}%
  \BibitemOpen
  \bibfield  {author} {\bibinfo {author} {\bibfnamefont {Justin}\ \bibnamefont
  {Foley}} \emph {et~al.},\ }\bibfield  {title} {\enquote {\bibinfo {title}
  {{Practical all-to-all propagators for lattice QCD}},}\ }\href {\doibase
  10.1016/j.cpc.2005.06.008} {\bibfield  {journal} {\bibinfo  {journal}
  {Comput. Phys. Commun.}\ }\textbf {\bibinfo {volume} {172}},\ \bibinfo
  {pages} {145--162} (\bibinfo {year} {2005})},\ \Eprint
  {http://arxiv.org/abs/hep-lat/0505023} {arXiv:hep-lat/0505023} \BibitemShut
  {NoStop}%
\bibitem [{\citenamefont {Calvetti}\ \emph {et~al.}(1994)\citenamefont
  {Calvetti}, \citenamefont {Reichel},\ and\ \citenamefont
  {Sorensen}}]{Calvetti1994AnIR}%
  \BibitemOpen
  \bibfield  {author} {\bibinfo {author} {\bibfnamefont {Daniela}\ \bibnamefont
  {Calvetti}}, \bibinfo {author} {\bibfnamefont {Lothar}\ \bibnamefont
  {Reichel}}, \ and\ \bibinfo {author} {\bibfnamefont {Danny~C.}\ \bibnamefont
  {Sorensen}},\ }\bibfield  {title} {\enquote {\bibinfo {title} {An implicitly
  restarted {Lanczos} method for large symmetric eigenvalue problems},}\ \
  }(\bibinfo {year} {1994})\BibitemShut {NoStop}%
\bibitem [{\citenamefont {Clark}\ \emph
  {et~al.}(2018{\natexlab{b}})\citenamefont {Clark}, \citenamefont {Jung},\
  and\ \citenamefont {Lehner}}]{Clark:2017wom}%
  \BibitemOpen
  \bibfield  {author} {\bibinfo {author} {\bibfnamefont {M.~A.}\ \bibnamefont
  {Clark}}, \bibinfo {author} {\bibfnamefont {Chulwoo}\ \bibnamefont {Jung}}, \
  and\ \bibinfo {author} {\bibfnamefont {Christoph}\ \bibnamefont {Lehner}},\
  }\bibfield  {title} {\enquote {\bibinfo {title} {{Multi-Grid Lanczos}},}\
  }\href {\doibase 10.1051/epjconf/201817514023} {\bibfield  {journal}
  {\bibinfo  {journal} {EPJ Web Conf.}\ }\textbf {\bibinfo {volume} {175}},\
  \bibinfo {pages} {14023} (\bibinfo {year} {2018}{\natexlab{b}})},\ \Eprint
  {http://arxiv.org/abs/1710.06884} {arXiv:1710.06884 [hep-lat]} \BibitemShut
  {NoStop}%
\bibitem [{\citenamefont {Blum}\ \emph {et~al.}(2016)\citenamefont {Blum} \emph
  {et~al.}}]{Blum:2014tka}%
  \BibitemOpen
  \bibfield  {author} {\bibinfo {author} {\bibfnamefont {T.}~\bibnamefont
  {Blum}} \emph {et~al.} (\bibinfo {collaboration} {RBC, UKQCD}),\ }\bibfield
  {title} {\enquote {\bibinfo {title} {{Domain wall QCD with physical quark
  masses}},}\ }\href {\doibase 10.1103/PhysRevD.93.074505} {\bibfield
  {journal} {\bibinfo  {journal} {Phys. Rev.}\ }\textbf {\bibinfo {volume}
  {D93}},\ \bibinfo {pages} {074505} (\bibinfo {year} {2016})},\ \Eprint
  {http://arxiv.org/abs/1411.7017} {arXiv:1411.7017 [hep-lat]} \BibitemShut
  {NoStop}%
\bibitem [{\citenamefont {Jang}\ and\ \citenamefont
  {Jung}(2019)}]{Jang:2019roq}%
  \BibitemOpen
  \bibfield  {author} {\bibinfo {author} {\bibfnamefont {Yong-Chull}\
  \bibnamefont {Jang}}\ and\ \bibinfo {author} {\bibfnamefont {Chulwoo}\
  \bibnamefont {Jung}},\ }\bibfield  {title} {\enquote {\bibinfo {title}
  {{Split Grid and Block Lanczos Algorithm for Efficient Eigenpair
  Generation}},}\ }\bibfield  {booktitle} {\emph {\bibinfo {booktitle}
  {{Proceedings, 36th International Symposium on Lattice Field Theory (Lattice
  2018): East Lansing, MI, United States, July 22-28, 2018}}},\ }\href
  {\doibase 10.22323/1.334.0309} {\bibfield  {journal} {\bibinfo  {journal}
  {PoS}\ }\textbf {\bibinfo {volume} {LATTICE2018}},\ \bibinfo {pages} {309}
  (\bibinfo {year} {2019})}\BibitemShut {NoStop}%
\bibitem [{\citenamefont {Cutting}\ and\ \citenamefont {Cafarella}()}]{hadoop}%
  \BibitemOpen
  \bibfield  {author} {\bibinfo {author} {\bibfnamefont {Doug}\ \bibnamefont
  {Cutting}}\ and\ \bibinfo {author} {\bibfnamefont {Mike}\ \bibnamefont
  {Cafarella}},\ }\href@noop {} {\enquote {\bibinfo {title} {Apache hadoop
  software library},}\ }\bibinfo {howpublished}
  {\url{https://hadoop.apache.org/}}\BibitemShut {NoStop}%
\bibitem [{\citenamefont {Michael}(1985)}]{Michael:1985ne}%
  \BibitemOpen
  \bibfield  {author} {\bibinfo {author} {\bibfnamefont {Christopher}\
  \bibnamefont {Michael}},\ }\bibfield  {title} {\enquote {\bibinfo {title}
  {Adjoint sources in lattice gauge theory},}\ }\href {\doibase
  10.1016/0550-3213(85)90297-4} {\bibfield  {journal} {\bibinfo  {journal}
  {Nucl. Phys.}\ }\textbf {\bibinfo {volume} {B259}},\ \bibinfo {pages}
  {58--76} (\bibinfo {year} {1985})}\BibitemShut {NoStop}%
\bibitem [{\citenamefont {L\"uscher}\ and\ \citenamefont
  {Wolff}(1990)}]{Luscher:1990ck}%
  \BibitemOpen
  \bibfield  {author} {\bibinfo {author} {\bibfnamefont {Martin}\ \bibnamefont
  {L\"uscher}}\ and\ \bibinfo {author} {\bibfnamefont {Ulli}\ \bibnamefont
  {Wolff}},\ }\bibfield  {title} {\enquote {\bibinfo {title} {{How to calculate
  the elastic scattering matrix in two- dimensional quantum field theories by
  numerical simulation}},}\ }\href {\doibase 10.1016/0550-3213(90)90540-T}
  {\bibfield  {journal} {\bibinfo  {journal} {Nucl. Phys.}\ }\textbf {\bibinfo
  {volume} {B339}},\ \bibinfo {pages} {222--252} (\bibinfo {year}
  {1990})}\BibitemShut {NoStop}%
\bibitem [{\citenamefont {Blossier}\ \emph {et~al.}(2009)\citenamefont
  {Blossier}, \citenamefont {Della~Morte}, \citenamefont {von Hippel},
  \citenamefont {Mendes},\ and\ \citenamefont {Sommer}}]{Blossier:2009kd}%
  \BibitemOpen
  \bibfield  {author} {\bibinfo {author} {\bibfnamefont {Benoit}\ \bibnamefont
  {Blossier}}, \bibinfo {author} {\bibfnamefont {Michele}\ \bibnamefont
  {Della~Morte}}, \bibinfo {author} {\bibfnamefont {Georg}\ \bibnamefont {von
  Hippel}}, \bibinfo {author} {\bibfnamefont {Tereza}\ \bibnamefont {Mendes}},
  \ and\ \bibinfo {author} {\bibfnamefont {Rainer}\ \bibnamefont {Sommer}},\
  }\bibfield  {title} {\enquote {\bibinfo {title} {{On the generalized
  eigenvalue method for energies and matrix elements in lattice field
  theory}},}\ }\href {\doibase 10.1088/1126-6708/2009/04/094} {\bibfield
  {journal} {\bibinfo  {journal} {JHEP}\ }\textbf {\bibinfo {volume} {04}},\
  \bibinfo {pages} {094} (\bibinfo {year} {2009})},\ \Eprint
  {http://arxiv.org/abs/0902.1265} {arXiv:0902.1265 [hep-lat]} \BibitemShut
  {NoStop}%
\bibitem [{\citenamefont {Peardon}\ \emph {et~al.}(2009)\citenamefont {Peardon}
  \emph {et~al.}}]{Peardon:2009gh}%
  \BibitemOpen
  \bibfield  {author} {\bibinfo {author} {\bibfnamefont {Michael}\ \bibnamefont
  {Peardon}} \emph {et~al.},\ }\bibfield  {title} {\enquote {\bibinfo {title}
  {{A novel quark-field creation operator construction for hadronic physics in
  lattice QCD}},}\ }\href {\doibase 10.1103/PhysRevD.80.054506} {\bibfield
  {journal} {\bibinfo  {journal} {Phys. Rev.}\ }\textbf {\bibinfo {volume}
  {D80}},\ \bibinfo {pages} {054506} (\bibinfo {year} {2009})},\ \Eprint
  {http://arxiv.org/abs/0905.2160} {arXiv:0905.2160 [hep-lat]} \BibitemShut
  {NoStop}%
\bibitem [{\citenamefont {Dudek}\ \emph {et~al.}(2012)\citenamefont {Dudek},
  \citenamefont {Edwards},\ and\ \citenamefont {Thomas}}]{Dudek:2012gj}%
  \BibitemOpen
  \bibfield  {author} {\bibinfo {author} {\bibfnamefont {Jozef~J.}\
  \bibnamefont {Dudek}}, \bibinfo {author} {\bibfnamefont {Robert~G.}\
  \bibnamefont {Edwards}}, \ and\ \bibinfo {author} {\bibfnamefont
  {Christopher~E.}\ \bibnamefont {Thomas}},\ }\bibfield  {title} {\enquote
  {\bibinfo {title} {{$S$}- and {$D$}-wave phase shifts in isospin-2 $\pi\pi$
  scattering from lattice {QCD}},}\ }\href {\doibase
  10.1103/PhysRevD.86.034031} {\bibfield  {journal} {\bibinfo  {journal} {Phys.
  Rev.}\ }\textbf {\bibinfo {volume} {D86}},\ \bibinfo {pages} {034031}
  (\bibinfo {year} {2012})},\ \Eprint {http://arxiv.org/abs/1203.6041}
  {arXiv:1203.6041 [hep-ph]} \BibitemShut {NoStop}%
\bibitem [{\citenamefont {Stathopoulos}\ \emph {et~al.}(2013)\citenamefont
  {Stathopoulos}, \citenamefont {Laeuchli},\ and\ \citenamefont
  {Orginos}}]{Stathopoulos:2013aci}%
  \BibitemOpen
  \bibfield  {author} {\bibinfo {author} {\bibfnamefont {Andreas}\ \bibnamefont
  {Stathopoulos}}, \bibinfo {author} {\bibfnamefont {Jesse}\ \bibnamefont
  {Laeuchli}}, \ and\ \bibinfo {author} {\bibfnamefont {Kostas}\ \bibnamefont
  {Orginos}},\ }\bibfield  {title} {\enquote {\bibinfo {title} {{Hierarchical
  probing for estimating the trace of the matrix inverse on toroidal
  lattices}},}\ }\href@noop {} {\  (\bibinfo {year} {2013})},\ \Eprint
  {http://arxiv.org/abs/1302.4018} {arXiv:1302.4018 [hep-lat]} \BibitemShut
  {NoStop}%
\bibitem [{\citenamefont {{L\"uscher,
  Martin}}(1991{\natexlab{a}})}]{Luscher:1990ux}%
  \BibitemOpen
  \bibfield  {author} {\bibinfo {author} {\bibnamefont {{L\"uscher, Martin}}},\
  }\bibfield  {title} {\enquote {\bibinfo {title} {{Two particle states on a
  torus and their relation to the scattering matrix}},}\ }\href {\doibase
  10.1016/0550-3213(91)90366-6} {\bibfield  {journal} {\bibinfo  {journal}
  {Nucl. Phys.}\ }\textbf {\bibinfo {volume} {B354}},\ \bibinfo {pages}
  {531--578} (\bibinfo {year} {1991}{\natexlab{a}})}\BibitemShut {NoStop}%
\bibitem [{\citenamefont {{L\"uscher,
  Martin}}(1991{\natexlab{b}})}]{Luscher:1991cf}%
  \BibitemOpen
  \bibfield  {author} {\bibinfo {author} {\bibnamefont {{L\"uscher, Martin}}},\
  }\bibfield  {title} {\enquote {\bibinfo {title} {{Signatures of unstable
  particles in finite volume}},}\ }\href {\doibase
  10.1016/0550-3213(91)90584-K} {\bibfield  {journal} {\bibinfo  {journal}
  {Nucl. Phys.}\ }\textbf {\bibinfo {volume} {B364}},\ \bibinfo {pages}
  {237--251} (\bibinfo {year} {1991}{\natexlab{b}})}\BibitemShut {NoStop}%
\bibitem [{\citenamefont {Rummukainen}\ and\ \citenamefont
  {Gottlieb}(1995)}]{Rummukainen:1995vs}%
  \BibitemOpen
  \bibfield  {author} {\bibinfo {author} {\bibfnamefont {K.}~\bibnamefont
  {Rummukainen}}\ and\ \bibinfo {author} {\bibfnamefont {Steven~A.}\
  \bibnamefont {Gottlieb}},\ }\bibfield  {title} {\enquote {\bibinfo {title}
  {{Resonance scattering phase shifts on a nonrest frame lattice}},}\ }\href
  {\doibase 10.1016/0550-3213(95)00313-H} {\bibfield  {journal} {\bibinfo
  {journal} {Nucl. Phys.}\ }\textbf {\bibinfo {volume} {B450}},\ \bibinfo
  {pages} {397--436} (\bibinfo {year} {1995})},\ \Eprint
  {http://arxiv.org/abs/hep-lat/9503028} {arXiv:hep-lat/9503028 [hep-lat]}
  \BibitemShut {NoStop}%
\bibitem [{\citenamefont {Briceno}\ \emph {et~al.}(2017)\citenamefont
  {Briceno}, \citenamefont {Dudek}, \citenamefont {Edwards},\ and\
  \citenamefont {Wilson}}]{Briceno:2016mjc}%
  \BibitemOpen
  \bibfield  {author} {\bibinfo {author} {\bibfnamefont {Raul~A.}\ \bibnamefont
  {Briceno}}, \bibinfo {author} {\bibfnamefont {Jozef~J.}\ \bibnamefont
  {Dudek}}, \bibinfo {author} {\bibfnamefont {Robert~G.}\ \bibnamefont
  {Edwards}}, \ and\ \bibinfo {author} {\bibfnamefont {David~J.}\ \bibnamefont
  {Wilson}},\ }\bibfield  {title} {\enquote {\bibinfo {title} {{Isoscalar
  $\pi\pi$ scattering and the $\sigma$ meson resonance from QCD}},}\ }\href
  {\doibase 10.1103/PhysRevLett.118.022002} {\bibfield  {journal} {\bibinfo
  {journal} {Phys. Rev. Lett.}\ }\textbf {\bibinfo {volume} {118}},\ \bibinfo
  {pages} {022002} (\bibinfo {year} {2017})},\ \Eprint
  {http://arxiv.org/abs/1607.05900} {arXiv:1607.05900 [hep-ph]} \BibitemShut
  {NoStop}%
\bibitem [{\citenamefont {Bedaque}(2004)}]{Bedaque:2004kc}%
  \BibitemOpen
  \bibfield  {author} {\bibinfo {author} {\bibfnamefont {Paulo~F.}\
  \bibnamefont {Bedaque}},\ }\bibfield  {title} {\enquote {\bibinfo {title}
  {{Aharonov-Bohm effect and nucleon nucleon phase shifts on the lattice}},}\
  }\href {\doibase 10.1016/j.physletb.2004.04.045} {\bibfield  {journal}
  {\bibinfo  {journal} {Phys. Lett.}\ }\textbf {\bibinfo {volume} {B593}},\
  \bibinfo {pages} {82--88} (\bibinfo {year} {2004})},\ \Eprint
  {http://arxiv.org/abs/nucl-th/0402051} {arXiv:nucl-th/0402051 [nucl-th]}
  \BibitemShut {NoStop}%
\bibitem [{\citenamefont {Detmold}\ and\ \citenamefont
  {Orginos}(2013)}]{Detmold:2012eu}%
  \BibitemOpen
  \bibfield  {author} {\bibinfo {author} {\bibfnamefont {William}\ \bibnamefont
  {Detmold}}\ and\ \bibinfo {author} {\bibfnamefont {Kostas}\ \bibnamefont
  {Orginos}},\ }\bibfield  {title} {\enquote {\bibinfo {title} {{Nuclear
  correlation functions in lattice QCD}},}\ }\href {\doibase
  10.1103/PhysRevD.87.114512} {\bibfield  {journal} {\bibinfo  {journal} {Phys.
  Rev.}\ }\textbf {\bibinfo {volume} {D87}},\ \bibinfo {pages} {114512}
  (\bibinfo {year} {2013})},\ \Eprint {http://arxiv.org/abs/1207.1452}
  {arXiv:1207.1452 [hep-lat]} \BibitemShut {NoStop}%
\bibitem [{\citenamefont {Vachaspati}\ and\ \citenamefont
  {Detmold}(2014)}]{Vachaspati:2014bda}%
  \BibitemOpen
  \bibfield  {author} {\bibinfo {author} {\bibfnamefont {Pranjal}\ \bibnamefont
  {Vachaspati}}\ and\ \bibinfo {author} {\bibfnamefont {William}\ \bibnamefont
  {Detmold}},\ }\bibfield  {title} {\enquote {\bibinfo {title} {Fast evaluation
  of multi-hadron correlation functions},}\ }\href {\doibase
  10.22323/1.214.0041} {\bibfield  {journal} {\bibinfo  {journal} {PoS}\
  }\textbf {\bibinfo {volume} {LATTICE2014}},\ \bibinfo {pages} {041} (\bibinfo
  {year} {2014})},\ \Eprint {http://arxiv.org/abs/1411.3691} {arXiv:1411.3691
  [hep-lat]} \BibitemShut {NoStop}%
\bibitem [{\citenamefont {Mackenzie}\ \emph {et~al.}(1988)\citenamefont
  {Mackenzie}, \citenamefont {Eichten}, \citenamefont {Hockney}, \citenamefont
  {Thacker}, \citenamefont {Atac}, \citenamefont {Cook}, \citenamefont
  {Fischler}, \citenamefont {Gaines}, \citenamefont {Husby},\ and\
  \citenamefont {Nash}}]{Mackenzie:1987fc}%
  \BibitemOpen
  \bibfield  {author} {\bibinfo {author} {\bibfnamefont {Paul~B.}\ \bibnamefont
  {Mackenzie}}, \bibinfo {author} {\bibfnamefont {E.}~\bibnamefont {Eichten}},
  \bibinfo {author} {\bibfnamefont {G.}~\bibnamefont {Hockney}}, \bibinfo
  {author} {\bibfnamefont {H.~B.}\ \bibnamefont {Thacker}}, \bibinfo {author}
  {\bibfnamefont {R.}~\bibnamefont {Atac}}, \bibinfo {author} {\bibfnamefont
  {A.}~\bibnamefont {Cook}}, \bibinfo {author} {\bibfnamefont {M.}~\bibnamefont
  {Fischler}}, \bibinfo {author} {\bibfnamefont {I.}~\bibnamefont {Gaines}},
  \bibinfo {author} {\bibfnamefont {D.}~\bibnamefont {Husby}}, \ and\ \bibinfo
  {author} {\bibfnamefont {T.}~\bibnamefont {Nash}},\ }\bibfield  {title}
  {\enquote {\bibinfo {title} {{ACPMAPS}: The {Fermilab} lattice supercomputer
  project},}\ }\href {\doibase 10.1016/0920-5632(88)90158-2} {\bibfield
  {journal} {\bibinfo  {journal} {Nucl. Phys. Proc. Suppl.}\ }\textbf {\bibinfo
  {volume} {4}},\ \bibinfo {pages} {580} (\bibinfo {year} {1988})}\BibitemShut
  {NoStop}%
\bibitem [{\citenamefont {Chen}\ \emph {et~al.}(1999)\citenamefont {Chen} \emph
  {et~al.}}]{Chen:1998cg}%
  \BibitemOpen
  \bibfield  {author} {\bibinfo {author} {\bibfnamefont {Dong}\ \bibnamefont
  {Chen}} \emph {et~al.},\ }\bibfield  {title} {\enquote {\bibinfo {title}
  {Status of the {QCDSP} project},}\ }\href {\doibase
  10.1016/S0920-5632(99)85238-4} {\bibfield  {journal} {\bibinfo  {journal}
  {Nucl. Phys. Proc. Suppl.}\ }\textbf {\bibinfo {volume} {73}},\ \bibinfo
  {pages} {898--900} (\bibinfo {year} {1999})},\ \Eprint
  {http://arxiv.org/abs/hep-lat/9810004} {arXiv:hep-lat/9810004 [hep-lat]}
  \BibitemShut {NoStop}%
\bibitem [{\citenamefont {Edwards}\ \emph {et~al.}()\citenamefont {Edwards},
  \citenamefont {Jo\'o},\ and\ \citenamefont {Winter}}]{ChromaWeb}%
  \BibitemOpen
  \bibfield  {author} {\bibinfo {author} {\bibfnamefont {R.~G.}\ \bibnamefont
  {Edwards}}, \bibinfo {author} {\bibfnamefont {B}~\bibnamefont {Jo\'o}}, \
  and\ \bibinfo {author} {\bibfnamefont {F.}~\bibnamefont {Winter}},\
  }\href@noop {} {\enquote {\bibinfo {title} {{The Chroma Code Web Page}},}\
  }\bibinfo {howpublished}
  {\url{http://jeffersonlab.github.io/chroma/}}\BibitemShut {NoStop}%
\bibitem [{\citenamefont {Jo\'o}\ \emph {et~al.}(2013)\citenamefont {Jo\'o},
  \citenamefont {Kalamkar}, \citenamefont {Vaidyanathan}, \citenamefont
  {Smelyanskiy}, \citenamefont {Pamnany}, \citenamefont {Lee}, \citenamefont
  {Dubey},\ and\ \citenamefont {Watson}}]{ISC13Phi}%
  \BibitemOpen
  \bibfield  {author} {\bibinfo {author} {\bibfnamefont {B\'alint}\
  \bibnamefont {Jo\'o}}, \bibinfo {author} {\bibfnamefont {Dhiraj~D.}\
  \bibnamefont {Kalamkar}}, \bibinfo {author} {\bibfnamefont {Karthikeyan}\
  \bibnamefont {Vaidyanathan}}, \bibinfo {author} {\bibfnamefont {Mikhail}\
  \bibnamefont {Smelyanskiy}}, \bibinfo {author} {\bibfnamefont {Kiran}\
  \bibnamefont {Pamnany}}, \bibinfo {author} {\bibfnamefont {VictorW.}\
  \bibnamefont {Lee}}, \bibinfo {author} {\bibfnamefont {Pradeep}\ \bibnamefont
  {Dubey}}, \ and\ \bibinfo {author} {\bibfnamefont {William}\ \bibnamefont
  {Watson}},\ }\bibfield  {title} {\enquote {\bibinfo {title} {Lattice {QCD} on
  {Intel\textregistered} {Xeon~Phi}\texttrademark coprocessors},}\ }in\ \href
  {\doibase 10.1007/978-3-642-38750-0_4} {\emph {\bibinfo {booktitle}
  {Supercomputing}}},\ \bibinfo {series} {Lecture Notes in Computer Science},
  Vol.\ \bibinfo {volume} {7905},\ \bibinfo {editor} {edited by\ \bibinfo
  {editor} {\bibfnamefont {Julian~Martin}\ \bibnamefont {Kunkel}}, \bibinfo
  {editor} {\bibfnamefont {Thomas}\ \bibnamefont {Ludwig}}, \ and\ \bibinfo
  {editor} {\bibfnamefont {Hans~Werner}\ \bibnamefont {Meuer}}}\ (\bibinfo
  {publisher} {Springer Berlin Heidelberg},\ \bibinfo {year} {2013})\ pp.\
  \bibinfo {pages} {40--54}\BibitemShut {NoStop}%
\bibitem [{\citenamefont {Projects}()}]{QPhiXLocation}%
  \BibitemOpen
  \bibfield  {author} {\bibinfo {author} {\bibfnamefont {Jefferson Lab~GitHub}\
  \bibnamefont {Projects}},\ }\href@noop {} {\enquote {\bibinfo {title} {{QPhiX
  Library}},}\ }\bibinfo {howpublished}
  {\url{https://github.com/jeffersonlab/qphix.git}}\BibitemShut {NoStop}%
\bibitem [{\citenamefont {Heybrock}\ \emph {et~al.}(2014)\citenamefont
  {Heybrock}, \citenamefont {Jo\'{o}}, \citenamefont {Kalamkar}, \citenamefont
  {Smelyanskiy}, \citenamefont {Vaidyanathan}, \citenamefont {Wettig},\ and\
  \citenamefont {Dubey}}]{Heybrock:2014:LQD:2683593.2683602}%
  \BibitemOpen
  \bibfield  {author} {\bibinfo {author} {\bibfnamefont {Simon}\ \bibnamefont
  {Heybrock}}, \bibinfo {author} {\bibfnamefont {B\'{a}lint}\ \bibnamefont
  {Jo\'{o}}}, \bibinfo {author} {\bibfnamefont {Dhiraj~D.}\ \bibnamefont
  {Kalamkar}}, \bibinfo {author} {\bibfnamefont {Mikhail}\ \bibnamefont
  {Smelyanskiy}}, \bibinfo {author} {\bibfnamefont {Karthikeyan}\ \bibnamefont
  {Vaidyanathan}}, \bibinfo {author} {\bibfnamefont {Tilo}\ \bibnamefont
  {Wettig}}, \ and\ \bibinfo {author} {\bibfnamefont {Pradeep}\ \bibnamefont
  {Dubey}},\ }\bibfield  {title} {\enquote {\bibinfo {title} {Lattice {QCD}
  with domain decomposition on {Intel}\textregistered {Xeon~Phi}\texttrademark
  co-processors},}\ }in\ \href {\doibase 10.1109/SC.2014.11} {\emph {\bibinfo
  {booktitle} {Proceedings of the International Conference for High Performance
  Computing, Networking, Storage and Analysis}}},\ \bibinfo {series and number}
  {SC '14}\ (\bibinfo  {publisher} {IEEE Press},\ \bibinfo {address}
  {Piscataway, NJ, USA},\ \bibinfo {year} {2014})\ pp.\ \bibinfo {pages}
  {69--80}\BibitemShut {NoStop}%
\bibitem [{\citenamefont {DeTar}\ \emph {et~al.}(2016)\citenamefont {DeTar},
  \citenamefont {Doerfler}, \citenamefont {Gottlieb}, \citenamefont {Jha},
  \citenamefont {Kalamkar}, \citenamefont {Li},\ and\ \citenamefont
  {Toussaint}}]{DeTar:2016ndn}%
  \BibitemOpen
  \bibfield  {author} {\bibinfo {author} {\bibfnamefont {Carleton}\
  \bibnamefont {DeTar}}, \bibinfo {author} {\bibfnamefont {Douglas}\
  \bibnamefont {Doerfler}}, \bibinfo {author} {\bibfnamefont {Steven}\
  \bibnamefont {Gottlieb}}, \bibinfo {author} {\bibfnamefont {Ashish}\
  \bibnamefont {Jha}}, \bibinfo {author} {\bibfnamefont {Dhiraj}\ \bibnamefont
  {Kalamkar}}, \bibinfo {author} {\bibfnamefont {Ruizi}\ \bibnamefont {Li}}, \
  and\ \bibinfo {author} {\bibfnamefont {Doug}\ \bibnamefont {Toussaint}},\
  }\bibfield  {title} {\enquote {\bibinfo {title} {{MILC staggered conjugate
  gradient performance on Intel KNL}},}\ }\href {\doibase 10.22323/1.256.0270}
  {\bibfield  {journal} {\bibinfo  {journal} {PoS}\ }\textbf {\bibinfo {volume}
  {LATTICE2016}},\ \bibinfo {pages} {270} (\bibinfo {year} {2016})},\ \Eprint
  {http://arxiv.org/abs/1611.00728} {arXiv:1611.00728 [hep-lat]} \BibitemShut
  {NoStop}%
\bibitem [{\citenamefont {Jo{\'{o}}}\ \emph {et~al.}(2016)\citenamefont
  {Jo{\'{o}}}, \citenamefont {Kalamkar}, \citenamefont {Kurth}, \citenamefont
  {Vaidyanathan},\ and\ \citenamefont
  {Walden}}]{DBLP:conf/supercomputer/JooKKVW16}%
  \BibitemOpen
  \bibfield  {author} {\bibinfo {author} {\bibfnamefont {B{\'{a}}lint}\
  \bibnamefont {Jo{\'{o}}}}, \bibinfo {author} {\bibfnamefont {Dhiraj~D.}\
  \bibnamefont {Kalamkar}}, \bibinfo {author} {\bibfnamefont {Thorsten}\
  \bibnamefont {Kurth}}, \bibinfo {author} {\bibfnamefont {Karthikeyan}\
  \bibnamefont {Vaidyanathan}}, \ and\ \bibinfo {author} {\bibfnamefont
  {Aaron}\ \bibnamefont {Walden}},\ }\bibfield  {title} {\enquote {\bibinfo
  {title} {Optimizing {Wilson-Dirac} operator and linear solvers for
  {Intel\textregistered} {KNL}},}\ }in\ \href {\doibase
  10.1007/978-3-319-46079-6\_30} {\emph {\bibinfo {booktitle} {High Performance
  Computing - {ISC} High Performance 2016 International Workshops, ExaComm,
  E-MuCoCoS, HPC-IODC, IXPUG, IWOPH, P{\^{}}3MA, VHPC, WOPSSS, Frankfurt,
  Germany, June 19-23, 2016, Revised Selected Papers}}}\ (\bibinfo  {publisher}
  {Springer International Publishing},\ \bibinfo {year} {2016})\ pp.\ \bibinfo
  {pages} {415--427}\BibitemShut {NoStop}%
\bibitem [{\citenamefont {Boyle}\ \emph {et~al.}(2017)\citenamefont {Boyle},
  \citenamefont {Chuvelev}, \citenamefont {Cossu}, \citenamefont {Kelly},
  \citenamefont {Lehner},\ and\ \citenamefont {Meadows}}]{Boyle:2017xcy}%
  \BibitemOpen
  \bibfield  {author} {\bibinfo {author} {\bibfnamefont {Peter}\ \bibnamefont
  {Boyle}}, \bibinfo {author} {\bibfnamefont {Michael}\ \bibnamefont
  {Chuvelev}}, \bibinfo {author} {\bibfnamefont {Guido}\ \bibnamefont {Cossu}},
  \bibinfo {author} {\bibfnamefont {Christopher}\ \bibnamefont {Kelly}},
  \bibinfo {author} {\bibfnamefont {Christoph}\ \bibnamefont {Lehner}}, \ and\
  \bibinfo {author} {\bibfnamefont {Lawrence}\ \bibnamefont {Meadows}},\
  }\bibfield  {title} {\enquote {\bibinfo {title} {{Accelerating HPC codes on
  Intel(R) Omni-Path architecture networks: From particle physics to machine
  learning}},}\ }\href@noop {} {\  (\bibinfo {year} {2017})},\ \Eprint
  {http://arxiv.org/abs/1711.04883} {arXiv:1711.04883 [cs.DC]} \BibitemShut
  {NoStop}%
\bibitem [{\citenamefont {Boyle}(2017)}]{Boyle:2017wul}%
  \BibitemOpen
  \bibfield  {author} {\bibinfo {author} {\bibfnamefont {Peter~A}\ \bibnamefont
  {Boyle}},\ }\bibfield  {title} {\enquote {\bibinfo {title} {Machines and
  algorithms},}\ }\href {\doibase 10.22323/1.256.0013} {\bibfield  {journal}
  {\bibinfo  {journal} {PoS}\ }\textbf {\bibinfo {volume} {LATTICE2016}},\
  \bibinfo {pages} {013} (\bibinfo {year} {2017})},\ \Eprint
  {http://arxiv.org/abs/1702.00208} {arXiv:1702.00208 [hep-lat]} \BibitemShut
  {NoStop}%
\bibitem [{\citenamefont {Strohmaier}\ \emph {et~al.}(2018)\citenamefont
  {Strohmaier}, \citenamefont {Simon}, \citenamefont {Dongarra},\ and\
  \citenamefont {Meuer}}]{Top500}%
  \BibitemOpen
  \bibfield  {author} {\bibinfo {author} {\bibfnamefont {E.}~\bibnamefont
  {Strohmaier}}, \bibinfo {author} {\bibfnamefont {H.}~\bibnamefont {Simon}},
  \bibinfo {author} {\bibfnamefont {J.}~\bibnamefont {Dongarra}}, \ and\
  \bibinfo {author} {\bibfnamefont {M.}~\bibnamefont {Meuer}},\ }\href@noop {}
  {\enquote {\bibinfo {title} {{Top 500 List, November 2018}},}\ }\bibinfo
  {howpublished} {\url{https://www.top500.org/lists/2018/11}} (\bibinfo {year}
  {2018})\BibitemShut {NoStop}%
\bibitem [{\citenamefont {NERSC}()}]{Perlmutter}%
  \BibitemOpen
  \bibfield  {author} {\bibinfo {author} {\bibnamefont {NERSC}},\ }\href@noop
  {} {\enquote {\bibinfo {title} {{Perlmutter Web Page}},}\ }\bibinfo
  {howpublished} {\url{http://www.nersc.gov/systems/perlmutter/}}\BibitemShut
  {NoStop}%
\bibitem [{\citenamefont {ALCF}({\natexlab{a}})}]{Aurora}%
  \BibitemOpen
  \bibfield  {author} {\bibinfo {author} {\bibnamefont {ALCF}},\ }\href@noop {}
  {\enquote {\bibinfo {title} {{Aurora}},}\ }\bibinfo {howpublished}
  {\url{https://aurora.alcf.anl.gov/}} ({\natexlab{a}})\BibitemShut {NoStop}%
\bibitem [{\citenamefont {OLCF}()}]{Frontier}%
  \BibitemOpen
  \bibfield  {author} {\bibinfo {author} {\bibnamefont {OLCF}},\ }\href@noop {}
  {\enquote {\bibinfo {title} {{Frontier: OLCF's Exascale Future}},}\ }\bibinfo
  {howpublished}
  {\url{https://www.olcf.ornl.gov/2018/02/13/frontier-olcfs-exascale-future}}\BibitemShut
  {NoStop}%
\bibitem [{\citenamefont {ALCF}({\natexlab{b}})}]{AuroraESP}%
  \BibitemOpen
  \bibfield  {author} {\bibinfo {author} {\bibnamefont {ALCF}},\ }\href@noop {}
  {\enquote {\bibinfo {title} {Alcf aurora 2021 early science program: Data and
  learning call for proposals},}\ }\bibinfo {howpublished}
  {\url{https://www.alcf.anl.gov/alcf-aurora-2021-early-science-program-data-and-learning-call-proposals}}
  ({\natexlab{b}})\BibitemShut {NoStop}%
\bibitem [{\citenamefont {{Kurth}}\ \emph {et~al.}(2018)\citenamefont
  {{Kurth}}, \citenamefont {{Treichler}}, \citenamefont {{Romero}},
  \citenamefont {{Mudigonda}}, \citenamefont {{Luehr}}, \citenamefont
  {{Phillips}}, \citenamefont {{Mahesh}}, \citenamefont {{Matheson}},
  \citenamefont {{Deslippe}}, \citenamefont {{Fatica}}, \citenamefont
  {{Prabhat}},\ and\ \citenamefont {{Houston}}}]{2018arXiv181001993K}%
  \BibitemOpen
  \bibfield  {author} {\bibinfo {author} {\bibfnamefont {T.}~\bibnamefont
  {{Kurth}}}, \bibinfo {author} {\bibfnamefont {S.}~\bibnamefont
  {{Treichler}}}, \bibinfo {author} {\bibfnamefont {J.}~\bibnamefont
  {{Romero}}}, \bibinfo {author} {\bibfnamefont {M.}~\bibnamefont
  {{Mudigonda}}}, \bibinfo {author} {\bibfnamefont {N.}~\bibnamefont
  {{Luehr}}}, \bibinfo {author} {\bibfnamefont {E.}~\bibnamefont {{Phillips}}},
  \bibinfo {author} {\bibfnamefont {A.}~\bibnamefont {{Mahesh}}}, \bibinfo
  {author} {\bibfnamefont {M.}~\bibnamefont {{Matheson}}}, \bibinfo {author}
  {\bibfnamefont {J.}~\bibnamefont {{Deslippe}}}, \bibinfo {author}
  {\bibfnamefont {M.}~\bibnamefont {{Fatica}}}, \bibinfo {author} {\bibnamefont
  {{Prabhat}}}, \ and\ \bibinfo {author} {\bibfnamefont {M.}~\bibnamefont
  {{Houston}}},\ }\bibfield  {title} {\enquote {\bibinfo {title} {Exascale deep
  learning for climate analytics},}\ }\href@noop {} {\  (\bibinfo {year}
  {2018})},\ \Eprint {http://arxiv.org/abs/1810.01993} {arXiv:1810.01993
  [cs.DC]} \BibitemShut {NoStop}%
\bibitem [{\citenamefont {Corporation}()}]{KNM}%
  \BibitemOpen
  \bibfield  {author} {\bibinfo {author} {\bibfnamefont {Intel}\ \bibnamefont
  {Corporation}},\ }\href@noop {} {\enquote {\bibinfo {title} {Intel unveils
  strategy for state-of-the-art artificial intelligence},}\ }\bibinfo
  {howpublished}
  {\url{https://newsroom.intel.com/news-releases/intel-ai-day-news-release/}}\BibitemShut
  {NoStop}%
\bibitem [{\citenamefont {Boyle}(2009)}]{Boyle2009TheBA}%
  \BibitemOpen
  \bibfield  {author} {\bibinfo {author} {\bibfnamefont {Peter~A.}\
  \bibnamefont {Boyle}},\ }\bibfield  {title} {\enquote {\bibinfo {title} {The
  {BAGEL} assembler generation library},}\ }\href@noop {} {\bibfield  {journal}
  {\bibinfo  {journal} {Comp. Phys. Commun.}\ }\textbf {\bibinfo {volume}
  {180}},\ \bibinfo {pages} {2739--2748} (\bibinfo {year} {2009})}\BibitemShut
  {NoStop}%
\bibitem [{\citenamefont {Pochinsky}()}]{MDWFWeb}%
  \BibitemOpen
  \bibfield  {author} {\bibinfo {author} {\bibfnamefont {A.~V.}\ \bibnamefont
  {Pochinsky}},\ }\href@noop {} {\enquote {\bibinfo {title} {{M\"obius domain
  wall fermion inverter}},}\ }\bibinfo {howpublished}
  {\url{http://www.mit.edu/~avp/mdwf}}\BibitemShut {NoStop}%
\bibitem [{\citenamefont {Carter~Edwards}\ \emph {et~al.}(2014)\citenamefont
  {Carter~Edwards}, \citenamefont {Trott},\ and\ \citenamefont
  {Sunderland}}]{CarterEdwards:2014:KOK:2841458.2841785}%
  \BibitemOpen
  \bibfield  {author} {\bibinfo {author} {\bibfnamefont {H.}~\bibnamefont
  {Carter~Edwards}}, \bibinfo {author} {\bibfnamefont {Christian~R.}\
  \bibnamefont {Trott}}, \ and\ \bibinfo {author} {\bibfnamefont {Daniel}\
  \bibnamefont {Sunderland}},\ }\bibfield  {title} {\enquote {\bibinfo {title}
  {Kokkos},}\ }\href {\doibase 10.1016/j.jpdc.2014.07.003} {\bibfield
  {journal} {\bibinfo  {journal} {J. Parallel Distrib. Comput.}\ }\textbf
  {\bibinfo {volume} {74}},\ \bibinfo {pages} {3202--3216} (\bibinfo {year}
  {2014})}\BibitemShut {NoStop}%
\bibitem [{\citenamefont {Jo\'o}()}]{MGPROTO}%
  \BibitemOpen
  \bibfield  {author} {\bibinfo {author} {\bibfnamefont {B.}~\bibnamefont
  {Jo\'o}},\ }\href@noop {} {\enquote {\bibinfo {title} {{mg\_proto: a
  prototype multi-grid library for QCD}},}\ }\bibinfo {howpublished}
  {\url{https://github.com/jeffersonlab/mg\_proto}}\BibitemShut {NoStop}%
\bibitem [{\citenamefont {Boyle}\ \emph {et~al.}(2015)\citenamefont {Boyle},
  \citenamefont {Yamaguchi}, \citenamefont {Cossu},\ and\ \citenamefont
  {Portelli}}]{Boyle:2015tjk}%
  \BibitemOpen
  \bibfield  {author} {\bibinfo {author} {\bibfnamefont {Peter}\ \bibnamefont
  {Boyle}}, \bibinfo {author} {\bibfnamefont {Azusa}\ \bibnamefont
  {Yamaguchi}}, \bibinfo {author} {\bibfnamefont {Guido}\ \bibnamefont
  {Cossu}}, \ and\ \bibinfo {author} {\bibfnamefont {Antonin}\ \bibnamefont
  {Portelli}},\ }\bibfield  {title} {\enquote {\bibinfo {title} {{Grid: A next
  generation data parallel C++ QCD library}},}\ }\href@noop {} {\  (\bibinfo
  {year} {2015})},\ \Eprint {http://arxiv.org/abs/1512.03487} {arXiv:1512.03487
  [hep-lat]} \BibitemShut {NoStop}%
\bibitem [{\citenamefont {Jin}\ and\ \citenamefont
  {Osborn}(2016)}]{Jin:2016ioq}%
  \BibitemOpen
  \bibfield  {author} {\bibinfo {author} {\bibfnamefont {Xiao-Yong}\
  \bibnamefont {Jin}}\ and\ \bibinfo {author} {\bibfnamefont {James~C.}\
  \bibnamefont {Osborn}},\ }\bibfield  {title} {\enquote {\bibinfo {title}
  {{QEX: a framework for lattice field theories}},}\ }\href {\doibase
  10.22323/1.282.0187} {\bibfield  {journal} {\bibinfo  {journal} {PoS}\
  }\textbf {\bibinfo {volume} {ICHEP2016}},\ \bibinfo {pages} {187} (\bibinfo
  {year} {2016})},\ \Eprint {http://arxiv.org/abs/1612.02750} {arXiv:1612.02750
  [hep-lat]} \BibitemShut {NoStop}%
\bibitem [{\citenamefont {Mehta}\ \emph {et~al.}(2018)\citenamefont {Mehta},
  \citenamefont {Bukov}, \citenamefont {Wang}, \citenamefont {Day},
  \citenamefont {Richardson}, \citenamefont {Fisher},\ and\ \citenamefont
  {Schwab}}]{Mehta:2018dln}%
  \BibitemOpen
  \bibfield  {author} {\bibinfo {author} {\bibfnamefont {Pankaj}\ \bibnamefont
  {Mehta}}, \bibinfo {author} {\bibfnamefont {Marin}\ \bibnamefont {Bukov}},
  \bibinfo {author} {\bibfnamefont {Ching-Hao}\ \bibnamefont {Wang}}, \bibinfo
  {author} {\bibfnamefont {Alexandre G.~R.}\ \bibnamefont {Day}}, \bibinfo
  {author} {\bibfnamefont {Clint}\ \bibnamefont {Richardson}}, \bibinfo
  {author} {\bibfnamefont {Charles~K.}\ \bibnamefont {Fisher}}, \ and\ \bibinfo
  {author} {\bibfnamefont {David~J.}\ \bibnamefont {Schwab}},\ }\bibfield
  {title} {\enquote {\bibinfo {title} {A high-bias, low-variance introduction
  to machine learning for physicists},}\ }\href@noop {} {\  (\bibinfo {year}
  {2018})},\ \Eprint {http://arxiv.org/abs/1803.08823} {arXiv:1803.08823
  [physics.comp-ph]} \BibitemShut {NoStop}%
\bibitem [{\citenamefont {Goodfellow}\ \emph {et~al.}(2016)\citenamefont
  {Goodfellow}, \citenamefont {Bengio},\ and\ \citenamefont
  {Courville}}]{Goodfellow-et-al-2016}%
  \BibitemOpen
  \bibfield  {author} {\bibinfo {author} {\bibfnamefont {Ian}\ \bibnamefont
  {Goodfellow}}, \bibinfo {author} {\bibfnamefont {Yoshua}\ \bibnamefont
  {Bengio}}, \ and\ \bibinfo {author} {\bibfnamefont {Aaron}\ \bibnamefont
  {Courville}},\ }\href@noop {} {\emph {\bibinfo {title} {Deep Learning}}}\
  (\bibinfo  {publisher} {MIT Press},\ \bibinfo {year} {2016})\ \bibinfo {note}
  {\url{http://www.deeplearningbook.org}}\BibitemShut {NoStop}%
\bibitem [{\citenamefont {Albergo}\ \emph {et~al.}(2019)\citenamefont
  {Albergo}, \citenamefont {Kanwar},\ and\ \citenamefont
  {Shanahan}}]{Albergo:2019eim}%
  \BibitemOpen
  \bibfield  {author} {\bibinfo {author} {\bibfnamefont {M.~S.}\ \bibnamefont
  {Albergo}}, \bibinfo {author} {\bibfnamefont {G.}~\bibnamefont {Kanwar}}, \
  and\ \bibinfo {author} {\bibfnamefont {P.~E.}\ \bibnamefont {Shanahan}},\
  }\bibfield  {title} {\enquote {\bibinfo {title} {{Flow-based generative
  models for Markov chain Monte Carlo in lattice field theory}},}\ }\href
  {\doibase 10.1103/PhysRevD.100.034515} {\bibfield  {journal} {\bibinfo
  {journal} {Phys. Rev.}\ }\textbf {\bibinfo {volume} {D100}},\ \bibinfo
  {pages} {034515} (\bibinfo {year} {2019})},\ \Eprint
  {http://arxiv.org/abs/1904.12072} {arXiv:1904.12072 [hep-lat]} \BibitemShut
  {NoStop}%
\bibitem [{\citenamefont {Shanahan}\ \emph {et~al.}(2018)\citenamefont
  {Shanahan}, \citenamefont {Trewartha},\ and\ \citenamefont
  {Detmold}}]{Shanahan:2018vcv}%
  \BibitemOpen
  \bibfield  {author} {\bibinfo {author} {\bibfnamefont {Phiala~E.}\
  \bibnamefont {Shanahan}}, \bibinfo {author} {\bibfnamefont {Daniel}\
  \bibnamefont {Trewartha}}, \ and\ \bibinfo {author} {\bibfnamefont {William}\
  \bibnamefont {Detmold}},\ }\bibfield  {title} {\enquote {\bibinfo {title}
  {{Machine learning action parameters in lattice quantum chromodynamics}},}\
  }\href@noop {} {\  (\bibinfo {year} {2018})},\ \Eprint
  {http://arxiv.org/abs/1801.05784} {arXiv:1801.05784 [hep-lat]} \BibitemShut
  {NoStop}%
\bibitem [{\citenamefont {Tanaka}\ and\ \citenamefont
  {Tomiya}(2017)}]{Tanaka:2017niz}%
  \BibitemOpen
  \bibfield  {author} {\bibinfo {author} {\bibfnamefont {Akinori}\ \bibnamefont
  {Tanaka}}\ and\ \bibinfo {author} {\bibfnamefont {Akio}\ \bibnamefont
  {Tomiya}},\ }\bibfield  {title} {\enquote {\bibinfo {title} {{Towards
  reduction of autocorrelation in HMC by machine learning}},}\ }\href@noop {}
  {\  (\bibinfo {year} {2017})},\ \Eprint {http://arxiv.org/abs/1712.03893}
  {arXiv:1712.03893 [hep-lat]} \BibitemShut {NoStop}%
\bibitem [{\citenamefont {Wang}(2017)}]{Wang:2017mzw}%
  \BibitemOpen
  \bibfield  {author} {\bibinfo {author} {\bibfnamefont {Lei}\ \bibnamefont
  {Wang}},\ }\bibfield  {title} {\enquote {\bibinfo {title} {{Exploring cluster
  Monte Carlo updates with Boltzmann machines}},}\ }\href {\doibase
  10.1103/PhysRevE.96.051301} {\bibfield  {journal} {\bibinfo  {journal} {Phys.
  Rev.}\ }\textbf {\bibinfo {volume} {E96}},\ \bibinfo {pages} {051301}
  (\bibinfo {year} {2017})},\ \Eprint {http://arxiv.org/abs/1702.08586}
  {arXiv:1702.08586 [physics.comp-ph]} \BibitemShut {NoStop}%
\bibitem [{\citenamefont {Beyl}\ \emph {et~al.}(2018)\citenamefont {Beyl},
  \citenamefont {Goth},\ and\ \citenamefont {Assaad}}]{Beyl:2017kwp}%
  \BibitemOpen
  \bibfield  {author} {\bibinfo {author} {\bibfnamefont {Stefan}\ \bibnamefont
  {Beyl}}, \bibinfo {author} {\bibfnamefont {Florian}\ \bibnamefont {Goth}}, \
  and\ \bibinfo {author} {\bibfnamefont {Fakher~F.}\ \bibnamefont {Assaad}},\
  }\bibfield  {title} {\enquote {\bibinfo {title} {Revisiting the hybrid
  quantum {Monte Carlo} method for {Hubbard} and electron-phonon models},}\
  }\href {\doibase 10.1103/PhysRevB.97.085144} {\bibfield  {journal} {\bibinfo
  {journal} {Phys. Rev.}\ }\textbf {\bibinfo {volume} {B97}},\ \bibinfo {pages}
  {085144} (\bibinfo {year} {2018})},\ \Eprint
  {http://arxiv.org/abs/1708.03661} {arXiv:1708.03661 [cond-mat.str-el]}
  \BibitemShut {NoStop}%
\bibitem [{\citenamefont {Xu}\ \emph {et~al.}(2017)\citenamefont {Xu},
  \citenamefont {Qi}, \citenamefont {Liu}, \citenamefont {Fu},\ and\
  \citenamefont {Meng}}]{Xu:2017vug}%
  \BibitemOpen
  \bibfield  {author} {\bibinfo {author} {\bibfnamefont {Xiao~Yan}\
  \bibnamefont {Xu}}, \bibinfo {author} {\bibfnamefont {Yang}\ \bibnamefont
  {Qi}}, \bibinfo {author} {\bibfnamefont {Junwei}\ \bibnamefont {Liu}},
  \bibinfo {author} {\bibfnamefont {Liang}\ \bibnamefont {Fu}}, \ and\ \bibinfo
  {author} {\bibfnamefont {Zi~Yang}\ \bibnamefont {Meng}},\ }\bibfield  {title}
  {\enquote {\bibinfo {title} {{Self-learning quantum Monte Carlo method in
  interacting fermion systems}},}\ }\href {\doibase 10.1103/PhysRevB.96.041119}
  {\bibfield  {journal} {\bibinfo  {journal} {Phys. Rev.}\ }\textbf {\bibinfo
  {volume} {B96}},\ \bibinfo {pages} {041119} (\bibinfo {year} {2017})},\
  \Eprint {http://arxiv.org/abs/1612.03804} {arXiv:1612.03804
  [cond-mat.str-el]} \BibitemShut {NoStop}%
\bibitem [{\citenamefont {{Yoon}}(2016)}]{2016arXiv160605560Y}%
  \BibitemOpen
  \bibfield  {author} {\bibinfo {author} {\bibfnamefont {Boram}\ \bibnamefont
  {{Yoon}}},\ }\bibfield  {title} {\enquote {\bibinfo {title} {{Estimation of
  matrix trace using machine learning}},}\ }\href@noop {} {\  (\bibinfo {year}
  {2016})},\ \Eprint {http://arxiv.org/abs/1606.05560} {arXiv:1606.05560
  [stat.ML]} \BibitemShut {NoStop}%
\bibitem [{\citenamefont {Yoon}(2018)}]{Yoon:2018krb}%
  \BibitemOpen
  \bibfield  {author} {\bibinfo {author} {\bibfnamefont {Boram}\ \bibnamefont
  {Yoon}},\ }\bibfield  {title} {\enquote {\bibinfo {title} {Machine learning
  estimators for lattice {QCD} observables},}\ }\href@noop {} {\  (\bibinfo
  {year} {2018})},\ \Eprint {http://arxiv.org/abs/1807.05971} {arXiv:1807.05971
  [hep-lat]} \BibitemShut {NoStop}%
\bibitem [{\citenamefont {Carrasquilla}\ and\ \citenamefont
  {Melko}(2017)}]{2017NatPh..13..431C}%
  \BibitemOpen
  \bibfield  {author} {\bibinfo {author} {\bibfnamefont {J.}~\bibnamefont
  {Carrasquilla}}\ and\ \bibinfo {author} {\bibfnamefont {R.~G.}\ \bibnamefont
  {Melko}},\ }\bibfield  {title} {\enquote {\bibinfo {title} {Machine learning
  phases of matter},}\ }\href {\doibase 10.1038/nphys4035} {\bibfield
  {journal} {\bibinfo  {journal} {Nature Phys.}\ }\textbf {\bibinfo {volume}
  {13}},\ \bibinfo {pages} {431--434} (\bibinfo {year} {2017})},\ \Eprint
  {http://arxiv.org/abs/1605.01735} {arXiv:1605.01735 [cond-mat.str-el]}
  \BibitemShut {NoStop}%
\bibitem [{\citenamefont {Torlai}\ and\ \citenamefont
  {Melko}(2016)}]{2016PhRvB..94p5134T}%
  \BibitemOpen
  \bibfield  {author} {\bibinfo {author} {\bibfnamefont {G.}~\bibnamefont
  {Torlai}}\ and\ \bibinfo {author} {\bibfnamefont {R.~G.}\ \bibnamefont
  {Melko}},\ }\bibfield  {title} {\enquote {\bibinfo {title} {Learning
  thermodynamics with {Boltzmann} machines},}\ }\href {\doibase
  10.1103/PhysRevB.94.165134} {\bibfield  {journal} {\bibinfo  {journal} {Phys.
  Rev.}\ }\textbf {\bibinfo {volume} {B94}},\ \bibinfo {pages} {165134}
  (\bibinfo {year} {2016})},\ \Eprint {http://arxiv.org/abs/1606.02718}
  {arXiv:1606.02718 [cond-mat.stat-mech]} \BibitemShut {NoStop}%
\bibitem [{\citenamefont {{Ch’ng}}\ \emph {et~al.}(2017)\citenamefont
  {{Ch’ng}}, \citenamefont {{Carrasquilla}}, \citenamefont {{Melko}},\ and\
  \citenamefont {{Khatami}}}]{2017PhRvX...7c1038C}%
  \BibitemOpen
  \bibfield  {author} {\bibinfo {author} {\bibfnamefont {K.}~\bibnamefont
  {{Ch’ng}}}, \bibinfo {author} {\bibfnamefont {J.}~\bibnamefont
  {{Carrasquilla}}}, \bibinfo {author} {\bibfnamefont {R.~G.}\ \bibnamefont
  {{Melko}}}, \ and\ \bibinfo {author} {\bibfnamefont {E.}~\bibnamefont
  {{Khatami}}},\ }\bibfield  {title} {\enquote {\bibinfo {title} {Machine
  learning phases of strongly correlated fermions},}\ }\href {\doibase
  10.1103/PhysRevX.7.031038} {\bibfield  {journal} {\bibinfo  {journal} {Phys.
  Rev.}\ }\textbf {\bibinfo {volume} {X7}},\ \bibinfo {pages} {031038}
  (\bibinfo {year} {2017})},\ \Eprint {http://arxiv.org/abs/1609.02552}
  {arXiv:1609.02552 [cond-mat.str-el]} \BibitemShut {NoStop}%
\bibitem [{\citenamefont {Li}\ \emph {et~al.}(2018)\citenamefont {Li},
  \citenamefont {Tan},\ and\ \citenamefont {Jiang}}]{Li:2017xaz}%
  \BibitemOpen
  \bibfield  {author} {\bibinfo {author} {\bibfnamefont {Chian-De}\
  \bibnamefont {Li}}, \bibinfo {author} {\bibfnamefont {Deng-Ruei}\
  \bibnamefont {Tan}}, \ and\ \bibinfo {author} {\bibfnamefont {Fu-Jiun}\
  \bibnamefont {Jiang}},\ }\bibfield  {title} {\enquote {\bibinfo {title}
  {{Applications of neural networks to the studies of phase transitions of
  two-dimensional Potts models}},}\ }\href {\doibase 10.1016/j.aop.2018.02.018}
  {\bibfield  {journal} {\bibinfo  {journal} {Annals Phys.}\ }\textbf {\bibinfo
  {volume} {391}},\ \bibinfo {pages} {312--331} (\bibinfo {year} {2018})},\
  \Eprint {http://arxiv.org/abs/1703.02369} {arXiv:1703.02369
  [cond-mat.dis-nn]} \BibitemShut {NoStop}%
\bibitem [{\citenamefont {Wetzel}\ and\ \citenamefont
  {Scherzer}(2017)}]{Wetzel:2017ooo}%
  \BibitemOpen
  \bibfield  {author} {\bibinfo {author} {\bibfnamefont {Sebastian~Johann}\
  \bibnamefont {Wetzel}}\ and\ \bibinfo {author} {\bibfnamefont {Manuel}\
  \bibnamefont {Scherzer}},\ }\bibfield  {title} {\enquote {\bibinfo {title}
  {Machine learning of explicit order parameters: From the {Ising} model to
  {SU(2)} lattice gauge theory},}\ }\href {\doibase 10.1103/PhysRevB.96.184410}
  {\bibfield  {journal} {\bibinfo  {journal} {Phys. Rev.}\ }\textbf {\bibinfo
  {volume} {B96}},\ \bibinfo {pages} {184410} (\bibinfo {year} {2017})},\
  \Eprint {http://arxiv.org/abs/1705.05582} {arXiv:1705.05582
  [cond-mat.stat-mech]} \BibitemShut {NoStop}%
\bibitem [{\citenamefont {Alexandru}\ \emph {et~al.}(2017)\citenamefont
  {Alexandru}, \citenamefont {Bedaque}, \citenamefont {Lamm},\ and\
  \citenamefont {Lawrence}}]{Alexandru:2017czx}%
  \BibitemOpen
  \bibfield  {author} {\bibinfo {author} {\bibfnamefont {Andrei}\ \bibnamefont
  {Alexandru}}, \bibinfo {author} {\bibfnamefont {Paulo~F.}\ \bibnamefont
  {Bedaque}}, \bibinfo {author} {\bibfnamefont {Henry}\ \bibnamefont {Lamm}}, \
  and\ \bibinfo {author} {\bibfnamefont {Scott}\ \bibnamefont {Lawrence}},\
  }\bibfield  {title} {\enquote {\bibinfo {title} {Deep learning beyond
  {Lefschetz} thimbles},}\ }\href {\doibase 10.1103/PhysRevD.96.094505}
  {\bibfield  {journal} {\bibinfo  {journal} {Phys. Rev.}\ }\textbf {\bibinfo
  {volume} {D96}},\ \bibinfo {pages} {094505} (\bibinfo {year} {2017})},\
  \Eprint {http://arxiv.org/abs/1709.01971} {arXiv:1709.01971 [hep-lat]}
  \BibitemShut {NoStop}%
\bibitem [{\citenamefont {Mori}\ \emph {et~al.}(2017)\citenamefont {Mori},
  \citenamefont {Kashiwa},\ and\ \citenamefont {Ohnishi}}]{Mori:2017pne}%
  \BibitemOpen
  \bibfield  {author} {\bibinfo {author} {\bibfnamefont {Yuto}\ \bibnamefont
  {Mori}}, \bibinfo {author} {\bibfnamefont {Kouji}\ \bibnamefont {Kashiwa}}, \
  and\ \bibinfo {author} {\bibfnamefont {Akira}\ \bibnamefont {Ohnishi}},\
  }\bibfield  {title} {\enquote {\bibinfo {title} {{Toward solving the sign
  problem with path optimization method}},}\ }\href {\doibase
  10.1103/PhysRevD.96.111501} {\bibfield  {journal} {\bibinfo  {journal} {Phys.
  Rev.}\ }\textbf {\bibinfo {volume} {D96}},\ \bibinfo {pages} {111501}
  (\bibinfo {year} {2017})},\ \Eprint {http://arxiv.org/abs/1705.05605}
  {arXiv:1705.05605 [hep-lat]} \BibitemShut {NoStop}%
\bibitem [{\citenamefont {Mori}\ \emph {et~al.}(2018)\citenamefont {Mori},
  \citenamefont {Kashiwa},\ and\ \citenamefont {Ohnishi}}]{Mori:2017nwj}%
  \BibitemOpen
  \bibfield  {author} {\bibinfo {author} {\bibfnamefont {Yuto}\ \bibnamefont
  {Mori}}, \bibinfo {author} {\bibfnamefont {Kouji}\ \bibnamefont {Kashiwa}}, \
  and\ \bibinfo {author} {\bibfnamefont {Akira}\ \bibnamefont {Ohnishi}},\
  }\bibfield  {title} {\enquote {\bibinfo {title} {{Application of a neural
  network to the sign problem via the path optimization method}},}\ }\href
  {\doibase 10.1093/ptep/ptx191} {\bibfield  {journal} {\bibinfo  {journal}
  {PTEP}\ }\textbf {\bibinfo {volume} {2018}},\ \bibinfo {pages} {023B04}
  (\bibinfo {year} {2018})},\ \Eprint {http://arxiv.org/abs/1709.03208}
  {arXiv:1709.03208 [hep-lat]} \BibitemShut {NoStop}%
\bibitem [{\citenamefont {Preskill}(2018)}]{Preskill2018quantumcomputingin}%
  \BibitemOpen
  \bibfield  {author} {\bibinfo {author} {\bibfnamefont {John}\ \bibnamefont
  {Preskill}},\ }\bibfield  {title} {\enquote {\bibinfo {title} {Quantum
  {C}omputing in the {NISQ} era and beyond},}\ }\href {\doibase
  10.22331/q-2018-08-06-79} {\bibfield  {journal} {\bibinfo  {journal}
  {{Quantum}}\ }\textbf {\bibinfo {volume} {2}},\ \bibinfo {pages} {79}
  (\bibinfo {year} {2018})}\BibitemShut {NoStop}%
\bibitem [{\citenamefont {Carlson}\ \emph {et~al.}(2018)\citenamefont
  {Carlson}, \citenamefont {Dean}, \citenamefont {Hjorth-Jensen}, \citenamefont
  {Kaplan}, \citenamefont {Preskill}, \citenamefont {Roche}, \citenamefont
  {Savage},\ and\ \citenamefont {Troyer}}]{Carlson:2018doe}%
  \BibitemOpen
  \bibfield  {author} {\bibinfo {author} {\bibfnamefont {Joseph}\ \bibnamefont
  {Carlson}}, \bibinfo {author} {\bibfnamefont {David~J.}\ \bibnamefont
  {Dean}}, \bibinfo {author} {\bibfnamefont {Morten}\ \bibnamefont
  {Hjorth-Jensen}}, \bibinfo {author} {\bibfnamefont {David}\ \bibnamefont
  {Kaplan}}, \bibinfo {author} {\bibfnamefont {John}\ \bibnamefont {Preskill}},
  \bibinfo {author} {\bibfnamefont {Kenneth}\ \bibnamefont {Roche}}, \bibinfo
  {author} {\bibfnamefont {Martin~J.}\ \bibnamefont {Savage}}, \ and\ \bibinfo
  {author} {\bibfnamefont {Matthias}\ \bibnamefont {Troyer}},\ }\bibfield
  {title} {\enquote {\bibinfo {title} {Quantum computing for theoretical
  nuclear physics},}\ }\href@noop {} {\  (\bibinfo {year} {2018})},\ \bibinfo
  {note}
  {\href{http://www.int.washington.edu/PROGRAMS/17-66W/QuantumComputing_NUCLEARPHYSICS_FINAL_pdf.pdf}{Institute
  for Nuclear Theory report 18-008}}\BibitemShut {NoStop}%
\end{thebibliography}%

\end{document}